\documentclass[a4paper,fleqn,usenatbib]{mnras}

\usepackage{float}
\usepackage[T1]{fontenc}
\usepackage{aecompl}
\usepackage{graphicx,color,bm}
\usepackage{latexsym}
\usepackage{epsfig}
\usepackage{amssymb}
\usepackage{amsmath}
\usepackage{wasysym}
\usepackage{pstricks}
\usepackage{enumitem}

\newcommand{\ramses}{{\sc ramses}}
\newcommand{\isis}{{\sc{isis}}}
\newcommand{\isisns}{{\sc{isis-nonstatic}}}
\newcommand{\ecosmog}{{\sc ecosmog}}
\newcommand{\dgpm}{{\sc dgpm}}
\newcommand{\mggadget}{{\sc mg-gadget}}
\newcommand{\gadget}{{\sc gadget2}}

\newcommand{\ahf}{{\sc ahf}}
\newcommand{\powmes}{{\sc powmes}}
\newcommand{\lcdm}{$\Lambda$CDM}
\newcommand{\pgadget}{{\sc p-gadget3}}
\newcommand{\bq}{\begin{eqnarray}}
\newcommand{\eq}{\end{eqnarray}}
\newcommand{\be}{\bq}
\newcommand{\ee}{\eq}

\title{Modified Gravity {\it N}-body Code Comparison Project}

\author[Winther et al.]{
\parbox{\textwidth}{Hans A. Winther$^1$, 
Fabian Schmidt$^2$, 
Alexandre Barreira$^{3,4}$,
Christian Arnold$^{5,6}$,
Sownak Bose$^{3}$,  
Claudio Llinares$^{3,7}$,
Marco Baldi$^{8,9,10}$,
Bridget Falck$^{11}$, 
Wojciech~A.~Hellwing$^{3,12}$,
Kazuya Koyama$^{11}$,
Baojiu Li$^{3}$,
David F. Mota$^7$,  
Ewald Puchwein$^{13}$,
Robert E. Smith$^{15}$ and
Gong-Bo Zhao$^{14,11}$}\vspace{0.4cm}\\\
\parbox{\textwidth}{${}^1$Astrophysics, University of Oxford, DWB, Keble Road, Oxford, OX1 3RH, UK\\
${}^2$Max-Planck-Institute for Astrophysics, D-85748 Garching, Germany\\
${}^3$Institute for Computational Cosmology, Department of Physics, Durham University, Durham DH1 3LE, U.K.\\
${}^4$Institute for Particle Physics Phenomenology, Department of Physics, Durham University, Durham DH1 3LE, U.K.\\
${}^5$Institute for Theoretical Physics, Heidelberg University, Philosophenweg 16, 69120 Heidelberg, Germany\\
${}^6$Heidelberg Institute for Theoretical Studies, Schloss-Wolfsbrunnenweg 35, 69118 Heidelberg, Germany\\
${}^7$Institute of Theoretical Astrophysics, University of Oslo, PO Box 1029 Blindern, 0315 Oslo, Norway\\
$^{8}$Dipartimento di Fisica e Astronomia, Alma Mater Studiorum Universit\`a di Bologna, viale Berti Pichat, 6/2, I-40127 Bologna, Italy\\
$^{9}$INAF - Osservatorio Astronomico di Bologna, via Ranzani 1, I-40127 Bologna, Italy\\
$^{10}$INFN - Sezione di Bologna, viale Berti Pichat 6/2, I-40127 Bologna, Italy\\
$^{11}$ Institute of Cosmology \& Gravitation, University of Portsmouth, Dennis Sciama Building, Portsmouth, PO1 3FX, United Kingdom\\
$^{12}$Interdisciplinary Centre for Mathematical and Computational Modeling (ICM), University of Warsaw, ul. Pawi\'nskiego 5a, Warsaw, Poland\\
$^{13}$Institute of Astronomy and Kavli Institute for Cosmology, University of Cambridge, Madingley Road, Cambridge CB3 0HA, UK\\
$^{14}$National Astronomy Observatories, Chinese Academy of Science, Beijing, 100012, P.R.China\\
$^{15}$Department of Physics and Astronomy, University of Sussex, Brighton BN1 9QH, UK}
}

\date{\today}
\pubyear{2015}

\begin{document}
\label{firstpage}
\pagerange{\pageref{firstpage}--\pageref{lastpage}} 
\maketitle

\begin{abstract}
Self-consistent {\it N}-body simulations of modified gravity models are a key
ingredient to obtain rigorous constraints on deviations from General
Relativity using large-scale structure observations.  This paper provides
the first detailed comparison of the results of different {\it N}-body codes for
the $f(R)$, DGP, and Symmetron models, starting from the same initial
conditions.  We find that the fractional deviation of the matter power 
spectrum from $\Lambda$CDM agrees to better than $1\%$ up to
$k \sim 5-10 h {\rm Mpc}^{-1}$ between the different codes. These codes are thus able to meet the stringent 
accuracy requirements of upcoming observational surveys. All codes are also in good
agreement in their results for the velocity divergence power spectrum, halo abundances and halo profiles.
We also test the quasi-static limit, which is employed in most modified gravity
{\it N}-body codes, for the Symmetron model for which the most significant non-static effects among the models considered are expected.  We conclude that 
this limit is a very good approximation for all of the observables considered here.
\end{abstract}



\section{Introduction}

Cosmology, and in particular observations of the large-scale-structure (LSS),
provide unique possibilities for testing General Relativity (GR) on length scales
that cannot be probed by any other means \citep[see e.g.][for reviews]{Jain/Khoury, 2015arXiv150404623K, 2015arXiv150107274B}.  Motivated
by the observed accelerating expansion of the universe \citep{1998AJ....116.1009R,2005ApJ...633..560E,2013ApJS..208...20B,2015arXiv150201589P}, a number of 
theories have been proposed, in which the acceleration is explained by deviations from GR on large scales
\citep{Carroll/etal,2000PhLB..485..208D,Nicolis2009PhRvD..79f4036N,2010PhRvL.104w1301H}, \citep[see][for a comprehensive review]{Clifton2012}. In almost all cases,
these theories add to the standard massless spin-2 graviton of GR a new light scalar degree of freedom $\varphi$.  
Thus, the Einstein and fluid equations of standard cosmology are augmented by an equation of motion for $\varphi$.  All these models are faced with the challenge of producing sizeable effects on large scales -- the most desirable being a natural explanation of the accelerated expansion -- while at the same time passing the stringent local constraints on modifications to GR \citep{2014LRR....17....4W}.  This requires some form of screening mechanism, that is, a way of dynamically suppressing the effects of the fifth force mediated by $\varphi$ in high-density regions (compared to the cosmological average) where local experiments have tested GR to high precision.  Such an effect is typically realised by nonlinearities in the equation of motion of $\varphi$, which results in a violation of the superposition principle that suppresses the fifth force. The screening leads to a complex interplay between the large scale distribution of matter and the magnitude of the fifth force mediated by $\varphi$.  

The standard tools to compute large-scale structure observables in 
the nonlinear regime are {\it N}-body simulations. Consequently, in order to robustly test
 GR with cosmology, reliable {\it N}-body simulations of modified gravity
models are a necessity.  These simulations must solve the nonlinear equation
of $\varphi$ in conjunction with the Vlasov-Poisson system that is
solved in standard {\it N}-body simulations. The nonlinear nature of the scalar field equation
requires the implementation of novel numerical techniques, which is what makes {\it N}-body simulations of modified gravity so challenging.

To date, a number of codes have been developed to perform simulations of modified gravity. For instance, \cite{2008PhRvD..78l3523O} presented a code that simulates the Hu-Sawicki $f(R)$ model \citep{2007PhRvD..76f4004H} by solving the scalar field equation on a fixed mesh/grid (throughout we use the words mesh and grid interchangeably). Based on the work of \cite{2008PhRvD..78l3523O}, \cite{2009PhRvD..80d3001S,2009PhRvD..80l3003S} developed a code (which we call \dgpm{} here) that performed simulations of the Dvali-Gabadadze-Porrati (DGP) braneworld model \citep{2000PhLB..485..208D}, also on a fixed mesh. Fixed grid simulations of the DGP model were also performed in \cite{2009PhRvD..80j4005C} and \cite{2009PhRvD..80f4023K}. More recently, efforts have been made to simulate modified gravity cosmologies on adaptively refined meshes, which allow for better resolution on small scales, where the effects of the screening are most important. These efforts resulted in the development of a modified version \citep{2009PhRvD..80d4027L, 2011PhRvD..83d4007Z, Li:2010zw, 2011PhRvD..83b4007L} of the {\sc mlapm} code \citep{2001MNRAS.325..845K}, which is a serial {\it N}-body code. The implementation of modified gravity solvers on adaptive mesh refinement (AMR) parallelizable codes was achieved with the development of the \ecosmog{} \citep{2012JCAP...01..051L, 2013JCAP...05..023L, Li2013JCAP...11..012L}, \mggadget{} \citep{2013MNRAS.436..348P} and \isis{} \citep{2013PhRvL.110p1101L, 2014AA...562A..78L} {\it N}-body codes. While careful consistency checks have been performed by the authors of each code (for example, by solving test cases with known analytical solutions), no detailed comparison between codes has been performed so far. A main goal of this paper is precisely to provide a rigorous cross-check of the accuracy of the nontrivial algorithms of these codes.  This is particularly important in light of the stringent accuracy requirements demanded by current and future observational campaigns.

Here, we simulate the $f(R)$, DGP and Symmetron \citep{2010PhRvL.104w1301H} models with the \dgpm{}, \ecosmog{}, \mggadget{} and \isis{} codes. We start all simulations of the different codes from the same initial conditions and compare their results for the matter and velocity divergence power spectra, halo mass function, as well as density, force and velocity profiles of dark matter haloes. For the case of the Symmetron model, we also measure the impact of assuming the quasi-static limit in {\it N}-body simulations of modified gravity, which amounts to neglecting time derivatives of $\varphi$, by comparing with the results of a version of the \isis{} code that explicitly solves for the time evolution of $\varphi$.

An important consideration in such a comparison project relates to determining the target accuracy.  
To guide ourselves in the interpretation of the results, we use the expected accuracy of the next generation
of LSS surveys. For instance, for a mission such as that to be carried out by the 
Euclid satellite\footnote{\href{http://sci.esa.int/euclid/}{http://www.euclid-ec.org/}} \citep{2011arXiv1110.3193L,Amendola2013}, 
the nonlinear matter power spectrum up to a wavenumber $k \sim 5 h\:{\rm Mpc}^{-1}$ should be accurate to $1\%$\footnote{See for example \cite{2011MNRAS.416.1717K} for requirements on the error envelope around the non-linear $P(k)$ for cosmic shear tomography.}.
Since our goal here is to accurately calibrate modified gravity effects,
we aim for an agreement in the fractional change of the matter power
spectrum relative to $\Lambda$CDM of $1\%$ or better.  The agreement
of the different codes on the absolute $\Lambda$CDM predictions is 
not of primary concern here, as there are dedicated comparison projects
for this purpose \citep{2015arXiv150305920S}. We shall also aim for an accuracy of a few per cent
in the code results for velocity statistics, the halo mass function and halo profiles.

Before proceeding, we note that the results presented in this paper contribute to recent efforts in testing and comparing {\it N}-body codes and codes that extract observables from simulations. For instance, there have been detailed comparisons of standard GR {\it N}-body codes \citep{2012MNRAS.423.1726S, 2015arXiv150305920S} and also of codes that identify dark matter haloes \citep{2011MNRAS.415.2293K}, voids \citep{2008MNRAS.387..933C}, halo substructure \citep{2012MNRAS.423.1200O, 2013MNRAS.429.2739O, 2014MNRAS.438.3205P, 2014MNRAS.442.1197H}, galaxies \citep{2013MNRAS.428.2039K}, tidal debris \citep{2013MNRAS.433.1537E}, merger trees \citep{2013MNRAS.436..150S}, halo mock generation \citep{2014arXiv1412.7729C} and galaxy mass reconstruction \citep{2014MNRAS.441.1513O}, just to mention a few. See \cite{2013MNRAS.435.1618K} for a review on the current status of structure finding in {\it N}-body simulations. Comparison projects of these (often complex) numerical techniques are crucial to identify any worrying systematics in the theoretical predictions.

The rest of the paper is organised as follows. In Section~\ref{sect:theo}, we briefly introduce and review the $f(R)$, DGP and Symmetron models that are used in our numerical comparison. In Section~\ref{sec:simulations}, we outline the general numerical techniques that are used to solve the modified gravity equations in a {\it N}-body solver. In Section~\ref{sec:codedes}, we summarise the main features of each of the codes and specific details in how they tackle the equations of the different models. In Section~\ref{sec:results}, we present our comparison results for the matter power spectrum (\ref{sec:power}), the velocity divergence spectrum (\ref{sec:velpower}), the halo mass function (\ref{sec:HMF}), and the density, force and velocity profiles of haloes (\ref{sec:profiles}). We summarise our findings and draw our conclusions in Section~\ref{sec:concl}.


\section{Modified Gravity Theory}
\label{sect:theo}

Alternative models to $\Lambda$CDM (like the modified gravity theories studied here) are numerous \citep[see e.g.][]{Amendola2010,Clifton2012,2015arXiv150404623K}, as are the problems with which they must struggle. Some models are plagued by theoretical instabilities and others require at least some degree of fine-tuning of the model parameters in order to meet observational constraints. One particularly simple extension of GR is the inclusion of a single scalar field $\varphi$ to the standard GR Einstein-Hilbert action. However, when coupled to matter, the scalar field gives rise to an additional gravitational interaction, which is often referred to as a {\it fifth force} \citep[see e.g.][]{Amendola2013,Mota2006,2009PhRvD..80h3522H, Hellwing2013}. This fifth force can be quantified by $\gamma\equiv |{\bf F}_{\rm Fifth}|/|{\bf F}_{\rm N}|$ where ${\bf F}_{\rm N}$ is the standard Newtonian gravitational force that we obtain in the weak-field limit of GR. Several experiments \citep[see e.g.][]{Adelberger2002cls..conf....9A,Berotti2003Natur.425..374B,Williams2004PhRvL..93z1101W, 2014LRR....17....4W} have constrained $\gamma \ll 1$ on Earth and in the Solar System. This seems to leave us with two possible explanations: either the fifth force  is zero on all scales, i.e., $\gamma = 0$, or $\gamma$ is not a constant but instead varies in space and/or time.

Models where $\gamma$ is space-dependent are dubbed \textit{screened modified gravity models}, as one typically desires the fifth force to be screened in high-density environments (like the Solar System). Next, we follow \cite{2015PhR...568....1J} in the classification of the different types of screening mechanisms. A general Lagrangian density for the scalar field can be written schematically as 
\be
{\cal L} = -\frac{1}{2} Z^{\mu \nu} (\varphi, \partial \varphi, \partial^2 \varphi) \partial_{\mu} \varphi \partial_{\nu} \varphi -V(\varphi) 
+ \beta(\varphi) T^{\mu}_{\mu},
\ee
where $Z^{\mu \nu}$ represents derivative self-interactions of the scalar field, $V(\varphi)$ is a potential, $\beta(\varphi)$ is a coupling function and $T^{\mu}_{\mu}$ is the trace of the matter energy-momentum tensor. For nonrelativistic matter fields,  $T^{\mu}_{\mu} = -\rho_m$, the dynamics of $\varphi$ therefore depend on the local density of the system, $\rho_m$. Around the background $\bar{\varphi}$, the dynamics of the fluctuations of $\varphi$ are determined by three parameters: the mass $m({\varphi})$ (roughly given by the curvature of the effective potential), the coupling $\beta({\varphi})$ and the kinetic function $Z^{\mu \nu}({\varphi})$. Screening can be realised mainly in three different ways utilising these three parameters:

\begin{itemize}
\setlength\itemsep{1em}
\item  {Large mass}\\
If the mass of the fluctuations $m^2({\varphi})$ is large in dense environments, then the scalar field cannot propagate beyond its Compton wavelength $m({\varphi})^{-1}$ and the fifth force mediated by the scalar field is suppressed. On the other hand, in low density environments such as the cosmological background, the mass can be light and the scalar field mediates a sizeable fifth force. This idea characterises the so-called {\it Chameleon} type of screening \citep{Khourya, Khoury};

\item {Large kinetic term}\\
If the kinetic function $Z^{\mu \nu}({\varphi})$ is large in dense environments, the coupling to matter is suppressed. One can either make the first or the second derivative of the scalar field large in dense environments. The former case is realised in the k-mouflage \citep{2009IJMPD..18.2147B, 2014PhRvD..90b3507B} and D-BIonic type of screening \citep{Burrage2014arXiv1403.6120B}, while the latter case characterises the Vainshtein screening mechanism \citep{Vainshtein1972PhLB...39..393V};

\item {Small coupling}\\
If the coupling to matter $\beta({\varphi})$ is small in the region of high density, the strength of the fifth force ${\bf  F}_{\rm Fifth}$ is weak and the modifications to gravity are suppressed. On the other hand, in low density environments, the size of the fifth force can be of the same order as standard gravity ($\gamma \sim 1$). This idea is realised in the dilaton \citep{2010PhRvD..82f3519B} and Symmetron mechanisms \citep{2010PhRvL.104w1301H}.
\end{itemize}

In this code comparison project, we take $f(R)$, DGP and Symmetron gravity as our working example models that screen the fifth force via large mass, large kinetic terms and small coupling strengths, respectively.  Thus, while we do not consider every individual modified gravity model proposed in the literature, our simulations do cover all classes of models.  Throughout, we work with the perturbed Friedmann-Robertson-Walker (FRW) spacetime metric in the Newtonian gauge
\be
{\rm d}s^2 = -(1 + 2 \Phi) {\rm d}t^2 + a^2 (1 - 2 \Psi) \delta_{ij} {\rm d}x^i {\rm d}x^j,
\ee
where $\Psi$ and $\Phi$ represent the two gravitational potentials. The dynamics of nonrelativistic matter is governed by $\Phi$, whereas the bending of light is determined by the lensing potential $\Phi_+ = (\Phi+ \Psi)/2$.  
The modified gravity simulations employed here assume the same weak-field and non-relativistic limit as standard GR simulations, i.e. higher order terms in the dark matter velocities, as well as dynamically generated vector and tensor modes are neglected.

In addition, unless otherwise specified, we assume the \emph{quasi-static limit} for the modified gravity field equation.
This refers to neglecting the time derivatives of the perturbed fields, as in $\dot{\varphi} = \dot{\bar{\varphi}} + \dot{\delta\varphi} \approx \dot{\bar{\varphi}}$, where $\delta\varphi$ is the fluctuation of the scalar field.  
In this paper, we shall assess the validity of the quasi-static limit in the {\it N}-body simulations of the Symmetron model. 

We now describe the specific modified gravity models considered in this paper.  We will often refer back to the quasi-Newtonian potential $\Phi_N$ which is defined through the Poisson equation,
\bq\label{eq:poissonN} 
\nabla^2\Phi_N = 4\pi Ga^2\delta\rho_m \,,
\eq
where $\delta\rho_m$ is the matter density perturbation.


\subsection{{\bf ${f(R)}$} gravity}

$f(R)$ gravity is arguably the most well-studied modified gravity model in the nonlinear regime of cosmological structure formation. In this model, one adds a function of the Ricci scalar $R$ to the Einstein-Hilbert action
\be
S=\int {\rm d}^4 x \sqrt{-g} \frac{1}{16\pi G}\Big(R + f(R) \Big)+ S_m(g_{\mu \nu}, \psi_i), 
\ee
where $g$ is the determinant of the metric $g_{\mu\nu}$ and $S_m$ is the action of the matter fields $\psi_i$. In the quasi-static and weak-field limits, the relevant equations for nonlinear structure formation can be written as
\bq
\label{eq:poisson-fr}\nabla^2\Phi &=& \frac{16\pi G}{3}a^2\delta\rho_m + \frac{1}{6}a^2\delta R, \\
\label{eq:eom-fr}\nabla^2 f_R &=& -\frac{a^2}{3}\left[\delta R + 8\pi G\delta\rho_m\right],
\eq
where $\delta\rho_m = \rho_m - \bar{\rho}_m$ and $\delta R = R - \bar{R}$ are the density and Ricci scalar perturbations, respectively (overbars denote background averaged quantities), and $f_R = {\rm d}f(R)/{\rm d}R$.  In this formulation, $f_R$ plays the role of the scalar degree of freedom $\varphi$ that determines the fifth force and that is solved for by the numerical codes. 

We specialise to the Hu-Sawicki model \citep{2007PhRvD..76f4004H}, which is characterised by
\bq
\label{eq:hs}f(R) &=& -m^2\frac{c_1(R/m^2)^n}{c_2(R/m^2)^n + 1}, \\
\label{eq:hs-diff}f_R &=& -\frac{c_1}{c_2^2}\frac{-n\Big(-R/m^2\Big)^{n-1}}{\Big[\Big(-R/m^2\Big)^n + 1\Big]^2},
\eq
where $m^2 = H_0^2\Omega_m$ is a mass scale (not to be confused with the mass of the scalar fluctuations relevant for the Chameleon mechanism discussed above) and $c_1$, $c_2$ and $n$ are model parameters. Note that the field value is negative ($f_R < 0$) which is necessary to ensure a positive mass of the scalar degree of freedom and hence stability of the theory.  Since
\bq
\label{eq:barR}-\bar{R} \approx 8\pi G\bar{\rho}_m - 2\bar{f}(R) = 3m^2\Big(a^{-3} + \frac{2c_1}{3c_2}\Big),
\eq
one recovers a $\Lambda$CDM expansion history by setting $c_1/c_2 = 6\Omega_\Lambda/\Omega_m$. For values $\left(\Omega_m, \Omega_\Lambda\right) =\left(0.269, 0.731\right)$ (as we consider in the simulations of this paper), then $-\bar{R} \gg m^2$ and one can write
\bq
\label{eq:hs-diff-approx}f_R = -n \frac{c_1}{c_2^2}\left(\frac{m^2}{-R}\right)^{n+1}.
\eq
For the simulations of this paper we always consider $n = 1$, and hence, the remaining free parameter is $c_2$. However, in previous studies it has become more common to specify the Hu-Sawicki model not in terms of $c_2$, but in terms of the equivalent value of $\bar{f}_R$ at the present day, $\bar{f}_{R0}$. Note that by making use of Eqs.~(\ref{eq:barR}) and (\ref{eq:hs-diff-approx}), one can eliminate $\delta R$ in favour of $f_R$ in Eqs.~(\ref{eq:poisson-fr}) and (\ref{eq:eom-fr}). For completeness, we note that the modified Poisson equation, equation~(\ref{eq:poisson-fr}), can also be written as
\bq\label{eq:poissonmod-fr} 
\nabla^2\Phi = \nabla^2\Phi_N - \frac{1}{2}\nabla^2f_R .
\eq
This makes it explicit that in $f(R)$ models the total gravitational force is governed by a modified gravitational potential $\Phi = \Phi_N - \frac{1}{2}f_R$.

We note that the modified gravitational equations defined above hold only for the dynamical potential of the model. The lensing potential $\Phi_+$ in $f(R)$ models (which are equivalent to scalar-tensor theories with a conformal coupling to matter) is not affected by the extra degree of freedom \citep{2008PhRvD..78j4021B} in the weak-field limit.

The term $\delta R$ on the right-hand side of equation~(\ref{eq:poisson-fr}) depends nonlinearly on $f_R$ (cf.~equation~(\ref{eq:hs-diff-approx})). The nonlinearity is what gives rise to the Chameleon screening mechanism. The screening of the fifth force is determined by the depth of the gravitational potential $\Phi_N$. A spherically symmetric object is screened if the {\it thin shell} condition 
\be
|f_{R \infty} - f_{Rs}| < \frac{2}{3} |\Phi_N|
\ee
is satisfied, where $f_{R \infty}$ is the $f_R$ field away from the object and $f_{R s}$ is that inside the object. In order to satisfy the Solar System constraint, the Milky Way galaxy with the potential $|\Phi_N| \sim 10^{-6}$ needs to be screened. This imposes the constraint $|f_{R0}| < 10^{-6}$, if one assumes that the Milky Way galaxy is an isolated object in the cosmological background \citep{2007PhRvD..76f4004H}. 

For the simulations presented in this paper, we consider models with $|\bar{f}_{R0}|=10^{-5}$ (F5) and  $|\bar{f}_{R0}|=10^{-6}$ (F6). Although the former parameter value may already be in tension with Solar System tests, we choose to simulate this model anyway, since it gives rise to larger fifth forces and places the screening threshold for halos at mass scales that are well resolved in the simulation\footnote{See Fig.~6 in \cite{Gronke:2014gaa}. For $|f_{R0}| = 10^{-5}$ halos with mass $M \gtrsim 3\times 10^{13} M_{\astrosun}/h$ are screened while the smallest halos we can resolve have $M\sim 10^{12}M_{\astrosun}/h$.}. Our main goal in this paper is to compare the different code predictions for the fifth force, and not so much to study the observational viability of the models. 
 
 
\subsection{DGP}

The DGP model is an example of a braneworld model. In this model, matter is confined to live in a four-dimensional brane, embedded in a five-dimensional bulk spacetime. The action is given by

\bq\label{eq:dgpaction}
S &=& \int_{\rm brane}\!\!\! {\rm d}^4x \sqrt{-g} \left(\frac{R}{16\pi G} \right) 
+ \int {\rm d}^5x \sqrt{-g^{(5)}} \left(\frac{R^{(5)}}{16\pi G^{(5)}}\right) \nonumber\\
&& +S_m(g_{\mu \nu}, \psi_i),
\eq
where $g^{(5)}$ denotes the five-dimensional metric in the bulk, with $R^{(5)}$ being the Ricci scalar for $g^{(5)}$, while $g$ and $R$ are the induced metric on the brane and its Ricci scalar, respectively.  $G^{(5)}$ and $G$ denote the five- and four-dimensional gravitational constants.  The matter fields $\psi_i$ are confined to the four-dimensional brane. The relative sizes of the two gravitational strengths is a parameter of the model known as the crossover scale, $r_c$,
\bq\label{eq:rc}
r_c = \frac{1}{2}\frac{G^{(5)}}{G},
\eq
below which gravity looks four-dimensional, and above which the five-dimensional aspects become important. The cosmological solutions of this model are characterised by two branches of solutions. The {\it normal branch} requires a dark energy term to be added to the four-dimensional part of the action to explain the accelerated expansion of the Universe \citep{SahniShtanov,LueStarkman,2009PhRvD..80l3003S}; the more appealing {\it self-accelerating} branch does not require a dark energy field, but it is in tension with CMB and supernovae data \citep{2008PhRvD..78j3509F} and is also plagued by problems associated with the propagation of ghosts (degrees of freedom whose energy is unbounded from below) \citep{2003JHEP...09..029L,2004JHEP...06..059N, 2007CQGra..24R.231K}. In this paper, we focus on the normal branch of the DGP model. The dark energy component on the brane is adjusted to precisely yield a flat $\Lambda$CDM background cosmology \citep{2009PhRvD..80l3003S}. 

The modifications to the gravitational law in this model are determined by a scalar field, $\varphi$, which is associated with the bending modes of the 4D brane. The brane-bending mode influences the dynamics of particles through
the dynamical potential $\Phi$, which, assuming the same boundary
conditions for $\Phi$ and $\varphi$, is given by
\be
\Phi = \Phi_N + \frac{1}{2} \varphi.
\label{eq:psi}
\ee
The equation for $\varphi$ reads, in the quasi-static and weak-field limits \citep{2007PhRvD..75h4040K},
\be
\nabla^2 \varphi + \frac{r_c^2}{3\beta\,a^2} \left[ (\nabla^2\varphi)^2
- (\nabla_i\nabla_j\varphi)^2 \right] = \frac{8\pi\,G\,a^2}{3\beta} \delta\rho_m,
\label{eq:phiQS}
\ee
where $\left(\nabla_i\nabla_j\varphi\right)^2 = (\nabla_i\nabla_j\varphi)(\nabla^i\nabla^j\varphi)$,
and the function $\beta(a)$ is given by
\be
\beta(a) = 1 + 2 H(a)\, r_c \left ( 1 + \frac{\dot H(a)}{3 H^2(a)} \right )\,,
\label{eq:beta}
\ee
where we have assumed the normal branch of the DGP model already. The quasi-static approximation in the DGP models was tested for self-consistency in Section~IV~C of \cite{2009PhRvD..80d3001S} and was also recently shown to be an excellent approximation in \cite{2014PhRvD..90l4035B, 2015arXiv150503539W}. 

As in $f(R)$ gravity, the propagation of photons, determined by the lensing potential $\Phi_+$, is not directly affected by $\varphi$. In models like DGP it is the Vainshtein screening mechanism that provides the chance to pass Solar System tests. For simplicity, we focus on spherically symmetric configurations to illustrate how the screening works. Writing down equation~(\ref{eq:phiQS}) in spherical coordinates and integrating once as $\int r^2{\rm d}r$, one gets
\bq
\frac{2r_c^2}{3\beta}\left(\frac{\varphi,_r}{r}\right)^2 + \left(\frac{\varphi,_r}{r}\right) = \frac{2}{3\beta}\frac{GM(r)}{r^3},
\eq 
where $M(r)$ is the mass enclosed inside a radius $r$, and a comma denotes partial differentiation. The solution to the last equation is given by
\bq
\varphi,_r = \frac{4}{3\beta}\left(\frac{r}{r_V}\right)^3\left[-1 + \sqrt{1 + \left(\frac{r_V}{r}\right)^3}\right]\frac{GM(r)}{r^2},
\eq
where we define the distance scale
\bq\label{eq:rv}
r_V(r) = \left(\frac{16r_c^2GM(r)}{9\beta^2}\right)^{1/3},
\eq
which is known as the Vainshtein radius. This radius defines the distance from the centre of the spherical overdensity below which the spatial gradient of $\varphi$ becomes suppressed (and hence the fifth force effects become negligible). Explicitly, for a top-hat density profile of radius $R_{\rm th}$ and mass $M_{\rm th}$, if $r \gg r_V> R_{\rm th}$, then
\bq
\frac{\varphi,_r}{2} = \frac{1}{3\beta}\frac{GM_{\rm th}}{r^2} = \frac{1}{3\beta} \Phi_{\rm N},_r,
\eq
i.e., the fifth force becomes a sizeable fraction of the force in GR (cf.~equation~(\ref{eq:psi})). On the other hand, if $R_{\rm th} < r \ll r_V$, then $\varphi,_r \rightarrow 0$.

In the simulations of this paper, we consider two parameter values, $r_cH_0 = 1$ and $r_cH_0 = 5$. These were chosen to roughly match the F5 and F6 models, respectively, in terms of the values of $\sigma_8$ at $z=0$.


\subsection{Symmetron}
The third model that we consider is the Symmetron model \citep{2010PhRvL.104w1301H} (see also \cite{2005PhRvD..72d3535P,2008PhRvD..77d3524O}), whose action is given by
\bq\label{eq:action_symm}
S &=& \int {\rm d}x^4 \sqrt{-g}\left[\frac{R}{16\pi G} - \frac{1}{2}(\partial\varphi)^2 -
V(\varphi)\right]\nonumber\\
&+& S_m(\tilde{g}_{\mu\nu},\psi).
\eq
The matter fields, $\psi$, couple to the Jordan frame metric $\tilde{g}_{\mu\nu}$ which is given by a conformal rescaling of the Einstein frame metric $g_{\mu\nu}$
\be\label{eq:conformal_coupling}
\tilde{g}_{\mu\nu} = A^2(\varphi)g_{\mu\nu}.
\ee
In the Symmetron model, the coupling function $A(\varphi)$ is given by
\be\label{eq:coupling_function}
A(\varphi) = 1 + \frac{1}{2}\left(\frac{\varphi}{M}\right)^2,
\ee
where $M$ is a mass scale. This coupling function determines the fifth force and the total gravitational force is given by
\be
{\bf F} = \nabla\left(\Phi_N + \frac{1}{2}\frac{\varphi^2}{M^2}\right) = \nabla\Phi_N + \frac{\varphi\nabla\varphi}{M^2}.
\ee
The potential is taken to be of the symmetry breaking form
\bq\label{potential}
V(\varphi) = V_0 -\frac{1}{2}\mu^2\varphi^2 + \frac{1}{4}\lambda\varphi^4.
\eq
With these choices for $A(\varphi)$ and $V(\varphi)$, the model becomes invariant under the symmetry $\varphi\to - \varphi$. The value of $V_0$ is determined by the condition that the model gives rise to the observed accelerated expansion of the Universe \citep{2011PhRvD..84j3521H}. The field equation for $\varphi$ follows from the variation of the action, equation~(\ref{eq:action_symm}), with respect to $\varphi$ and reads
\be\label{eq:eom_symm}
\square\varphi = V_{\rm eff,\varphi},
\ee
where, for nonrelativistic matter, the effective potential is given by
\be\label{eq:veff_symm}
V_{\rm eff}(\varphi) =  V_0 + \frac{1}{2}\left(\frac{\rho_m}{M^2}-\mu^2\right)\varphi^2 + \frac{1}{4}\lambda\varphi^4.
\ee
In working with the model, it is convenient to define a matter density scale for symmetry breaking, $\rho_{\rm SSB}$, and its associated scale factor, $a_{\rm SSB}$, where $\bar\rho(a_{\rm SSB}) = \rho_{\rm SSB}$, as
\be\label{eq:rho_crit_symm}
\rho_{\rm SSB} \equiv \mu^2 M^2 = 3H_0^2M_{\rm pl}^2\Omega_m/ a_{\rm SSB}^3.
\ee
Other useful quantities are
\be
\beta_0 = \frac{\varphi_0 M_{\rm Pl}}{M^2}\ ; \ \ \lambda_0 = \frac{1}{\sqrt{2}\mu}\ ; \ \ \ \varphi_0 = \frac{\mu}{\sqrt{\lambda}},
\ee
where $\beta_0$ is the coupling strength of the unscreened fifth force, $\lambda_0$ is the Compton length (giving the range of the fifth force) and $\varphi_0$ is the symmetry breaking vacuum expectation value (VEV) of $\varphi$ (when $\rho_m = 0$). In the simulations of this paper we consider $\beta_0 = 1$, $\lambda_0 = 1 h^{-1} {\rm Mpc}$ and $a_{\rm SSB} = 0.5$ which lie on the boundary of the allowed $\{a_{\rm SSB},\lambda_0\}$ parameter space coming from local constraints (see the discussion below equation~(\ref{eq:symm_screen})). These parameter values have been previously simulated in \citet{2012ApJ...748...61D,2012JCAP...10..002B,2014AA...562A..78L,2014PhRvD..89h4023L}.

In the quasi-static limit, equation (\ref{eq:eom_symm}) becomes
\be
\nabla^2\chi = \frac{a^2}{2\lambda_0^2}\left( \frac{\rho_m}{\rho_{\rm SSB}} - 1 + \chi^2\right)\chi,
\ee
where $\chi = {\varphi}/{\varphi_0}$. The full field equation, without applying the quasi-static limit, is discussed in Section~\ref{sect:nonstaticsims}.

During the cosmological evolution \citep{2011PhRvD..84l3524B,2012PhRvD..86d4015B,2011PhRvD..84j3521H,2012ApJ...748...61D} the field sits close to the global minimum of the effective potential at $\varphi = 0$ for $a < a_{\rm SSB}$. For $a > a_{\rm SSB}$ the effective potential develops two minima at $\varphi = \pm \varphi_0\sqrt{1 - a_{\rm SSB}^3/a^3}$, to which the field at $\varphi = 0$ (now a maximum) evolves, thereby spontanously breaking the $\varphi\to -\varphi$ symmetry. Since the field can choose different minima ($+$ or $-$ branches) in different parts of the Universe, this model therefore leads to the formation of domain walls. The properties of these domain walls have been studied beyond the quasi-static approximation \citep{2013PhRvL.110p1101L,2014PhRvD..90l4041L,2014PhRvD..90l5011P}.

Screening in the Symmetron model is very similar to that in the Chameleon/$f(R)$ cases in the sense that the condition for screening is determined by the local gravitational potential. There is however the important difference that the coupling $\beta(\varphi) = \frac{\beta_0 \varphi}{\varphi_0}$, which is constant for $f(R)$ gravity, now depends on the local field value. In high density regions, $\rho_m  > \rho_{\rm SSB}$, the field moves towards $\varphi = 0$, and since the coupling is proportional to $\varphi$, the fifth force is suppressed. In the Symmetron model, the condition for the thin-shell effect is given by
\be\label{eq:symm_screen}
\left|\frac{\varphi_{s} - \varphi_{\infty}}{\varphi_0}\right| \ll \Phi_N \beta(\varphi_\infty),
\ee
where $\varphi_{\infty}$ is the $\varphi$ field far away from the object and $\varphi_{s}$ is that inside the object. In order to satisfy Solar System bounds, we get the constraint $\left(\frac{\lambda_0}{h^{-1} {\rm Mpc}}\right)^2a_{\rm SSB}^{-3} \lesssim \mathcal{O}(1)$ by assuming that the Milky Way galaxy is an isolated object in the cosmological background \citep{2012PhRvD..86d4015B}.


\section{Modified Gravity Simulations}
\label{sec:simulations}

\subsection{General force calculation}

Cosmological dark-matter {\it N}-body simulations for standard gravity are characterised by the following two equations. First, we have the Poisson equation (\ref{eq:poissonN})
\be\label{eq:poisson}
\nabla^2\Phi_N = 4\pi G a^2 \delta\rho_m,
\ee
which determines the Newtonian potential, $\Phi_N$, given the density fluctuations $\delta\rho_m$. Second, we have the geodesic equation 
\be\label{eq:geodesic}
\ddot{{\bf x}} + 2H \dot{{\bf x}} = -{\bf \nabla} \Phi,
\ee
which tells the particles how to move. At every timestep in the simulation, one (i) computes the density field from the particle positions; (ii) uses it in the Poisson equation to solve for the potential; and (iii) plugs $\Phi$ into the geodesic equation to move the particles. This process is repeated from some initial redshift, $z = z_i$, until typically $z=0$.

As we have seen in the previous section, modified gravity models alter this picture by modifying the Poisson equation that governs the total gravitational potential (like in $f(R)$ and DGP gravity), or by adding extra terms to the right-hand side of the geodesic equation (like the term $\propto\nabla A(\varphi)$ in the Symmetron model)\footnote{Note that any extra term in the geodesic equation can always be absorbed into the definition of a modified gravitational potential. This essentially illustrates the equivalence between the Jordan and Einstein frames.}. In general, modified gravity models can also differ in the background expansion rate of the Universe. In this paper, however, we shall always consider models with the same \lcdm{} background evolution.

The bulk of the computing time in modified gravity simulations is spent solving the nonlinear equation that governs $\varphi$ (see the previous section), which can be cast in the general form
\be\label{eq:generalform}
L[\varphi] = S(\delta\rho_m,\varphi),
\ee
where $L$ is some nonlinear operator that acts on $\varphi$, and $S$ is a source term that depends on the matter density fluctuations and possibly on the scalar field. The exact functional form of $L$ and $S$ varies from theory to theory, but as we have discussed in the previous section, this operator should possess some degree of nonlinearity to ensure the presence of screening effects. The nonlinearity in the equations, however, is what makes {\it N}-body simulations of these models so challenging. On the other hand, equation~(\ref{eq:poisson}) is a linear elliptic partial differential equation (PDE), which means that it can be solved with efficient fast Fourier transform (FFT) methods.  This is in general not possible in modified gravity models, which typically have nonlinear equations. This difficulty can be overcome by employing a FFT-relaxation method \citep{2009PhRvD..80j4005C}, if the equations are to be solved on a regular grid. However, this method does not work on irregularly-shaped refinements. The codes we compare in this study solve equation~(\ref{eq:generalform}) via direct discretisation and relaxation on such an irregular grid.

In the rest of this section, we briefly outline the relaxation algorithm, describing also the main idea behind {\it multigrid} acceleration methods. The latter significantly improves the efficiency of the relaxation algorithms.


\subsection{Iterative methods with multigrid acceleration}\label{sec:multigrid}

With the exception of the \isisns{} code (see next section), all codes make use of multigrid acceleration to speed up numerical convergence of the partial differential equations for $\varphi$. Here, we briefly review the main aspects of these techniques and refer the reader to \cite{Brandt77, Wesseling92, Trottenberg} and to the code papers (cf.~Table \ref{table:codes}) for further details on their implementation.


\subsubsection{Gauss-Seidel iterations}

The goal is to solve a differential equation that can be written in the form of equation~(\ref{eq:generalform}). The basic algorithm consists in discretising the equation on a grid and using an iterative scheme to obtain improved solutions given an initial guess. All codes assume periodic boundary conditions on the domain (unrefined) grid. The codes that include grid refinements use fixed boundary conditions on the boundary of the refinements, obtained by interpolating from the next coarser refinement level. 
Upon discretisation, the solution to equation~(\ref{eq:generalform}) is given by the solution of the large set of algebraic equations
\be\label{eq:generalform-discrete}
L^l[\varphi^l] = S^l, 
\ee
where $L^l$ and $S^l$ are the discretised versions of the $L$ and $S$ operators and $\varphi^l$ is the field solution we aim to determine. The index $l$ labels the refinement level of the grid. The discretisation of the equation consists in writing each of the derivatives that appear in $L$ as a combination of the values of $\varphi$ on the grid cells. For instance, the codes employed here use the 3- and 4-point stencils
\begin{eqnarray}\label{eq:discexample}
\partial_x^2\varphi_{i,j,k} &=& \frac{1}{h^2} \left ( \varphi_{i+1,j,k} + \varphi_{i-1,j,k} - 2\varphi_{i,j,k} \right ) \\
\partial_x\partial_y\varphi_{i,j,k} &=& \frac{1}{4h^2} \left (\varphi_{i+1,j+1,k} - \varphi_{i+1,j-1,k} \right. \nonumber \\ 
&& \ \ \ \ \ \ \ \ \ \left. - \varphi_{i-1,j+1,k} + \varphi_{i-1,j-1,k} \right ) ,
\end{eqnarray}
where $\left\{i, j, k\right\}$ labels each grid cell. The iterations can be made in two different ways. If the operator $L$ is linear, then it is best to perform {\it explicit} iterations, in which one rearranges the discretised equation analytically to solve directly for $\varphi_{ijk}$ in each cell (this is, for instance, how the standard \ramses{} code solves the Poisson equation). For nonlinear problems (as those in modified gravity), an implicit iteration scheme is more suitable. In this case, one can use the {\it Newton-Raphson} method to solve the equation
\be
T^l[\varphi^l] = L^l[\varphi^l] - S^l = 0.
\ee
The resulting value of $\varphi$ can be written as
\be
\bar{\varphi}^l = \varphi^l - \frac{T^l}{\partial T^l/\partial \varphi^l},
\ee
where the barred field corresponds to the updated value. Finally, one needs to specify the way in which sweeps are made across the grid in each iteration step. The simplest sweeping strategy consists in following a lexicographic ordering, in which the calculation on the cell $\{i,j,k\}$ is followed by the calculation on the cell $\{i+1,j,k\}$, and so on.  A more widespread strategy which has better convergence and parallelisation properties is the so-called two colours scheme, in which the calculation is performed alternatively in cells of the same {\it colour}, as in the colour scheme of a chess board. In the first half-sweep all {\it black} cells are updated, and the second half-sweep takes care of the remaining {\it white} cells. Further generalisations of this scheme exist with four or even eight colours.

The iterations proceed until a certain convergence criterion is fulfilled.  There are several criteria in the literature. These typically involve computing the residual, $\epsilon^l$, of the solution defined as
\be
\epsilon^l =  L^l[\varphi^l] - S^l . 
\ee
The convergence criterion is then given by $\parallel\epsilon^l\parallel < \epsilon_{\rm converged}$, 
where $\parallel.\parallel$ is a norm (typically $L_2$) that is taken over the entire grid and $\epsilon_{\rm converged}$ is a (small) predefined constant that we use to define convergence. Exact solutions of the algebraic equation have $\epsilon^l=0$. However, owing to truncation errors (the error inevitably introduced by discretising a continuous equation), exact solutions of the algebraic equation are not equal to the solutions of the discretised equation. The iterations are then assumed to have converged once the residual falls below a predetermined fraction of the truncation error.


\subsubsection{Multigrid acceleration}

In a relaxation method such as the one described above, during the first few iterations the residual decays very efficiently. However, the convergence becomes considerably slower as one approaches the true solution. This slowdown of the convergence is attributed to the fact that the components of the residual whose Fourier wavelength modes are longer than the size of the grid cell decay much more slowly than those modes whose wavelength is comparable to the grid size. The goal of multigrid methods is to speed up the convergence of these longer wavelength modes by using a hierarchy of coarser grids. In short, when the convergence of the solution starts to slow down, one interpolates the equation onto the next coarser grid of the hierarchy and solves it there. This makes the longer wavelength modes decay faster, therefore bringing the solution closer to its true value. This {\it coarsening} scheme can proceed up to several coarser grids. The coarser solutions can then be interpolated back to the finer (original) level.

To be more concrete, the typical way to arrange different resolutions is to use a set of grids whose resolutions are half, one quarter and so on of the target resolution. A two grid scheme is defined in the following way. One starts by performing a given number of iterations on the target grid $l$.  Then, when the convergence slows down, one moves to the next coarser grid $l-1$ and performs iterations for an equation whose solution corresponds to the error $\delta\varphi^{l-1}$ of the previous solution. In the case of a linear PDE (we will turn to the nonlinear case below), the equation that must be solved on the coarser grid is
\be
L^{l-1}[\delta\varphi^{l-1}] = R(\epsilon^l),
\ee
where $L$ is now a linear operator while $R$ is a restriction operator that is chosen according to the problem and translates information from the fine grid (the target grid) to the coarse grid. Once the coarse grid iterations are done, one corrects the fine grid solution as 
\be
\bar{\varphi}^l = \varphi^l + P(\delta\varphi^{l-1}),
\ee
where $P$ is now a prolongation operator which translates information from the coarse to the fine grid.  In general, one uses more than one coarse grid and these processes of going up and down in resolution are called V-cycles. After one V-cycle, if convergence is not yet achieved on the target grid, then further V-cycles are performed. All the codes analysed here use V-cycles, although other arrangements are possible such as W-cycles (in these, one can move in between coarser levels several times before returning to the target grid).

In the case of nonlinear equations however, this multigrid algorithm requires some changes, as it relies on the linear superposition of solutions from different grids.  In the nonlinear case, instead of solving following the solution for the errors on the coarse grids, one obtains improved approximations of the solution (not the error) itself. In this case, the coarse grid iterations are made according to
\be
L^{l-1}[\varphi^{l-1}] = -R(\epsilon^l(\varphi^l, S^l)) + \epsilon^{l-1}(R(\varphi^l), R(S^l))
\ee
and the coarse grid correction of the fine grid solution is given by
\be
\bar{\varphi}^l = \varphi^l + P(\varphi^{l-1} - R(\varphi^l)).
\ee


\section{Code and algorithm description}\label{sec:codedes}

In this section, we briefly introduce the different {\it N}-body codes compared in this paper and comment on some aspects of the numerical handling of the specific model equations. We shall keep our description simple and refer the interested reader to the code papers for the details. Some of the key features of the codes are summarised in Table \ref{table:codes}.

\begin{table*}
\scriptsize
\caption{Key features of the {\it N}-body codes compared in this paper}
\begin{tabular}{@{}lccccccccccc}
\hline\hline
\\
Code  & \dgpm{} & \ecosmog{} & \mggadget{} & \isis{} & \isisns{}\ \ 
\\
\hline
\\
Code paper  & \cite{2009PhRvD..80d3001S} &  \cite{2012JCAP...01..051L, 2013JCAP...05..023L} & \cite{2013MNRAS.436..348P} & \cite{2014AA...562A..78L} & \cite{2014PhRvD..89h4023L}\ \ 
\\
Base code  & \cite{2008PhRvD..78l3523O} &  \ramses{} &  \pgadget  & \ramses{} & \ramses{}\ \ 
\\
Density assignment & CIC & CIC/TSC & CIC & CIC & CIC &\ \ 
\\
Force assignment &CIC & CIC/TSC & effective mass & CIC & CIC \ \ 
\\
Adaptive refinement? & No & Yes & Yes & Yes & No &\ \ 
\\
Timestep &Fixed&Adaptive&Adaptive&Adaptive&Adaptive&\ \ 
\\
MG solver & Multigrid & Multigrid & Multigrid & Multigrid & Leapfrog &\ \
\\
Gravity solver & Multigrid & Multigrid & TreePM & Multigrid & Multigird &\ \
\\
Parallelisation & {\sc OpenMP}& {\sc MPI} & {\sc MPI} &  {\sc MPI} & {\sc MPI} &\ \ 
\\
Programming language & {\sc C++} & {\sc Fortran} &  {\sc C} & {\sc Fortran} & {\sc Fortran} &\ \ 
\\
Models simulated & DGP & $f(R)$/DGP &$f(R)$  &$f(R)$/Symmetron  & Symmetron &\ \ 
\\
\\
\hline
\hline
\\
\end{tabular} \\
\label{table:codes}
\end{table*}

\subsection{Code summary}


\subsubsection{\dgpm{}}

The DGPM code \citep{2009PhRvD..80d3001S,2009PhRvD..80l3003S} is a fixed-grid particle-mesh code that solves the Vainshtein-type equation of motion (\ref{eq:phiQS}).  Based on the fixed-grid $f(R)$ code presented in \cite{2008PhRvD..78l3523O}, it employs a second-order leapfrog scheme with fixed step size $\Delta a$ in scale factor to advance particles.  Densities are interpolated onto the grid using cloud-in-cell (CIC) interpolation, which is also used to evaluate derivatives on the grid.  The Poisson equation for the Newtonian potential is solved using FFT on the fixed grid. The Gauss-Seidel relaxation is then performed using the Newton-Raphson method and multigrid acceleration as described in Section~\ref{sec:multigrid}.  At each multigrid level, $5$ to $10$ relaxation sweeps are performed. Convergence to an RMS residual of less than $10^{-10}$ (where typical values of $\varphi$ are of order $10^{-5}$) is usually reached within $3$ V-cycles. The step size used for this paper is $\Delta a = 0.02$, which results in 490 steps from $z=49$ (when all our simulations start) to $z=0$.  


\subsubsection{\ecosmog{}}
The \ecosmog{} code \citep{2012JCAP...01..051L} is built on top of the publicly available adaptive mesh refinement (AMR) {\it N}-body code \ramses{} \citep{2002A&A...385..337T}. The code can be compiled to work with CIC (like in \ramses{}) or triangular shaped cloud (TSC) schemes for the interpolation of the density and force fields. Unless otherwise specified, the \ecosmog{} results shown in this paper are for TSC. The time evolution is performed with a second-order leapfrog algorithm with adaptive timesteps (set by the AMR grid, as in \ramses{}). The Gauss-Seidel relaxations of the scalar field equation are performed using the Newton-Raphson methods with multigrid acceleration on all levels of the AMR grid. For the simulations in this study, the grid was set to be refined whenever the particle number exceeded 8 inside a grid cell.

In addition to the $f(R)$ and DGP gravity results shown in this paper, the \ecosmog{} code has also been used to simulate dilaton \citep{2011PhRvD..83j4026B}, Symmetron \citep{2012ApJ...748...61D,Brax2013a}, Cubic Galileon \citep{2013JCAP...10..027B} and Quartic Galileon \citep{Li2013JCAP...11..012L} gravity cosmologies, as well as a general parametrization of Chameleon theories \citep{2012JCAP...10..002B}. Different versions of the code differ in the detailed way the scalar field equations are solved. The performance of the relaxation algorithm is also slightly different, although, for all these models, the residuals always reach a value of $\lesssim 10^{-3}$ times the truncation error after $5-10$ V-cycles. The extensions made to \ramses{} to develop \ecosmog{} can also be straightforwardly coupled to the hydrodynamic modules of the base code, although to date such a project has never been undertaken.  In case of the DGP model, \ecosmog{} solves a different version of the scalar field equation as \dgpm{}, which will be discussed in Section~\ref{sec:dgpsims} below.


\subsubsection{\mggadget{}}
The \mggadget{} code \citep{2013MNRAS.436..348P} is an extension of the cosmological hydrodynamical TreePM+SPH simulation code \textsc{p-gadget3} which is itself based on \textsc{gadget2} \citep{2005MNRAS.364.1105S}. It features the baryonic physics modules of \textsc{p-gadget3} as well as a modified gravity solver. In addition, \mggadget{} allows the inclusion of massive neutrinos in simulations of modified gravity \citep{2014MNRAS.440...75B} by making use of the particle-based massive neutrino module \citep{2010JCAP...06..015V} which is implemented in \textsc{p-gadget3}.

In contrast to \ramses , \pgadget\ does not intrinsically possess an AMR grid. To overcome this, \mggadget\ constructs an adaptively-refining grid that covers the whole simulation volume by appropriately choosing nodes from the oct-tree structure of \pgadget 's Poisson solver. This grid is then used to solve for the scalar degree of freedom using the method described in Section~\ref{sec:multigrid}, i.e. using CIC density assignment and multigrid-accelerated Gauss-Seidel relaxation on the different levels of the AMR grid. 

So far the code has been used to simulate the \cite{2007PhRvD..76f4004H} $f(R)$ gravity model, both in collisionless (DM only) and hydrodynamical simulations \citep{Arnold2014MNRAS.440..833A, 2015MNRAS.448.2275A}. For hydrodynamical simulations the fluid equations are solved using the same entropy-conserving SPH scheme \citep{2002MNRAS.333..649S} as \textsc{p-gadget3}. 


\subsubsection{\isis{}}

The \isis{} code \citep{2014AA...562A..78L}, like \ecosmog{}, is a modified version of \ramses{}.  To date, \isis{} has been used to simulate $f(R)$ gravity \citep{2014AA...562A..78L}, the Symmetron model in both the quasi-static \citep{2014AA...562A..78L} and non-static limits \citep{2014PhRvD..89h4023L}, the non-static disformal gravity model \citep{Koivisto:2012za} in its pure disformal limit \citep{llinares_disformal}, the non-static disformally coupled Symmetron model \citep{2015arXiv150407142H} and the Cubic Galileon / DGP model \citep{2014arXiv1403.6492W}. In \cite{2015MNRAS.449.3635H} and \cite{Hammami:2015ela}, \isis{} has also been used to study hydrodynamic effects in simulations of $f(R)$ and Symmetron models.

The static version of \isis{} solves the equation of motion of the scalar field using the multigrid methods outlined in the previous section. The code uses a CIC scheme to interpolate the density from the particles to the grid, and the time steps of each particle are determined by the AMR grid (as in standard \ramses{}). In our simulations of the static \isis{} code, each grid cell was refined whenever the particle number contained in it exceeded 8 (as in the \ecosmog{} simulations).

The non-static version of \isis{} (the version that goes beyond the quasi-static limit) uses a leapfrog scheme to evolve the scalar field in time.  In this case, the code includes two time steps:  a coarse one for the particles, which is determined by the domain grid (the non-static version does not admit refinements) and a finer one for the time evolution of the scalar field. The total number of scalar field time steps is typically three to four orders of magnitude larger than the number of particle time steps.


\subsection{Model algorithms}

In the remainder of this section, we briefly outline the strategy employed by the different codes to solve the equations of the three modified gravity models we consider.

\subsubsection{$f(R)$ simulations}
\label{sec:fr_algorithms}

The simulations of the $f(R)$ Hu-Sawicki model were performed with \ecosmog{}, \mggadget{} and \isis{}. All these codes discretise and relax the scalar field equation of motion, equation~(\ref{eq:eom-fr}), in a similar way. 
Instead of solving for $f_R$ directly, the scalar field is redefined in terms of $u \equiv \ln (f_R /\bar{f}_R(a))$ which is then numerically computed. 
Using the variable $u$ has considerable advantages in terms of numerical stability as it implicitly avoids unphysical positive values of $f_R$ when performing the Newton-Raphson iterations.  

Once $f_R$ is found, the three codes compute the total force in slightly different ways. 
In \ecosmog{}  the code uses the solution for $f_R$ to compute the $\delta R$ term on the right-hand side of equation~(\ref{eq:poisson-fr}).  
The code then determines the total potential $\Phi$ by solving the modified Poisson equation in a similar way to the standard gravity solver in \ramses{}. The total (modified) force, $\nabla\Phi$, is finally interpolated from the mesh to the particle positions (like in standard \ramses{}).

In \mggadget{}, the total force is also obtained by solving the modified Poisson equation, but by making use of the standard tree+particle-mesh gravity algorithm of the base code. 
In order to do so, the modified Poisson equation is rewritten in terms of an effective mass density: $\nabla \Phi = 4 \pi G ( \delta \rho + \delta \rho_{\rm eff})$, where $\delta\rho_{\rm eff} = \frac{1}{3}\delta\rho - \frac{1}{24\pi G}\delta R$.  
Adding the effective to the real mass density, the values of the scalar field can be directly used in the highly optimized and efficient TreePM gravity algorithm of \pgadget{} to compute the total force. 

\begin{figure}
\includegraphics[width=1\columnwidth]{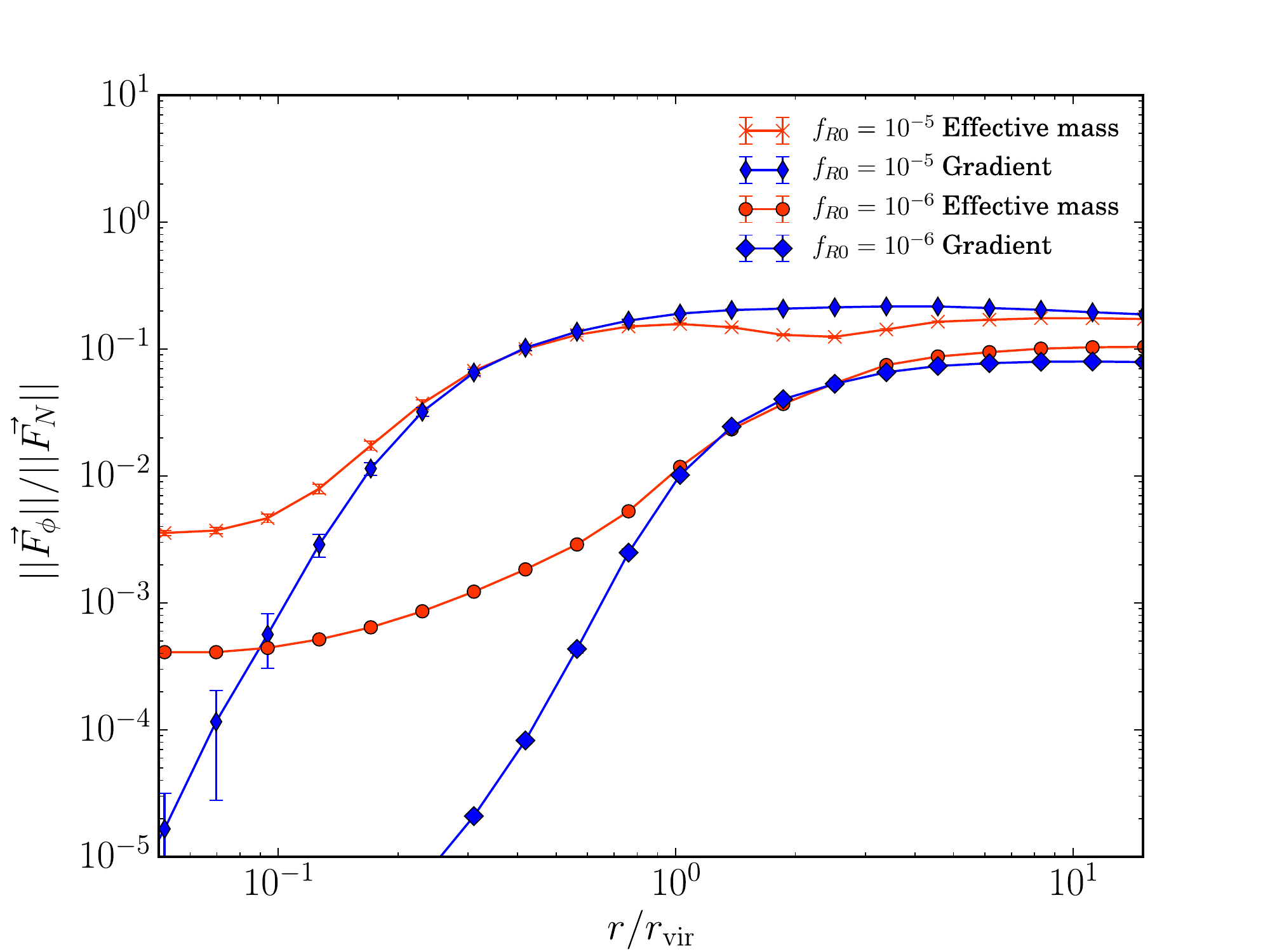}
\caption{Fifth to Newtonian force ratio profile, $F_\phi / F_N$, obtained from interpolating the scalar field gradient from the grid to the particle positions (blue) and from the effective matter density method using the tree force calculation (red) in \mggadget{} for the $f(R)$ simulations. The profiles shown are the average for haloes with masses $M\in[1\times 10^{14},5\times 10^{14}]\,M_{\astrosun}/h$. The errorbars are the variance of this average.}
\label{fig:ratio_mggadget_comp}
\end{figure}

Although these two methods for solving the modified Poisson equation are mathematically equivalent, they can yield different numerical accuracies. In addition to the convenience of using the standard Poisson solver, \mggadget{}'s effective mass algorithm avoids an interpolation of the scalar field gradient (the fifth force) from the adaptive mesh to the particle positions, as the tree force is directly computed there. It might, nevertheless, be somewhat less accurate in highly screened regions than the method used in \ecosmog{} due to numerical summation errors in the tree gravity, which can result in less precise screening of the fifth force. This causes somewhat larger random force errors for the individual particles while the integrated effects are expected to average out. Fig.~\ref{fig:ratio_mggadget_comp} displays profiles of the ratio of fifth to Newtonian force in dark matter haloes obtained via interpolation of the gradient from the grid (blue) and via the effective density scheme (red). Although it is noticeable that there are significant differences between the two methods on small scales, this occurs only in a regime where the fifth force is already highly screened ($\lesssim 1\%$ of normal gravity). As force errors of around one percent also occur in the standard tree gravity algorithm, these errors are expected to be negligible. In the region where screening just sets in (which several observables might be sensitive to) the curves almost perfectly match each other. Consequently, the mentioned inaccuracies in the fifth force calculation will not change the total force significantly and will therefore only have a very minor impact on observables (as we shall see in more detail in the next sections). 

Finally, in \isis{}, the code computes the gradient of the scalar field and interpolates it onto the particle positions. The GR gravitational potential, $\Phi_N$, is solved exactly as in \ramses{}, and the total force at the particle positions is given by $\nabla\Phi = \nabla\Phi_N - \nabla f_R/2$ (cf.~equation~(\ref{eq:poissonmod-fr})).

We refer the reader to the respective code papers for more details about how \ecosmog{}, \mggadget{} and \isis{} solve the modified gravity equations, code tests, and explicit discretisation of the equations.


\subsubsection{DGP simulations}\label{sec:dgpsims}

The simulations of the DGP braneworld model shown here were performed with the \dgpm{} and \ecosmog{} codes. In \dgpm{}, the scalar field equation is discretised as it is written in equation~(\ref{eq:phiQS}). Some of the simulations using DGPM presented in \cite{2009PhRvD..80d3001S,2009PhRvD..80l3003S} employed a Gaussian smoothing of the density field in order to improve convergence. For the analysis of this paper, however, no such smoothing was performed. 

In the strategy employed by the \ecosmog{} code, equation~(\ref{eq:phiQS}) is manipulated analytically before being discretised. This manipulation is called the {\it operator-splitting trick} \citep{2009PhRvD..80j4005C, 2013JCAP...05..023L, Li2013JCAP...11..012L}, which we describe next. Equation (\ref{eq:phiQS}) can be cast as
\be\label{eq:phiQS-1}
\left(1-w\right)\left(\nabla^2\varphi\right)^2 + \alpha\nabla^2\varphi - \Sigma = 0,
\ee
where
\bq
\alpha &=& \frac{3\beta(a)a^2}{r_c^2}, \\
\Sigma &=& \left(\nabla_i\nabla_j\varphi\right)^2 - w\left(\nabla^2\varphi\right)^2 + \frac{8\pi G a^4}{r_c^2}\delta\rho,
\eq
and $w$ is a constant numerical factor. Equation (\ref{eq:phiQS-1}) can be solved once to yield
\bq\label{eq:phiQS-2}
\nabla^2\varphi = \frac{\alpha \pm \sqrt{\alpha^2 + 4(1-w)\Sigma}}{2(1-w)}.
\eq
Decomposing the term $\nabla_i\nabla_j\varphi$ into its trace and traceless part (this is the operator-splitting trick),
\bq
\nabla_i\nabla_j\varphi = \frac{1}{3}\gamma_{ij}\nabla^2\varphi + \hat{\nabla}_i\hat{\nabla}_j\varphi,
\eq
it is possible to show, after a bit of algebra, that
\bq
\Sigma = \left(\hat{\nabla}_i\hat{\nabla}_j\varphi\right)^2 + \frac{8\pi G a^4}{r_c^2}\delta\rho,\ \ \ {\rm if}\ w = \frac{1}{3}.
\eq
That is, if $w = 1/3$, then this cancels out the term $\nabla^2\varphi$ in $\Sigma$. The reason why this is useful is because, upon discretisation, the traceless part $\hat{\nabla}_i\hat{\nabla}_j\varphi$ does not depend on the grid cell value, $\varphi_{ijk}$, but only on its neighbours. Hence, in the equation that \ecosmog{} solves, which is equation~(\ref{eq:phiQS-2}), $\varphi_{ijk}$ appears only on the left-hand side, and not inside the square-root. This is found to improve considerably the performance of the code. Moreover, by taking out $\varphi_{ijk}$ from inside the square-root one also avoids potential problems associated with imaginary square-roots caused by some bad initial guess for $\varphi_{ijk}$. The sign of the square-root in equation~(\ref{eq:phiQS-2}) is chosen to be the same as the sign of the $\alpha$ function. This is the solution which corresponds to the physical (linear theory) result that $\nabla^2\varphi \rightarrow 0$, when $\delta\rho \rightarrow 0$. 

Once $\varphi$ is found on every grid cell,  both \dgpm{} and \ecosmog{} compute the total force, $\nabla\Phi$, as the sum of normal gravity and the fifth force, $\nabla\Phi = \nabla\Phi_N + \nabla\varphi/2$.

For completeness, we point out that if $\delta\rho$ becomes negative (as it does in voids), then there is the risk that the argument of the square-root in equation~(\ref{eq:phiQS-2}) may become negative. This does not happen for the DGP model, but similar Vainshtein screening models such as the Cubic \citep{2013JCAP...10..027B} and Quartic Galileon \citep{Li2013JCAP...11..012L, 2013JCAP...11..056B} do suffer from imaginary square-root problems in low-density regions (see \cite{2015arXiv150503539W} for a discussion about the meaning of these imaginary square-root problems).


\subsubsection{Quasi-static and non-static simulations of the Symmetron model}\label{sect:nonstaticsims}

The simulations of the $f(R)$ and DGP models are performed under the quasi-static approximation. To go beyond this approximation means to explicitly take into account the time derivative terms of the scalar field in the equations. This way, the solution for the scalar field at a given time depends also on its past evolution, as opposed to depending only on the matter configuration at that given time. To date, non-static cosmological simulations of modified gravity have been performed for the Symmetron and disformal gravity models using the explicit leap-frog method \citep{2013PhRvL.110p1101L,2014PhRvD..89h4023L,2015arXiv150407142H}, for $f(R)$ gravity using the implicit Newton-Gauss-Seidel method \citep{2015JCAP...02..034B} and for the DGP and the Cubic Galileon models using both of the methods mentioned above, but only in a spherical symmetric spacetime \citep{2015arXiv150503539W}.

For the Symmetron model, the full Klein-Gordon equation~(\ref{eq:eom_symm}) reads
\be
\label{eq_motion_chi}
\ddot{\chi} + 3H\dot{\chi} - \frac{\nabla^2\chi}{a^2} = -\frac{1}{2\lambda_0^2} \left[ \frac{a_{SSB}^3}{a^3}\chi\eta - \chi + \chi^3  \right], 
\ee
where $\eta$ is the matter density in units of the background value. In \isisns{} (the modified version of \isis{} that relaxes the quasi-static approximation), the second order equation of motion of the scalar field is decomposed into a system of two first order equations as
\bq
 \dot{\chi} &=& \frac{q}{a^3},\\
\label{hamilton_2}
\dot{q} &=& a \nabla^2\chi - \frac{a^3}{2\lambda_0^2} \left[ \frac{a_{SSB}^3}{a^3}\chi\eta - \chi + \chi^3  \right],
\eq
which are used to propagate both $\chi$ and $q$ using a leapfrog algorithm. The "position" $\chi$ and "velocity" $q$ are displaced from each other by $1/2$ timestep and the discretised equations become
\bq
\chi_{n} &=& \chi_{n-1} + \dot{\chi}_{n-1/2}\Delta t,\\
q_{n+1/2} &=& q_{n-1/2} + \dot{q}_{n}\Delta t,
\eq
where $f_n \equiv f(t_n)$. The spatial derivatives of $\chi$ in the formulas above are calculated from the grid using a 5-point stencil, see \citet{2014PhRvD..89h4023L} for a detailed description of the implementation of the scheme.

The quasi-static simulations of the Symmetron model are performed by discretising equation~(\ref{eq_motion_chi}) and neglecting the first two terms on the left-hand side. In this case, one does not need to explicitly evolve the scalar field. Instead, given the matter distribution, $\eta$, at a given time step, the code relaxes the equation that contains only the scalar field and its second spatial derivative (not time derivatives), using the multigrid methods described in Section \ref{sec:multigrid}.


\section{Results}
\label{sec:results}

In this section we present the main results of this code comparison project. We present and discuss the different code results for the matter and velocity divergence power spectra and halo mass function, as well as the halo profiles of the scalar degree of freedom, forces, and density.


\subsection{Simulation setup}

All the simulations performed in this study have used the same initial conditions, which were generated using 2LPT \citep{2006MNRAS.373..369C} from a \lcdm{} cosmology with $\Omega_m = 0.269$, $\Omega_\Lambda = 0.731$, $h=0.704$, $n_s = 0.966$ and $\sigma_8 = 0.8$. The simulations have $N=512^3$ particles in a box of size $B=250 h^{-1} {\rm Mpc}$ and they start at redshift $z=49$. All modified gravity models simulated here have the same expansion history as a \lcdm{} model with the above parameters {and the evolution of density perturbations at high redshifts ($z\gtrsim 10$) is almost identical to that of the $\Lambda$CDM model justifying the use of the same initial conditions.}

For simulations with a \ramses{} based code, we used a coarse-level grid with refinement level $l_{\rm min} = 9$ corresponding to $512^3$ coarse cells. Each cell was refined if the number of particles contained in it exceeded 8. The maximum level of refinement corresponded to $l_{\rm max}=15$ for F6 and $r_cH_0 = 5$ and $l_{\rm max}=16$ for F5 and $r_cH_0 = 1$. For the \mggadget{} simulations the relative tree opening criterion was used where the tree is open when the relative force acceleration error is larger than $0.0025$, the force softening was $18.75$ kpc$/h$ and for the long-ranged forces a particle mesh grid with $512^3$ grid cells was used. The force softening corresponds to the grid spacing at level $ \simeq 14$ in the AMR hierarchy.

The codes compared in this project differ in two aspects: (i) the exact way in which the equations of the scalar field are solved and (ii) the force calculation and time stepping, which were taken from the original codes they were built from (e.g. \pgadget{} and \ramses{}). To separate the differences that arise from these two aspects, simulations of the standard \lcdm{} model were also performed. This way, differences between the code predictions for the absolute value of a measured quantity (e.g. $P_k$) are affected by both aspects. On the other hand, comparisons of the code predictions for the relative difference to \lcdm{} (e.g. $\Delta P_k/P_{k, \Lambda{\rm CDM}}$) should be mostly determined by aspect (i) and not so much by aspect (ii).

The run-time of the modified gravity simulations we have performed was about $5-10$ times that of the corresponding $\Lambda$CDM simulation.


\subsection{Matter power spectrum}
\label{sec:power}

As a consistency check, we have measured the matter power spectrum, $P_k$, from the simulations using three independent codes. One is the \powmes{} code \citep{2009MNRAS.393..511C}, which uses fourier transforms with folding methods to compute $P_k$. Another is a code written by one of the authors of this paper (RS), which works similarly to \powmes{} except that it does not apply folding methods. Both these codes deconvolve the window function of the density assignment and shot-noise is subtracted. Lastly, we have also measured the power spectrum using a density field obtained with the {\it Delanuay Tesselation Field Estimator} (DTFE) method of \cite{Schaap_vdWeygaert2000}. We have found that the results from these three codes agree very well (to the $1\%$ level \footnote{For this comparison we have ignored the four largest Fourier modes where cosmic variance is significant since the different codes use different methods to estimate the power on these scales which can lead to quite large differences.}) and when considering $\frac{\Delta P}{P_{\rm LCDM}}$ the agreement was below $0.1-0.5\%$ for all scales of interest. All the plots shown in this paper are those obtained using the code made by RS mentioned above. We show the power-spectrum out to the particle Nyquist frequency, $k_{\rm max} =  \pi N_{\rm particles}^{1/3}B_0^{-1} = 6.4 h \text{Mpc}^{-1}$, of the simulations. 

Next, we present our results for the matter power spectrum, discussing separately the results obtained for $\Lambda$CDM, $f(R)$, DGP and Symmetron models.


\subsubsection{$\Lambda$CDM}

\begin{figure*}
\includegraphics[width=2.1\columnwidth]{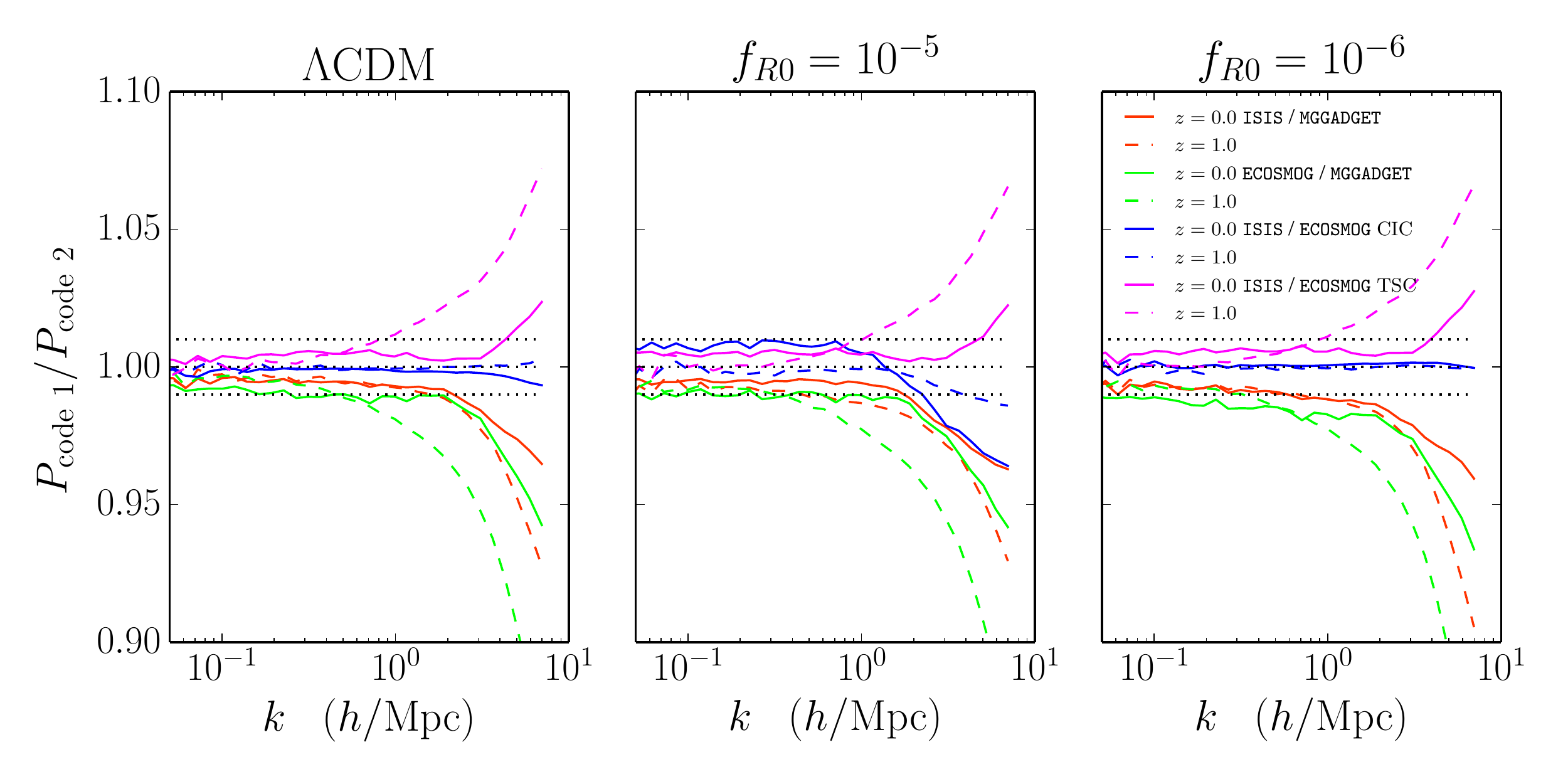}
\caption{Comparison of the matter power spectrum results of the \ecosmog{}, \mggadget{} and \isis{} codes for the \lcdm{} (left), F5 (middle) and F6 (right) models, as labelled. When comparing the two \ramses{}-based codes (\isis{} and \ecosmog{}), we show the \ecosmog{} results from simulations run with CIC and with TSC interpolation schemes, as labelled.}
\label{fig:pofk_LCDM_F5_F6}
\end{figure*}

Before comparing the results for the modified gravity models, it is instructive to have a look at how the codes compare for \lcdm{}. This is shown in the left panel of Fig.~\ref{fig:pofk_LCDM_F5_F6} for $z = 0$ (solid) and $z = 1$ (dashed). The red and green lines show, respectively, the ratio of the \isis{} and \ecosmog{} result to that of \mggadget{}. One notes that the two \ramses{}-based codes predict less power ($4-5\%$ at $z = 0, k \approx 7h {\rm Mpc}^{-1}$) than the \mggadget{} simulations. This can be attributed to the different ways the base codes compute the gravitational force on small scales. In {\sc p-gadget3}, the force on large scales is computed using a particle-mesh method, just like in \ramses{}. On small scales, however, {\sc p-gadget3} switches to a tree method, whereas in \ramses{} the calculation remains as on large scales. The accuracy of the \ramses{} code on small scales depends also on the criteria to trigger a refinement of the AMR grid. In all \ramses{}-based AMR simulations of this paper, the grid refines itself whenever the particle number inside a given cell exceeds 8. Due to these differences between the force calculation on small scales, one should therefore not expect perfect agreement between the \ramses{}-based and {\sc gadget}-based codes. We refer the reader to \cite{2015arXiv150305920S} for a more detailed comparison study of the performance of the \ramses{}, {\sc gadget} and also {\sc Pkdgrav3} codes in $\Lambda$CDM simulations.

The result shown by the blue and pink lines illustrates the impact of using different schemes for interpolating the density and the force between the particle positions and the grid in the two \ramses{}-based codes. The blue lines show that the \lcdm{} results obtained with \isis{} and \ecosmog{} are in very good agreement ($< 1\%$ error for all scales and times shown) if both codes use the same interpolation scheme, in this case CIC. This shows that the modifications made to \ramses{} to develop \ecosmog{} and \isis{} do not introduce any systematics in the way the codes work for GR \footnote{We note that the versions of the \isis{} and \ecosmog{} used in this paper are not built on the exact same release of \ramses{}, which explains why their $\Lambda{\rm CDM}$ results are not 'exactly' the same.}. On the other hand, compared to the \ecosmog{} run with TSC, the power in \isis{} is higher by $\approx 3\%, (7\%)$ for $z = 0, (z = 1)$, for $k\approx 7h {\rm Mpc}^{-1}$. This is because the CIC interpolation distributes the mass of each particle into fewer grid cells, which results in higher peaks in the density field, compared to TSC. We note that the size of the differences between \isis{} and \ecosmog{} to \mggadget{} are comparable to the differences induced by different interpolation schemes.

\subsubsection{$f(R)$}

\begin{figure*}
\includegraphics[width=2.1\columnwidth]{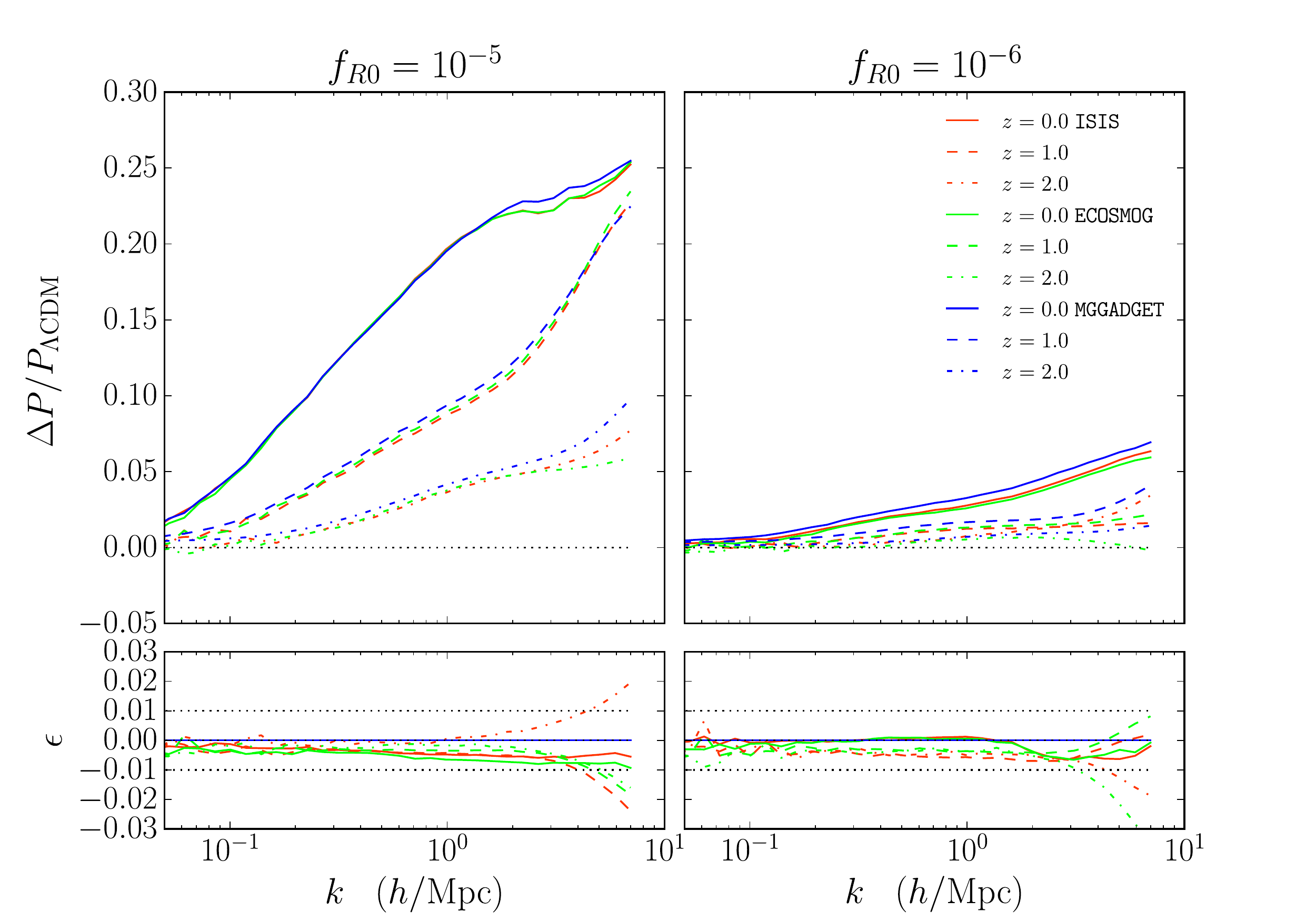}
\caption{Fractional difference of the matter power spectrum with respect to \lcdm{} from the simulations of the F5 (left) and F6 (right) models performed with the \ecosmog{} (TSC), \mggadget{} and \isis{} codes, as labelled. In the lower panel insets, $\epsilon = \left(P/P_{\Lambda{\rm CDM}}\right)_{\rm code}/\left(P/P_{\Lambda{\rm CDM}}\right)_{\rm ref} - 1$, with \mggadget{} being the reference code. At $z = 0$, the codes are all accurate to within $1\%$ on all scales shown.}
\label{fig:dpofk_F5_F6}
\end{figure*}

The $f(R)$ model was simulated with \ecosmog{}, \mggadget{} and \isis{}. In all codes, the relaxation of the scalar field equation was performed on an AMR grid.  The results are shown in the middle and right panels of Fig.~\ref{fig:pofk_LCDM_F5_F6} for the F5 and F6 $f(R)$ models, respectively.  Fig.~\ref{fig:dpofk_F5_F6} shows the relative difference to \lcdm{} in F5 (left) and F6 (right) for $z = 0, 1, 2$.  This depicts the known result that the modifications to gravity in this model boost structure formation on small scales and that these effects are stronger in the F5 than in the F6 model. It is perhaps also interesting to note that, for the F5 model at $z = 0$, the relative difference to \lcdm{} does not flatten out on scales $k \lesssim 0.1h {\rm Mpc}^{-1}$, which are scales on which the growth is scale-independent in $\Lambda{\rm CDM}$. This shows that in modified gravity, the naive expectation (inspired from \lcdm{}) for the scales on which the growth of structure is scale-independent can be misleading \citep{Hellwing2013_clustering}.  

The differences observed in the right panels of Fig.~\ref{fig:pofk_LCDM_F5_F6} are similar in shape and size to those in the left panel for \lcdm{}, which suggests that the calculation of the fifth force is consistent in between the three codes. This is confirmed by the result depicted in Fig.~\ref{fig:dpofk_F5_F6}, which shows the good agreement between the three codes for all scales and times shown \footnote{Note that \mggadget{} agrees well with the two \ramses{}-based codes on small scales. This shows that any errors arising from summation errors in the tree algorithm in {\sc gadget} do not translate into discrepant power spectrum results.}. In particular, at $z = 0$ ($z = 2$), all codes agree to within $1\%$ for scales $k < 7h {\rm Mpc}^{-1}$ ($k \lesssim 5h {\rm Mpc}^{-1}$). From this, we can conclude that  any differences between these three modified gravity codes for $f(R)$ are driven almost exclusively by the differences in their main codes (in this case \ramses{} and {\sc gadget}), and not by the extra modules that solve for the effects of the fifth force. This is a very reassuring result.

\subsubsection{DGP}

\begin{figure*}
\includegraphics[width=1\columnwidth]{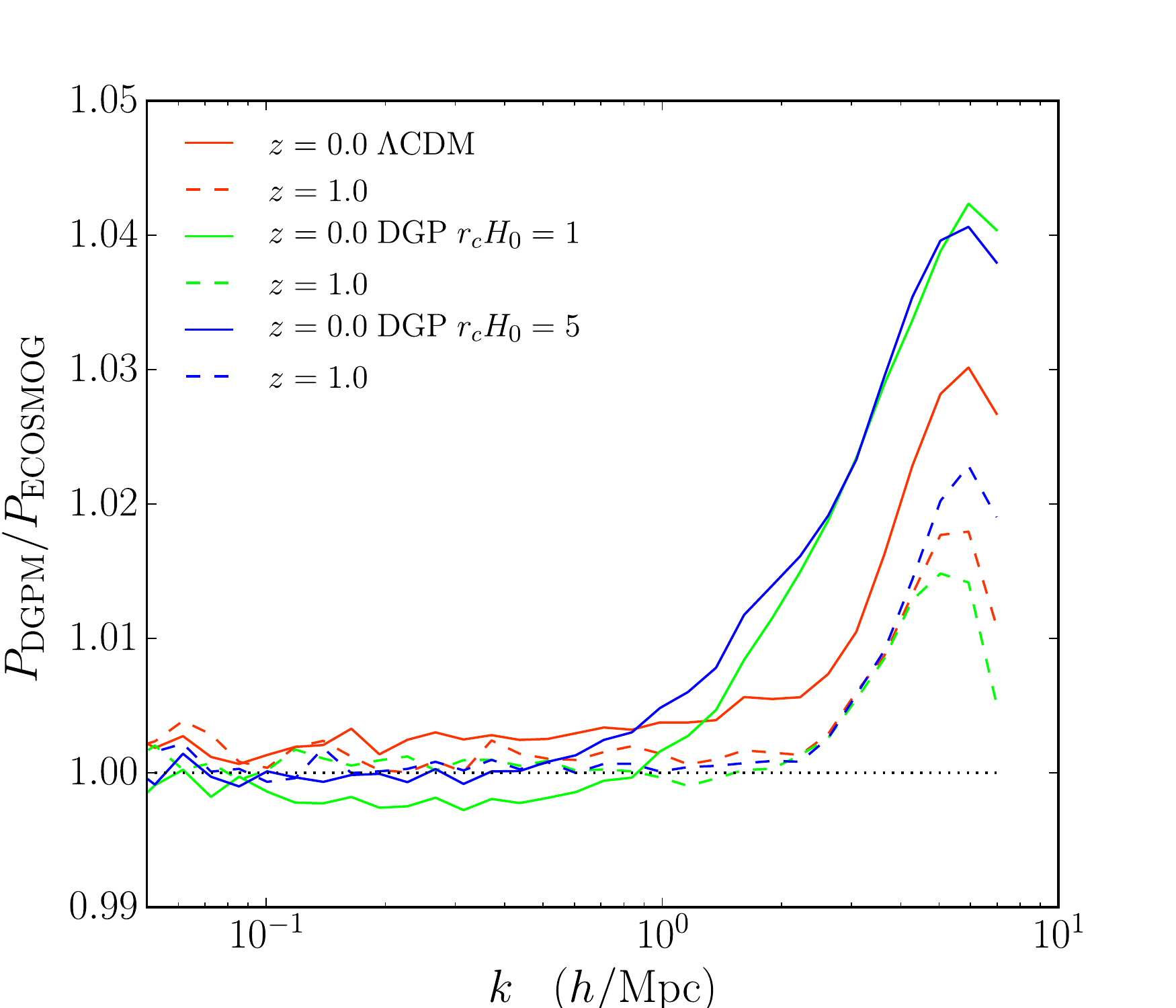}
\includegraphics[width=2\columnwidth]{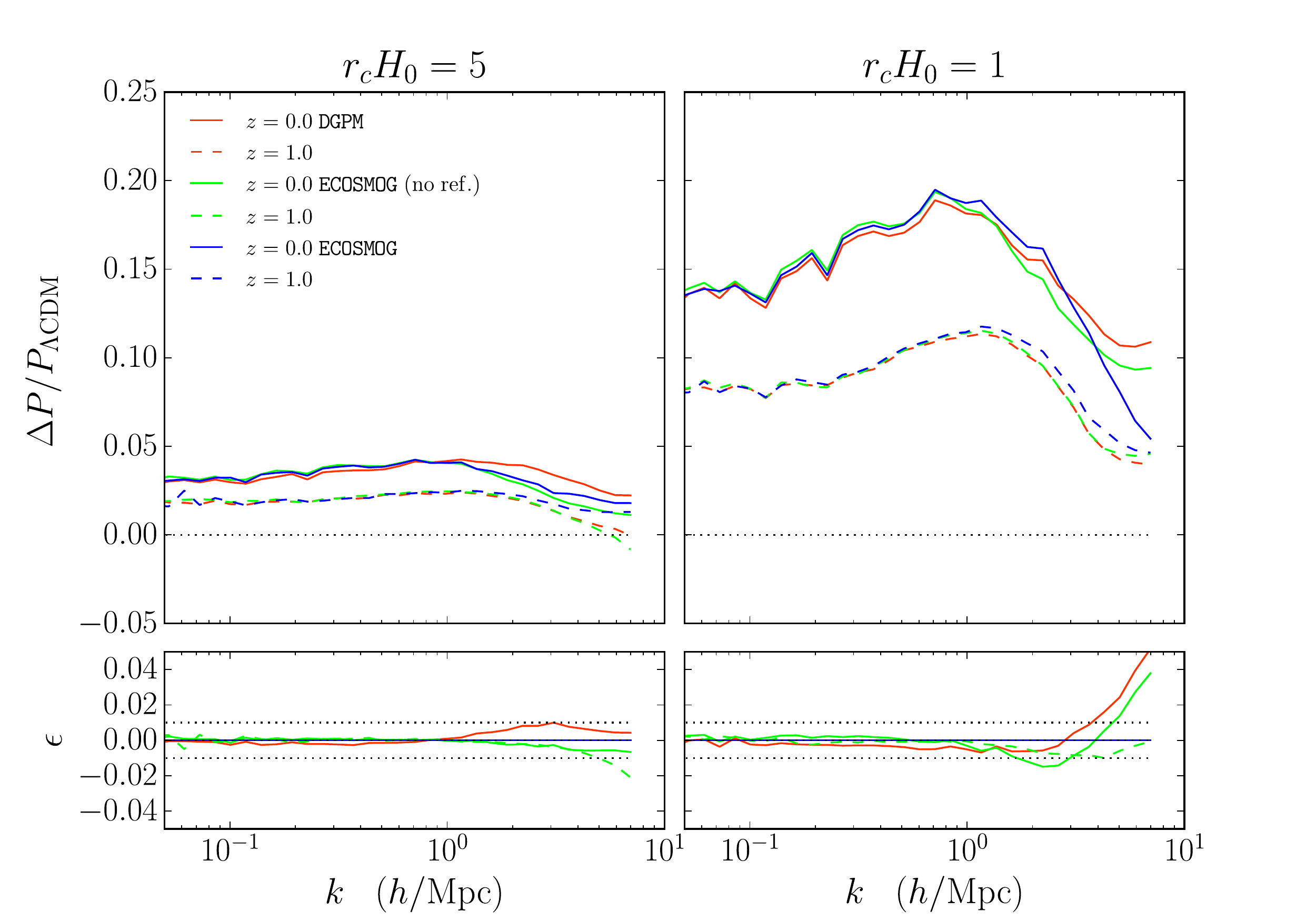}
\caption{The top panel shows the comparison of the matter power spectrum results of the \dgpm{} and \ecosmog{} (fixed grid) simulations of the \lcdm{}, $r_cH_0 = 5$ and $r_cH_0 = 1$ DGP models, as labelled. The lower left and lower right panels show the fractional difference with respect to \lcdm{} of the two codes for the $r_cH_0 = 5$ and $r_cH_0 = 1$ DGP models, respectively. The lower panels show the \ecosmog{} results both for fixed and refined grid simulations, as labelled. In the lower panel insets, $\epsilon = \left(P/P_{\Lambda{\rm CDM}}\right)_{\rm code}/\left(P/P_{\Lambda{\rm CDM}}\right)_{\rm ref} - 1$, with \ecosmog{} (refined grid) being the reference code.}
\label{fig:pofk_DGP_dPP}
\end{figure*}

The simulations of the DGP model were performed with the \dgpm{} and \ecosmog{} codes. The latter was run both with refinements and with a fixed grid to better compare with \dgpm{}, which is fixed grid only. The top panel of Fig.~\ref{fig:pofk_DGP_dPP} shows the ratio of the \lcdm{} power spectrum of the \dgpm{} to \ecosmog{} simulations without refinements for $z = 0$ and $z =1$. 
The lower panels of Fig.~\ref{fig:pofk_DGP_dPP} show the relative difference to \lcdm{} measured in the simulations of the $r_cH_0 = 5$ (left) and $r_cH_0 = 1$ (right) DGP models. We recover the known result that, in the DGP model, the amplitude of the power spectrum is boosted by a scale-independent factor on scales $k \lesssim 0.1h {\rm Mpc}^{-1}$. On mildly nonlinear scales, $0.1h {\rm Mpc}^{-1} \lesssim k \lesssim 1h {\rm Mpc}^{-1}$, the boost in the power spectrum is stronger than on linear scales due to mode-coupling. However, on nonlinear scales, $k \gtrsim 1h {\rm Mpc}^{-1}$ (halo size scales), the suppression effects of the Vainshtein screening mechanism are dominant, which effectively reduces the impact of the fifth force on the power spectrum \citep[e.g.,][]{2010PhRvD..81f3005S}.  

Fig.~\ref{fig:pofk_DGP_dPP} shows that the three codes agree very well (up to $1\%$) on scales $k \lesssim 1 h{\rm Mpc}^{-1}$. For $k \gtrsim 1 h{\rm Mpc}^{-1}$, however, the power in the \dgpm{} simulations is higher than in \ecosmog{}. This is due to the different interpolation schemes used in the codes. In particular, in \dgpm{}, the CIC interpolation yields a density field with higher peaks than the density field in \ecosmog, which is smoother because of the use of the TSC scheme. This is similar to the \ecosmog{} and \isis{} results in the left panel of Fig.~\ref{fig:pofk_LCDM_F5_F6} (blue and pink lines).  The lower panels of Fig.~\ref{fig:pofk_DGP_dPP} also show the \ecosmog{} result with refinements (blue). For the $r_cH_0 = 5$ model the three codes are, overall, in good agreement for all redshifts and scales shown ($< 1\%$ for $k \lesssim 5h {\rm Mpc}^{-1}$). However, the modifications to gravity in the $r_cH_0 = 5$ model are weaker than in the $r_cH_0 = 1$ case, and as a result, it is easier to interpret the code results for the $r_cH_0 = 1$ model. For this case, the three codes agree very well for $k \lesssim 1 h {\rm Mpc}^{-1}$. Note also that the \ecosmog{} results with refinements agree with its results for fixed grid on these large scales. For $k \gtrsim 1 h {\rm Mpc}^{-1}$, \dgpm{} is also in good agreement with the results from \ecosmog{} for fixed grid. Recall that the two codes solve the DGP scalar field equation in substantially different ways (cf.~Section~\ref{sec:dgpsims}), so this is a nontrivial test.  On these small scales, the agreement of the fixed grid codes with the \ecosmog{} code with refinements gets worse, but this is expected due to the gain in resolution in the latter. It is also interesting to note that, at $z = 0$ and for $k \gtrsim 4h {\rm Mpc}^{-1}$, the enhancement in the power is smaller in the \ecosmog{} simulations with refinements compared to the fixed grid cases. The explanation here is that the AMR nature of the grid allows it to resolve better the higher density peaks that exist on these smaller scales. This results in the code capturing better the suppression effects of the screening, and hence, the boost in the power relative to \lcdm{} becomes less pronounced. This is more pronounced at $z = 0$ compared to $z=1$ because at earlier times the density field is less evolved, hence the screening efficiency is also weaker.  

In summary, we conclude that the two available N-body implementations of the Vainshtein screeening agree well in the non-refining case, with the differences from the refined case appearing fully consistent with being due to the higher resolution of the latter.  In the future, it would be desirable to also test an independent implementation of Vainshtein screening with refinements.

\subsubsection{Symmetron}

\begin{figure*}
\includegraphics[width=2\columnwidth]{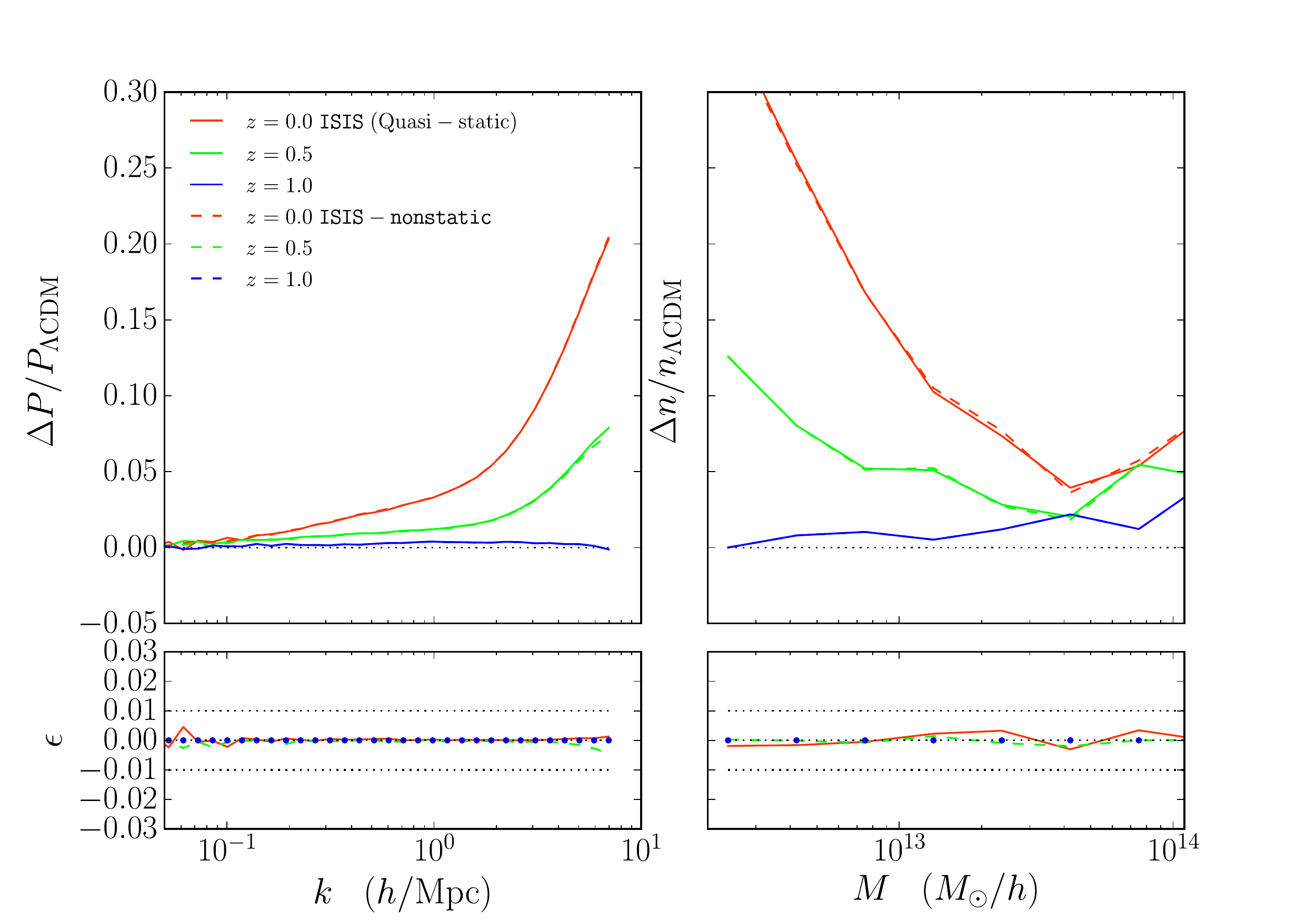}
\caption{Comparison of the matter power spectrum (left) and halo mass function (right) results of the Symmetron model simulations performed by the \isis{} and \isisns{} codes. The lower panels shows the error induced by applying the quasi-static approximation, $\epsilon =  P_{\rm Quasi-static} / P_{\rm Full}-1$, whereas the upper panels shows the fractional difference relative to \lcdm{}, as labelled. In all panels, the error induced by employing the quasi-static limit lies comfortably below $1\%$ for all the times and scales shown.}
\label{fig:pofk_qsa}
\end{figure*}

The left panel of Fig.~\ref{fig:pofk_qsa} shows the power spectrum results for the Symmetron model, which were obtained with the \isis{} and \isisns{} code, both run without refinements. The figure shows that the impact of the time-derivative terms in the equation of the Symmetron model, equation~(\ref{eq_motion_chi}), is below the $0.5\%$ level for all times and scales shown. Moreover, there seems to be no trend with scale. Hence, we can conclude that, in what concerns measurements of the nonlinear matter power spectrum from {\it N}-body simulations of the Symmetron model, the use of the quasi-static limit has virtually no impact on the results.

This result is not unexpected since the calculation of the matter power spectrum is dominated by high-density regions (haloes, if one thinks about it in the framework of the halo model), where the time derivatives are indeed expected to be negligible relative to the spatial ones. Consequently, it may be of interest to investigate whether the quasi-static assumption remains also a good approximation for observables which are more sensitive to lower density regions. Such an investigation is not explored in this study.

It would also have been good to have a comparison with \mggadget{} and \ecosmog{} for the symmetron model but we leave this to future work. A brief comparison of $P(k)$ for the (quasi-static) symmetron model and the codes \isis{}, \ecosmog{} and {\sc mlapm} can be found in \cite{2014AA...562A..78L}.

\subsection{Velocity divergence spectra}
\label{sec:velpower}

We have measured the power spectrum of the velocity divergence field defined as $P_{\theta\theta}(k)\equiv\langle\theta_k^2\rangle$, where $\theta(\mathbf{x})=H_0^{-1}\nabla\cdot v(\mathbf{x})$, with $v$ being the peculiar velocity field. In the linear regime, $\theta$ is related to the matter density contrast as $\theta \propto - \delta f$, where $f = {\rm dln}\delta/{\rm dln}a$ is the linear growth rate. We show only results for the divergence of the velocity field, but note that on small scales, where nonlinear processes become important, the vorticity (rotational component of $v(x)$) is nonnegligible and hence the whole velocity field cannot be described solely by $\theta$.

To measure $P_{\theta\theta}(k)$, we constructed a volume-weighted velocity field \citep{Bernardeau1996} with the DTFE method implemented in the publicly available code of \cite{2011arXiv1105.0370C}. We refer the reader to \cite{2013MNRAS.428..743L} for more details about our method to compute the velocity divergence field from the {\it N}-body particle positions and velocities. Next, we discuss our results for the $f(R)$ and DGP simulations. We have also measured $P_{\theta\theta}$ for the Symmetron simulations of the \isis{} and \isisns{} codes, but since there are virtually no differences between the full and quasi-static results, we refrain from showing them.


\subsubsection{$f(R)$}
\label{subsec:velpow_fR}
 
\begin{figure*}
\includegraphics[width=2\columnwidth]{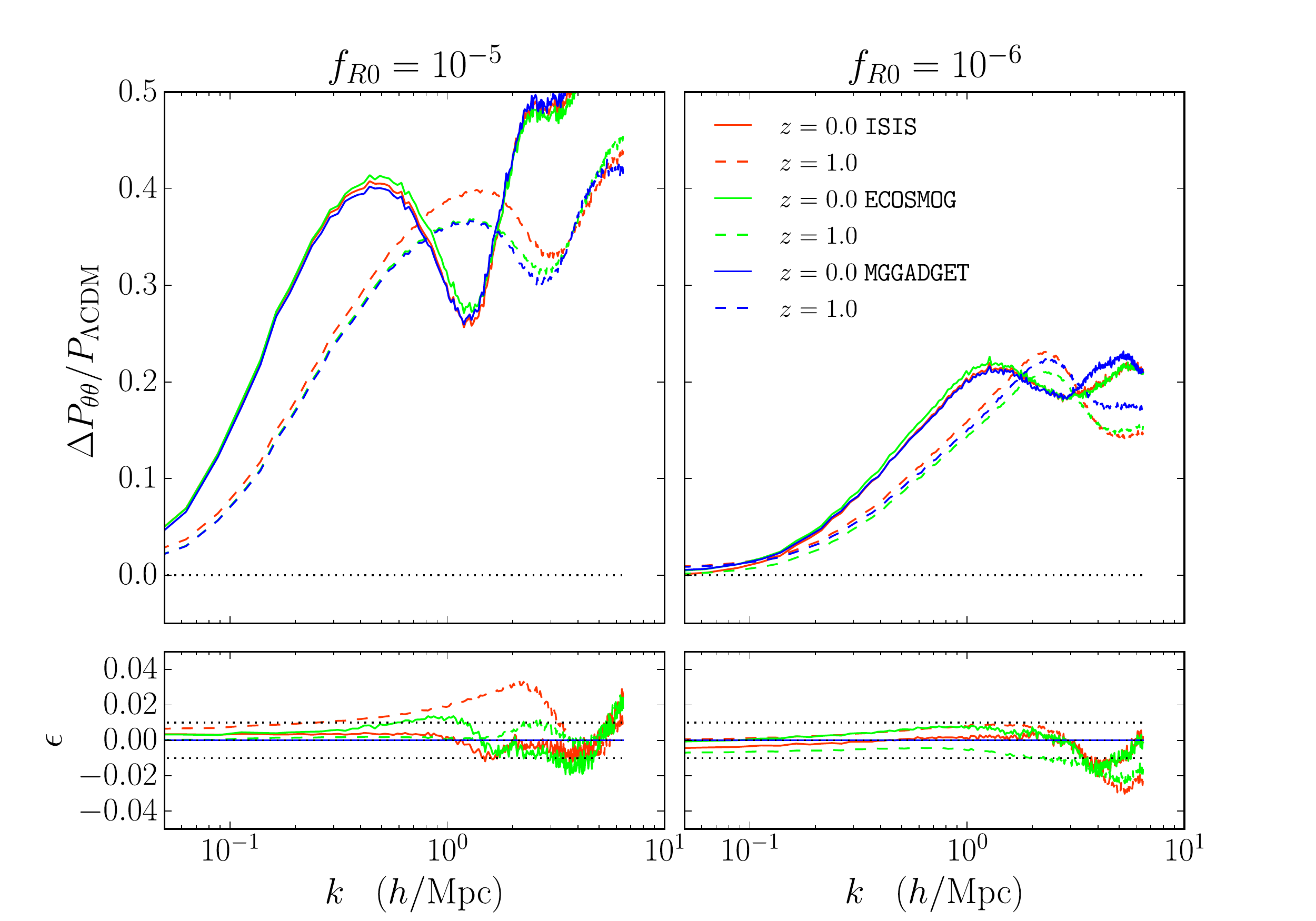}
\caption{Fractional difference of the velocity divergence power spectrum with respect to \lcdm{} from the simulations of the F5 (left) and F6 (right) models performed with the \ecosmog{} (CIC), \mggadget{} and \isis{} codes, as labelled. In the lower panel insets, $\epsilon = \left(P/P_{\Lambda{\rm CDM}}\right)_{\rm code}/\left(P/P_{\Lambda{\rm CDM}}\right)_{\rm ref} - 1$, with \mggadget{} being the reference code.}
\label{fig:dptt_F5_F6}
\end{figure*}

Fig.~\ref{fig:dptt_F5_F6} shows the fractional difference of the velocity divergence power spectrum with respect to \lcdm{} in the F5 (left) and F6 (right) models.  
The enhancement in the amplitude of $P_{\theta\theta}$ in $f(R)$ relative to \lcdm{} is noticeably larger than that seen for the matter power spectrum \citep[see also][]{2012MNRAS.425.2128J,2013MNRAS.428..743L,Li2013JCAP...11..012L,2014PhRvL.112v1102H}. In particular, for the F5 (F6) model at $z = 0$ and $k \approx 3h {\rm Mpc}^{-1}$, the amplitude of $P_{\theta\theta}$ is enhanced by $\approx 50\%$ ($\approx 20\%$), while the boost in the amplitude of $P(k)$ is kept at $20-25\%$ ($\approx 5\%$) only.  The velocity field is more sensitive than the density field to the modifications to gravity because it starts to be affected at earlier times, and the effects get accumulated throughout the history of structure growth.

As in the case of the matter power spectrum, the agreement between \ecosmog{}, \mggadget{} and \isis{} is notable. In particular, for both F5 and F6 at $z = 0$, the three codes agree up to $\approx 1\%$ down to $k\sim 3 h {\rm Mpc}^{-1}$, and for higher $k$-values the difference never exceeds $2\%$. The case of the \isis{} code for F5 at $z=1$ is an exception to this very good agreement, for which the difference with the other two codes is at the $2-4\%$ level for $k \sim 0.3 - 3h {\rm Mpc}^{-1}$. We have checked that at $z=2$ the agreement between the three codes is at the $1-2\%$ level, which suggests that the larger disagreement at $z=1$ could be due to a transient effect related, for instance, to differences in the time integration \footnote{For example, slight differences in the AMR structure and velocity field in the \ecosmog{} and \isis{} simulations can result in different time step sizes (determined by the same criteria as in standard \ramses{}).}. We leave a more thorough investigation of this difference for future work.

\subsubsection{DGP}
\label{subsec:velpow_DGP}

\begin{figure*}
\includegraphics[width=2\columnwidth]{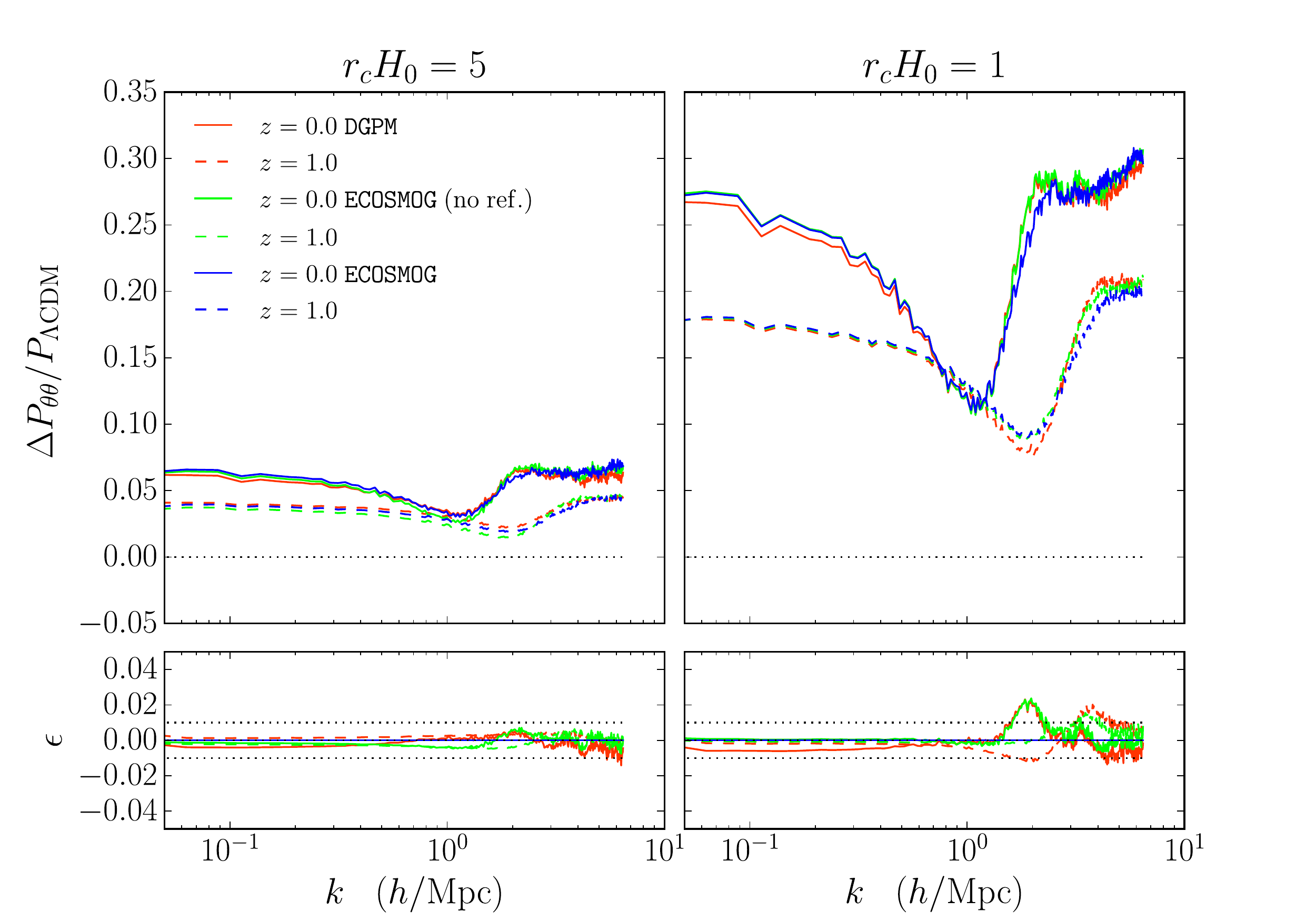}
\caption{Fractional difference of the velocity divergence power spectrum with respect to \lcdm{} from the simulations of the $r_cH_0 = 5$ (left) and $r_cH_0 = 1$ (right) DGP models performed with the \dgpm{} and \ecosmog{} codes. The two sets of \ecosmog{} results correspond to the results from simulations run with fixed and refined grids, as labelled. In the lower panel insets, $\epsilon = \left(P/P_{\Lambda{\rm CDM}}\right)_{\rm code}/\left(P/P_{\Lambda{\rm CDM}}\right)_{\rm ref} - 1$, with \ecosmog{} (refined grid) being the reference code.}
\label{fig:dptt_DGP}
\end{figure*}

Fig.~\ref{fig:dptt_DGP} shows the fractional difference of the velocity divergence power spectrum with respect to~\lcdm{} in the $r_cH_0 = 5$ (left) and $r_cH_0 = 1$ (right) DGP models. The agreement between \dgpm{} and \ecosmog{} is very good, with any differences in $r_cH_0 = 5$ predictions being below $\approx 1\%$ for all times and scales shown. For the $r_cH_0 = 1$ model the agreement between the codes worsens, but the difference is always below the $2\%$ level. Similarly to the case of the $f(R)$ model, we also note that the modifications to gravity in the DGP model affect the amplitude of $P_{\theta\theta}$ more than they affect the amplitude of $P(k)$.

In Fig.~\ref{fig:dptt_DGP}, it is interesting to note that on small scales ($k\gtrsim2-3h {\rm Mpc}^{-1}$), the difference between the fixed (red and green) and refined grid (blue) results is smaller than that seen in Fig.~\ref{fig:pofk_DGP_dPP} for the matter power spectrum. Given the gain in resolution when \ecosmog{} is run with refinements, one does not expect the results to fully agree with fixed grid simulations which cannot resolve small scale structures, and as a result, the agreement depicted in Fig.~\ref{fig:dptt_DGP} may seem surprising. We do not perform any detailed investigations of this result, but simply note that the density field used to compute $P(k)$ is mass-weighted, whereas the density field used to compute $P_{\theta\theta}$ is volume-weighted. This, together with the suppressed contribution of virial velocities to the velocity divergence, may help explain why the use of adaptively refined grids does not have a critical impact on the resulting $P_{\theta\theta}$. For completeness, we further note that, as found in \cite{2015arXiv150306673F}, peculiar velocities on small scales are also less affected by the Vainshtein mechanism, compared to the effects of the Chameleon mechanism.

\subsection{Halo mass function}
\label{sec:HMF}

The halo mass function, $n(M)$, defined as the number density of dark matter haloes with mass $M$, is an important statistic that is particularly sensitive to modifications to gravity. We used a modified version of the spherical overdensity {\sc Amiga's Halo Finder} (\ahf{}) code \citep{MHF,Amiga} to identify dark matter haloes and calculate their profiles. For this, simulation outputs from all codes were written in the standard \gadget\ format, but with new data blocks added, which contain the components of standard gravity and the fifth force as well as the Newtonian potential and the scalar field at particle positions. We outputted quantities at particle positions rather than leaf cells of the AMR grids or trees because, unlike the leaf cells, the particle IDs are the same in all simulations, making comparisons more straightforward. Note that this means that the scalar field and forces in underdense regions were not very well sampled, which however should not be a serious issue given that we are mostly interested in haloes. 

The major modifications to \ahf{} were threefold: (i) new routines to read the above data, which has the same format as the default \gadget{} data such as particle coordinates and velocities; (ii) new routines to compute the force and scalar field profiles in haloes, by averaging over their values at all particle positions in a given spherical shell (the radial binning scheme for this was the same as in the default \ahf{} code); (iii) the routine in \ahf{} which does the removal of unbound particles was also modified so that the code determined whether a particle was bound or not by comparing its velocity with the total gravitational potential instead of the standard Newtonian potential. Note that if (iii) is not properly done, then the mass function tends to be lower because more particles are considered as too fast to be bound. \cite{2010PhRvD..81j4047L} studied the effect of taking account the fifth force in the halo unbinding process for certain chameleon models using {\it N}-body simulations and found a noticeable difference in the resulting mass functions (see also \cite{2010MNRAS.408L.104H, Hellwing2013}).

We used \ahf{} with $\Delta_{\rm vir} = 200$ so that $M_{\rm AHF} = M_{200c}$. Next, we discuss our mass function results for the $f(R)$, DGP and Symmetron simulations.


\subsubsection{$f(R)$}

\begin{figure*}
\includegraphics[width=2.1\columnwidth]{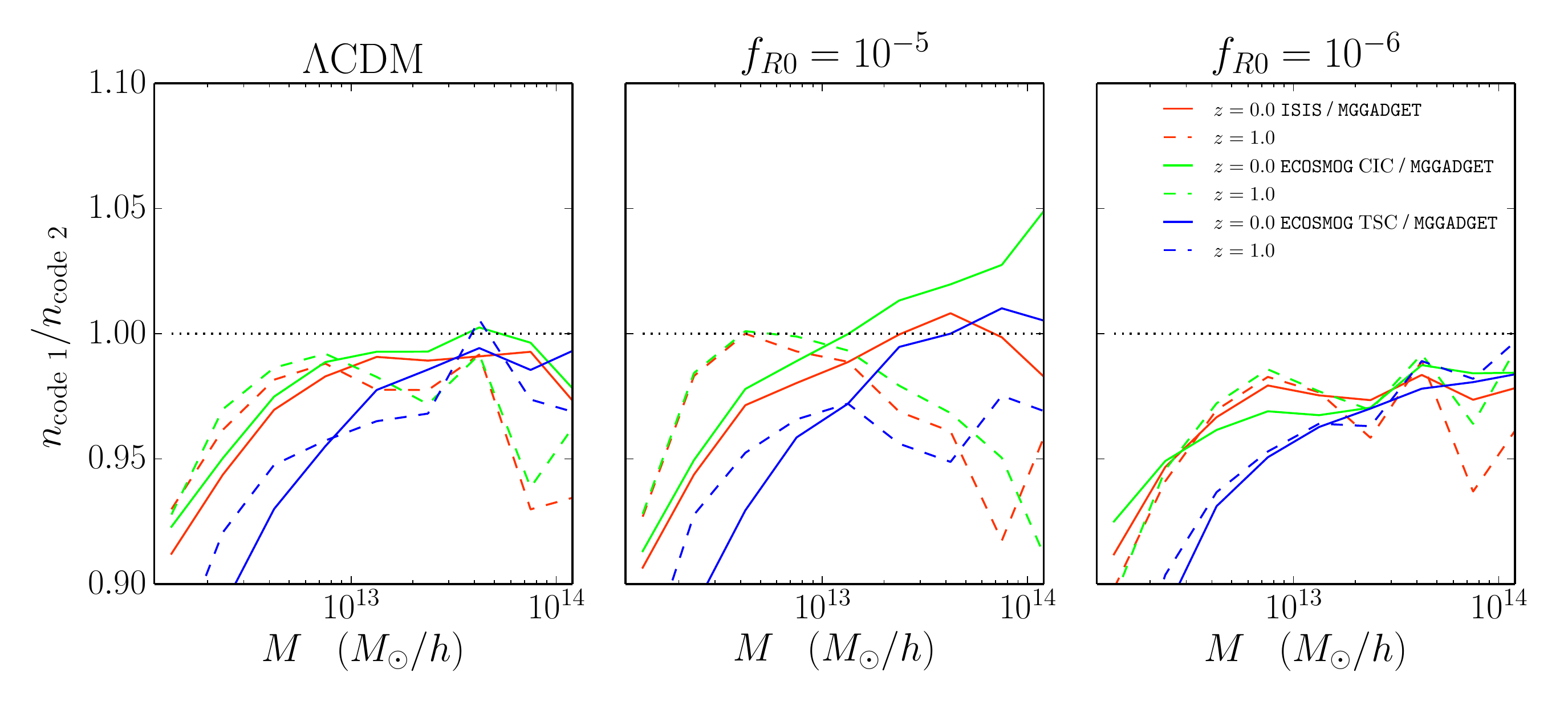}
\caption{Comparison of the halo mass function results of the \ecosmog{}, \mggadget{} and \isis{} codes for the \lcdm{} (left), F5 (middle) and F6 (right) models, as labelled. The two sets of \ecosmog{} results correspond to the results from simulations using the CIC and TSC density assignment, as labelled.}
\label{fig:nofm_F5_F6_LCDM}
\end{figure*}

\begin{figure*}
\includegraphics[width=2.1\columnwidth]{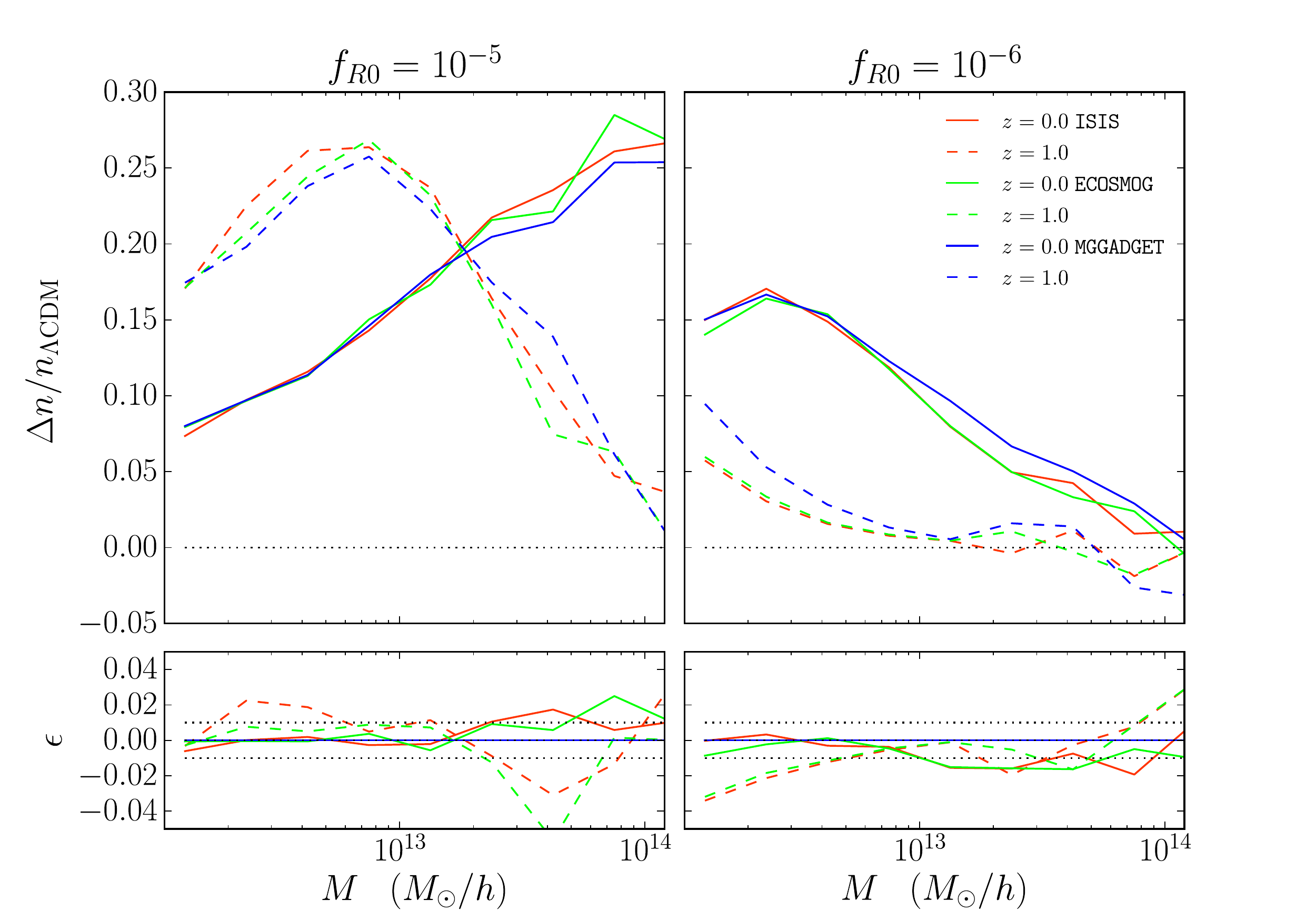}
\caption{Fractional difference of the halo mass function with respect to \lcdm{} from the simulations of the F5 (left) and F6 (right) models performed with the \ecosmog{} (TSC), \mggadget{} and \isis{} codes, as labelled. In the lower panel insets, $\epsilon = \left(n/n_{\Lambda{\rm CDM}}\right)_{\rm code}/\left(n/n_{\Lambda{\rm CDM}}\right)_{\rm ref} - 1$, with \mggadget{} being the reference code.}
\label{fig:dnofm_F5_F6}
\end{figure*}

Figs.~\ref{fig:nofm_F5_F6_LCDM}-\ref{fig:dnofm_F5_F6} show the mass function results of the \ecosmog{}, \isis{} and \mggadget{} simulations for $f(R)$.  As first shown quantitatively in \cite{2009PhRvD..79h3518S}, $f(R)$ models predict an enhancement in the abundance of haloes relative to \lcdm{}. The enhancement is more pronounced in the F5 model because the Chameleon screening is less efficient. This is particularly noticeable at the high-mass end for which, in the F6 model, the number density of haloes is almost the same as in \lcdm{}. The differences in the mass dependence of $\Delta n/n_{\Lambda{\rm CDM}}$ for different $f(R)$ model parameters and redshifts illustrates the complex interplay between the mass and time dependence of the Chameleon screening mechanism \citep[see e.g.][for studies of halo properties in $f(R)$]{2014JCAP...03..021L, 2015arXiv150301109S, 2015arXiv150507129G, Gronke:2014gaa, Gronke:2013mea, 2012ApJ...756..166W}.  

Fig.~\ref{fig:nofm_F5_F6_LCDM} shows the ratio of the \ecosmog{} and \isis{} results to those of \mggadget{} for the \lcdm{} (left panel), F5 (middle panel) and F6 (right panel) models. The three codes show varying levels of agreement (between 2-10\%) throughout the mass range probed by our simulations and for the two redshifts shown. In particular, at $z = 0$ and intermediate mass scales, $M \sim 10^{13}\,M_{\odot}/h$, \ecosmog{} and \isis{} agree with \mggadget{} at $\lesssim 4\%$, but the agreement worsens to $5-10\%$ for smaller mass scales.  At the high-mass end, $M \sim 10^{14}\,M_{\odot}/h$, the finite box size limits us to only a few halo samples, which is why the high mass end is noisy (as can be checked by the scatter between different mass bins). Overall, the trend is for the two \ramses{}-based codes to underpredict the abundance of smaller mass haloes, compared to \mggadget{}. This discrepancy can be linked to the way the base codes \ramses{} and {\sc p-gadget3} compute the force on small scales (see discussion in Sec.~\ref{sec:power}). Also, there is a marked improvement in the agreement between \ecosmog{} and \isis{} when the two codes are run with the same density assignment scheme (CIC in this case).

The amplitude and shape of the curves in Fig.~\ref{fig:nofm_F5_F6_LCDM} is similar for \lcdm{} and the two $f(R)$ models, which indicates that differences between the modified gravity codes are mostly driven by differences in the default codes. Indeed, this is again confirmed by the result of Fig.~\ref{fig:dnofm_F5_F6} which shows the fractional difference in the $f(R)$ mass functions relative to \lcdm{}. One notes that the three codes agree very well, especially at $z = 0$, echoing the results seen in the previous two sections for the matter and velocity divergence power spectra.

\subsubsection{DGP}

\begin{figure*}
\includegraphics[width=1\columnwidth]{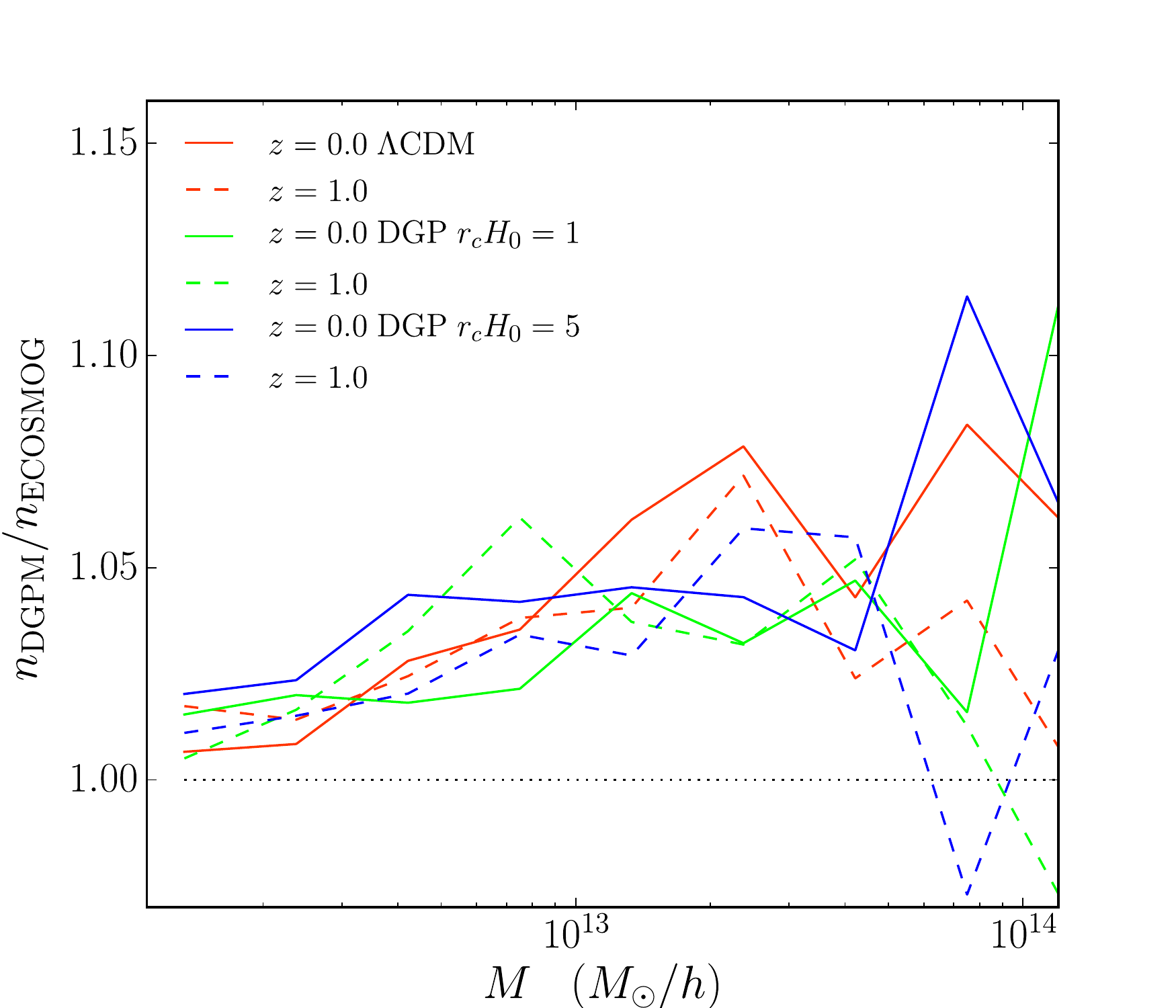}
\includegraphics[width=2.\columnwidth]{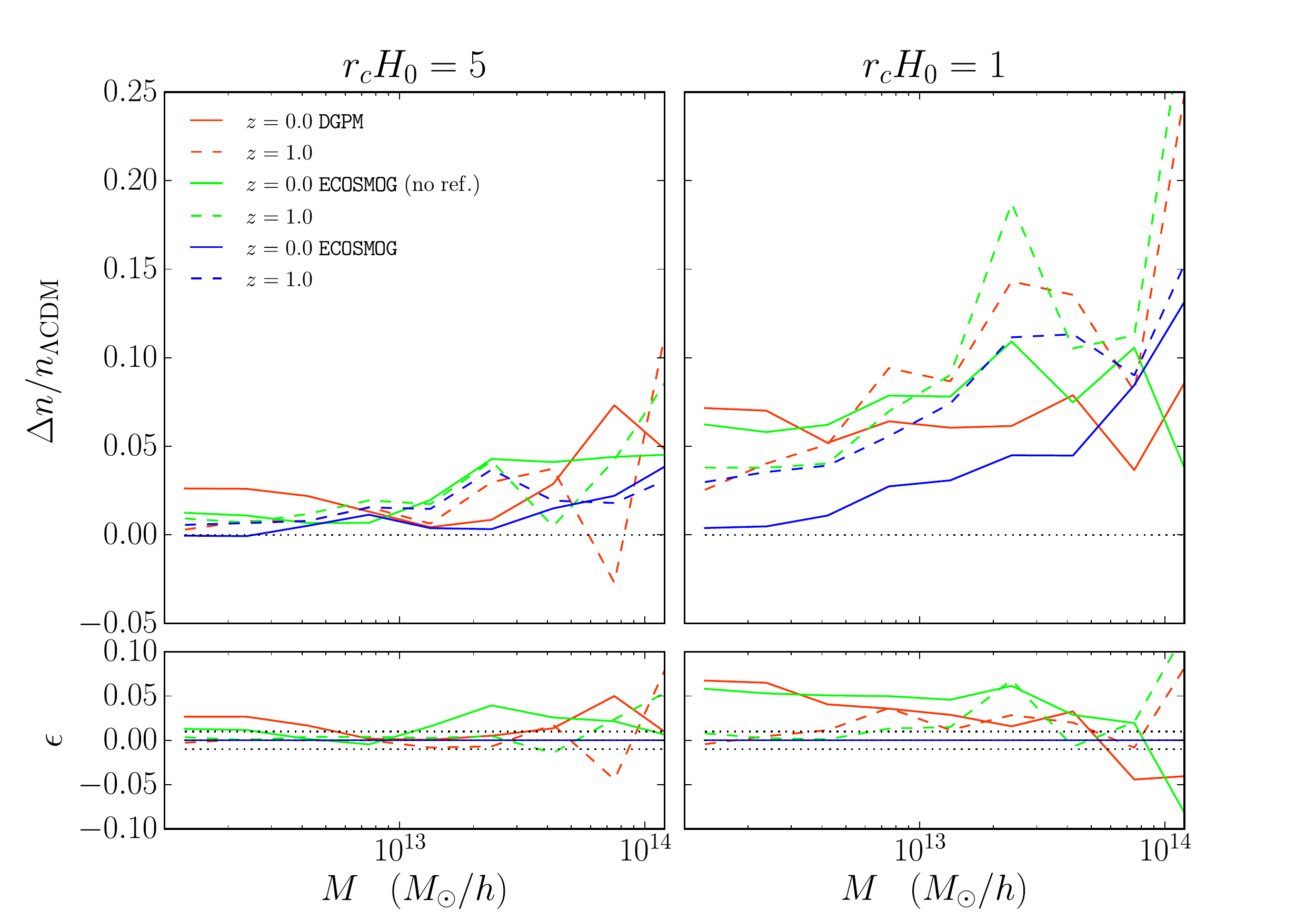}
\caption{The top panel shows the comparison of the halo mass function results of the \dgpm{} and \ecosmog{} (fixed grid) simulations of the \lcdm{}, $r_cH_0 = 5$ and $r_cH_0 = 1$ DGP models, as labelled. The lower left and lower right panels show the fractional difference with respect to \lcdm{} of the two codes for the $r_cH_0 = 5$ and $r_cH_0 = 1$ DGP models, respectivly. The lower panels show the \ecosmog{} results both for fixed and refined grid simulations, as labelled. In the lower panel insets, $\epsilon = \left(n/n_{\Lambda{\rm CDM}}\right)_{\rm code}/\left(n/n_{\Lambda{\rm CDM}}\right)_{\rm ref} - 1$, with \ecosmog{} (refined grid) being the reference code.}
\label{fig:nofm_DGP}
\end{figure*}

The top panel of Fig.~\ref{fig:nofm_DGP} shows the ratio of the mass function results obtained with the \dgpm{} code to those obtained with \ecosmog{} (both with and without refinements) for the simulations of the \lcdm{} and $r_cH_0 = 5$ and $r_cH_0 = 1$ DGP models. For the three models, there is a systematic trend for \dgpm{} to produce more massive haloes than \ecosmog{}, even when \ecosmog{} is run without refinements like \dgpm{}. This result can be linked to the different density assignments of the two codes (CIC for \dgpm{} vs TSC for \ecosmog{}). The lower panels of Fig.~\ref{fig:nofm_DGP} show that the agreement between the two codes improves when one looks at the fractional difference of the two DGP models to \lcdm{}. For the case of the $r_cH_0 = 5$ model, although there are still visible differences between the results of the two codes, these remain of the same order as the bin-to-bin scatter. The agreement worsens slightly for the $r_cH_0 = 1$ model. Overall, both \dgpm{} and \ecosmog{} are in good agreement in their predictions for $\Delta n/n_{\Lambda{\rm CDM}}$, although to a lesser extent than the agreement for the matter and velocity power spectra seen in the previous sections (Figs.~\ref{fig:pofk_DGP_dPP} and \ref{fig:dptt_DGP}).

\subsubsection{Symmetron}

The right panel of Fig.~\ref{fig:pofk_qsa} measures the impact of the quasi-static limit on the mass function of the Symmetron model, as predicted by the \isis{} and \isisns{} codes (both run without refinements). As seen in the case of the mass power spectrum (upper panels), relaxing the quasi-static approximation has no appreciable effect on halo abundances: differences are $\lesssim 0.5\%$ and have no clear dependence on mass. 


\subsection{Halo profiles}
\label{sec:profiles}

Finally, we turn our attention to the code results for the radial profiles of the density, velocity dispersion, force and scalar field around dark matter haloes. For this, we used the $z = 0$ \ahf{} halo catalogues of each simulation and binned them in mass according to: $[5\times 10^{12}, 1\times 10^{13}]$, $[1\times 10^{13}, 5\times 10^{13}]$, $[5\times 10^{13}, 1\times 10^{14}]$ and $[1\times 10^{14}, 5\times 10^{14}]$ $M_{\odot}/h$. For each mass bin, we scaled the \ahf{} haloes by their virial radii (determined by the \ahf{} code) and stacked them to compute average profiles of the different quantities. We take the variance of this average as the errorbars. The density, scalar field, and force profiles are calculated using all particles around a given \ahf{} halo centre, which includes particles that lie beyond the \ahf{} halo. On the other hand, the velocity dispersion profiles are calculated using only particles defined to be part of the \ahf{} halo, thus they only extend out to the virial radius. 

As before, we discuss our results for the halo profiles in turn for the $f(R)$, DGP and Symmetron models.


\subsubsection{$f(R)$}
\label{sec:fr_profiles}

\begin{figure*}
\includegraphics[width=2.2\columnwidth]{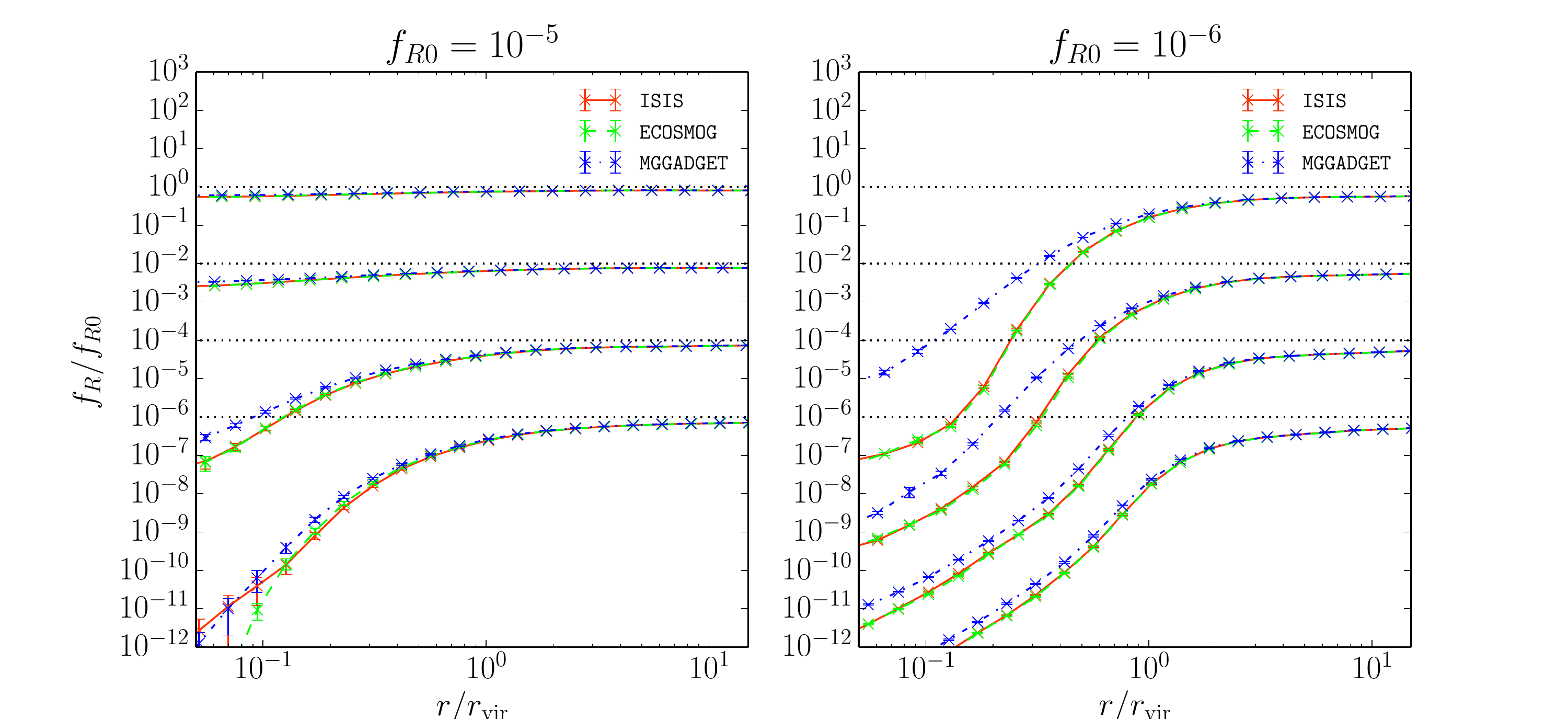}
\caption{Radial profiles of the $f_R$ scalar field in dark matter haloes found in the simulations of the \ecosmog{} (CIC), \mggadget{} and \isis{} codes for the F5 (left) and F6 (right) models. The result is shown for the following four halo mass bins: $M \in [1\times 10^{14},5\times 10^{14}]$, $M \in [5\times 10^{13},1\times 10^{14}]$, $M \in [1\times 10^{13},5\times 10^{13}]$ and $M \in [5\times 10^{12},1\times 10^{13}]$ $M_{\astrosun}/h$ (from bottom to top). For clarity, the profiles for each mass bin are displaced vertically, with the four horizontal lines (also displaced vertically) indicating $f_R / f_{R0} = 1$ of each mass bin.}
\label{fig:phi_fofr}
\end{figure*}

\begin{figure*}
\includegraphics[width=1.9\columnwidth]{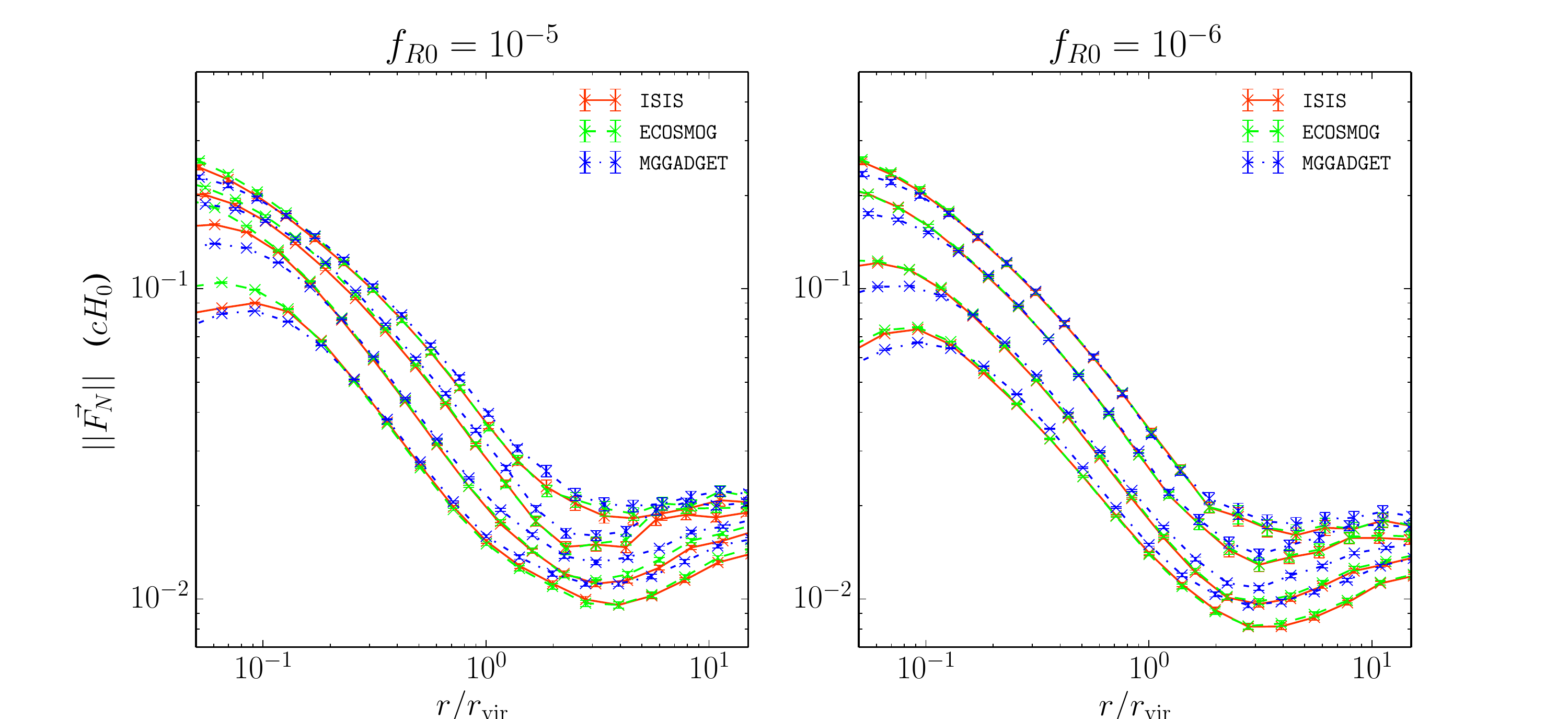}
\includegraphics[width=1.9\columnwidth]{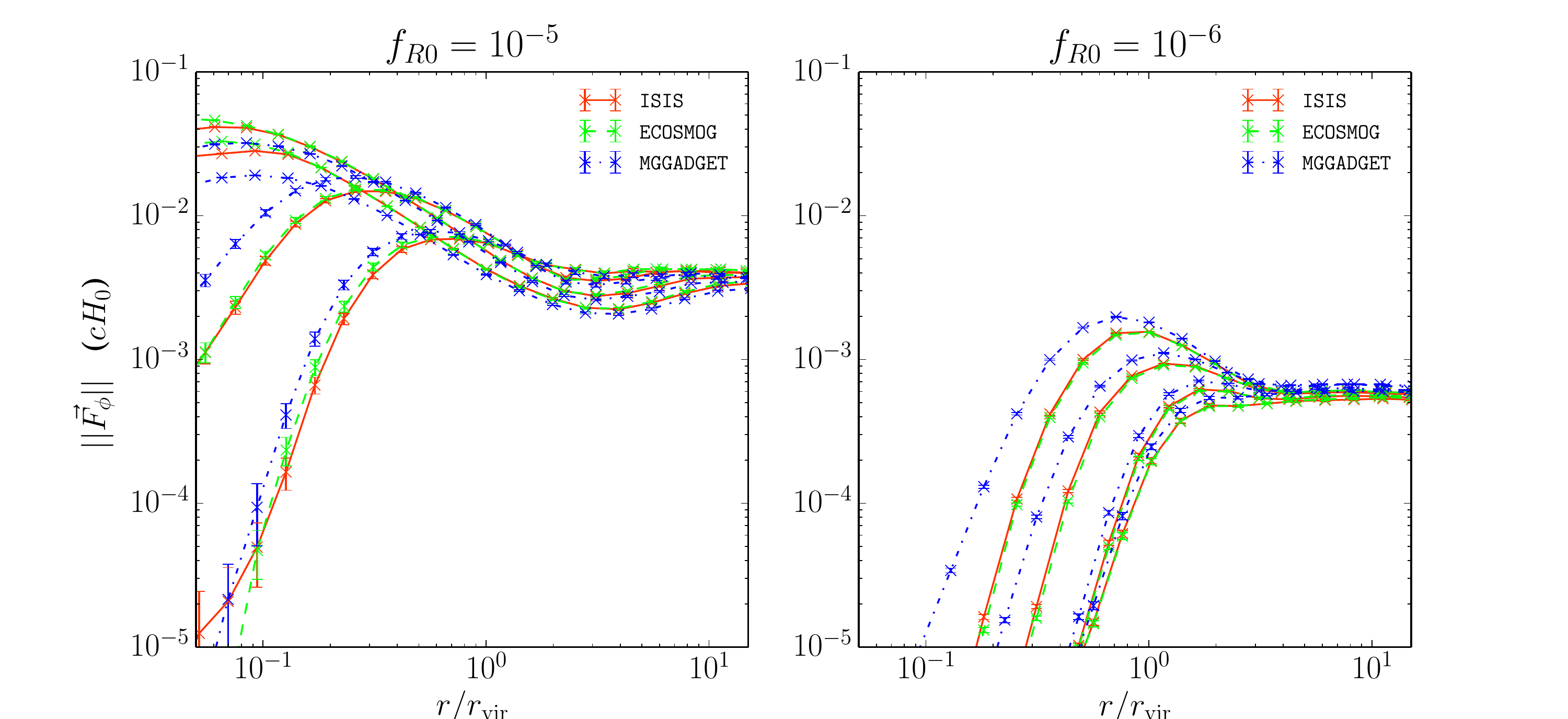}
\includegraphics[width=1.9\columnwidth]{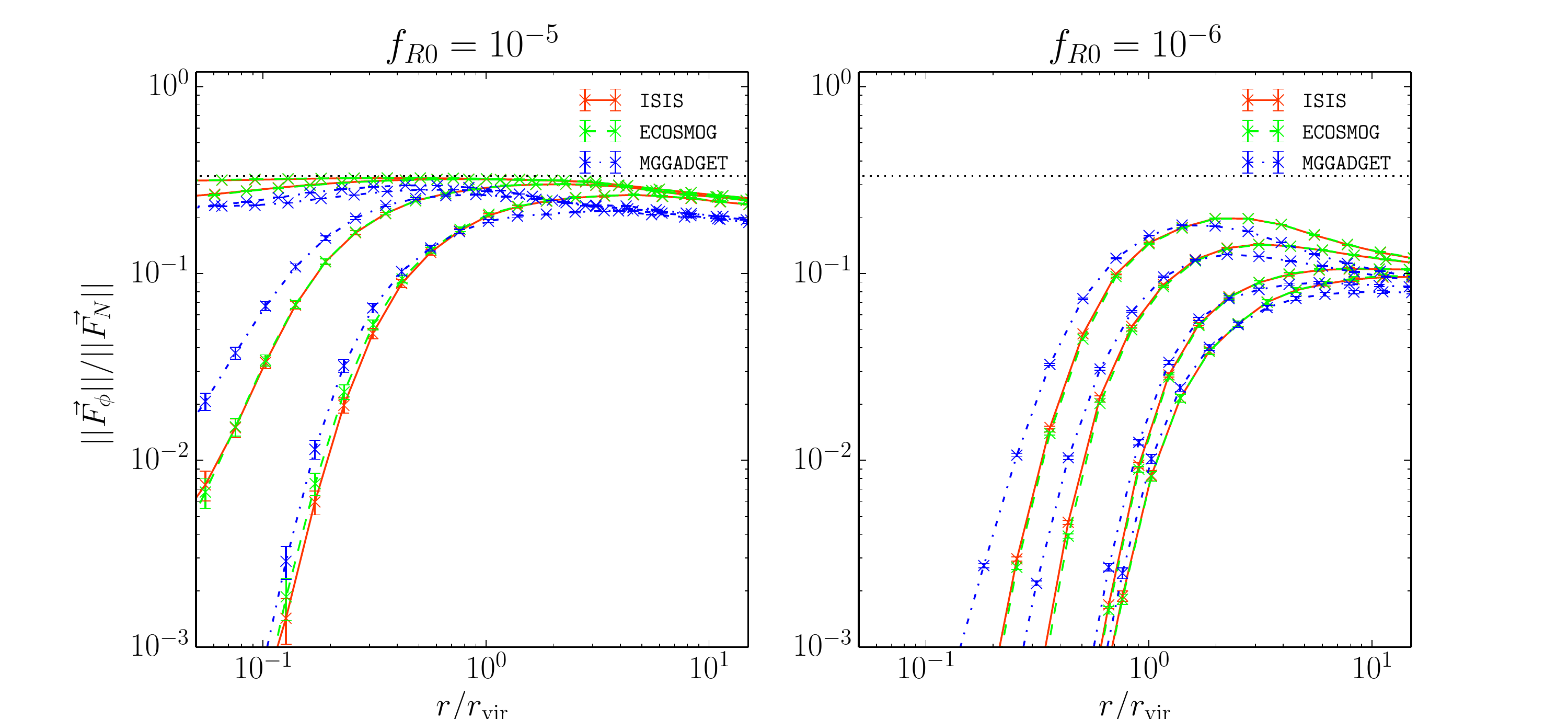}
\caption{Radial profiles of the Newtonian force (upper), fifth force (middle) and fifth to Newtonian force ratio (bottom) for haloes in the F5 (left) and F6 (right) simulations, performed with the \ecosmog{} (CIC), \mggadget{} and \isis{} codes, as labelled. The result is shown by splitting haloes into the same mass bins used in Fig.~\ref{fig:phi_fofr}. In this case, however, the results for each bin are not displaced vertically. In the lower panels, the horizontal lines show $F_{\phi} / F_N = {1}/{3}$, which is the expected value in the absence of screening. When determining the force profiles in each mass bin, what is averaged is the force modulii of the haloes, and not its radial component.}
\label{fig:force_fofr}
\end{figure*}

\begin{figure*}
\includegraphics[width=2.0\columnwidth]{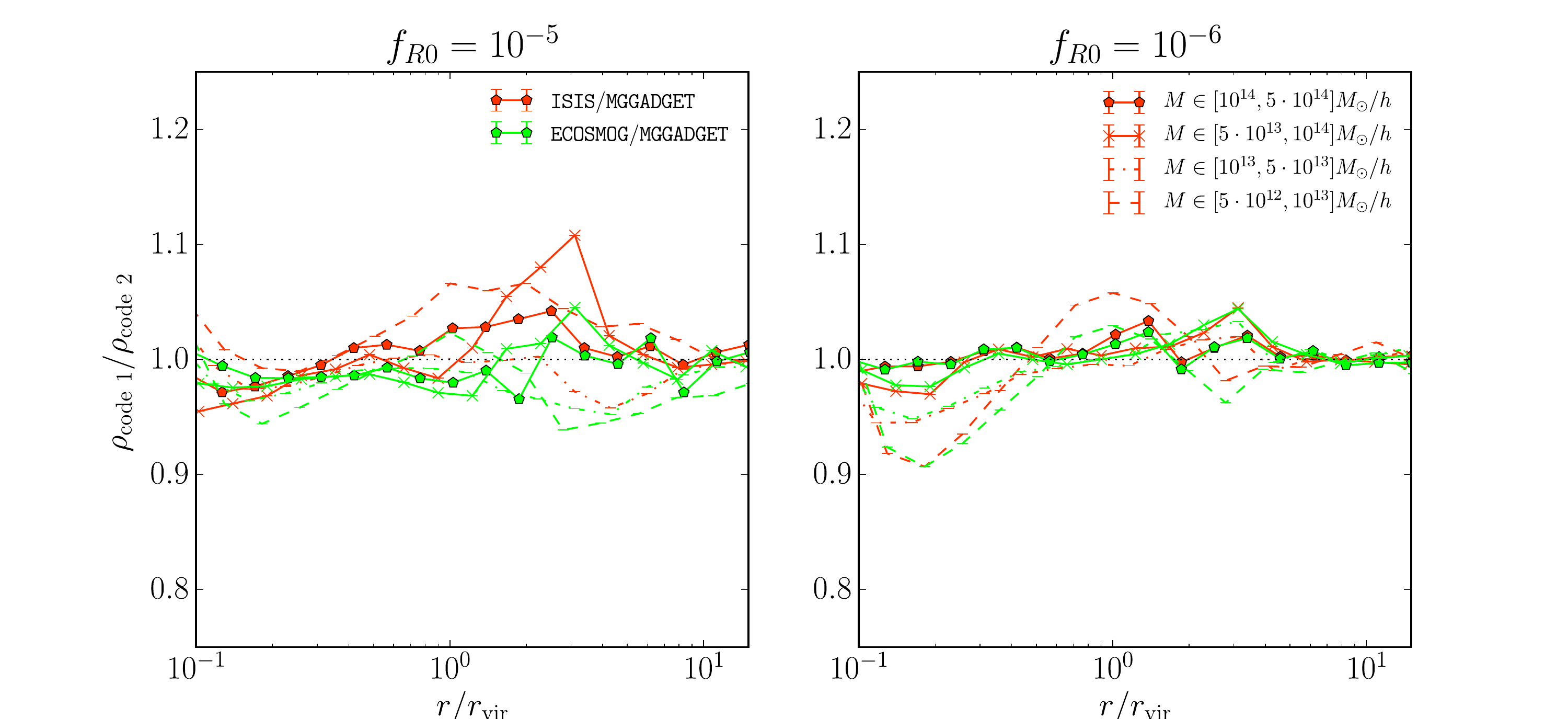}
\includegraphics[width=2.0\columnwidth]{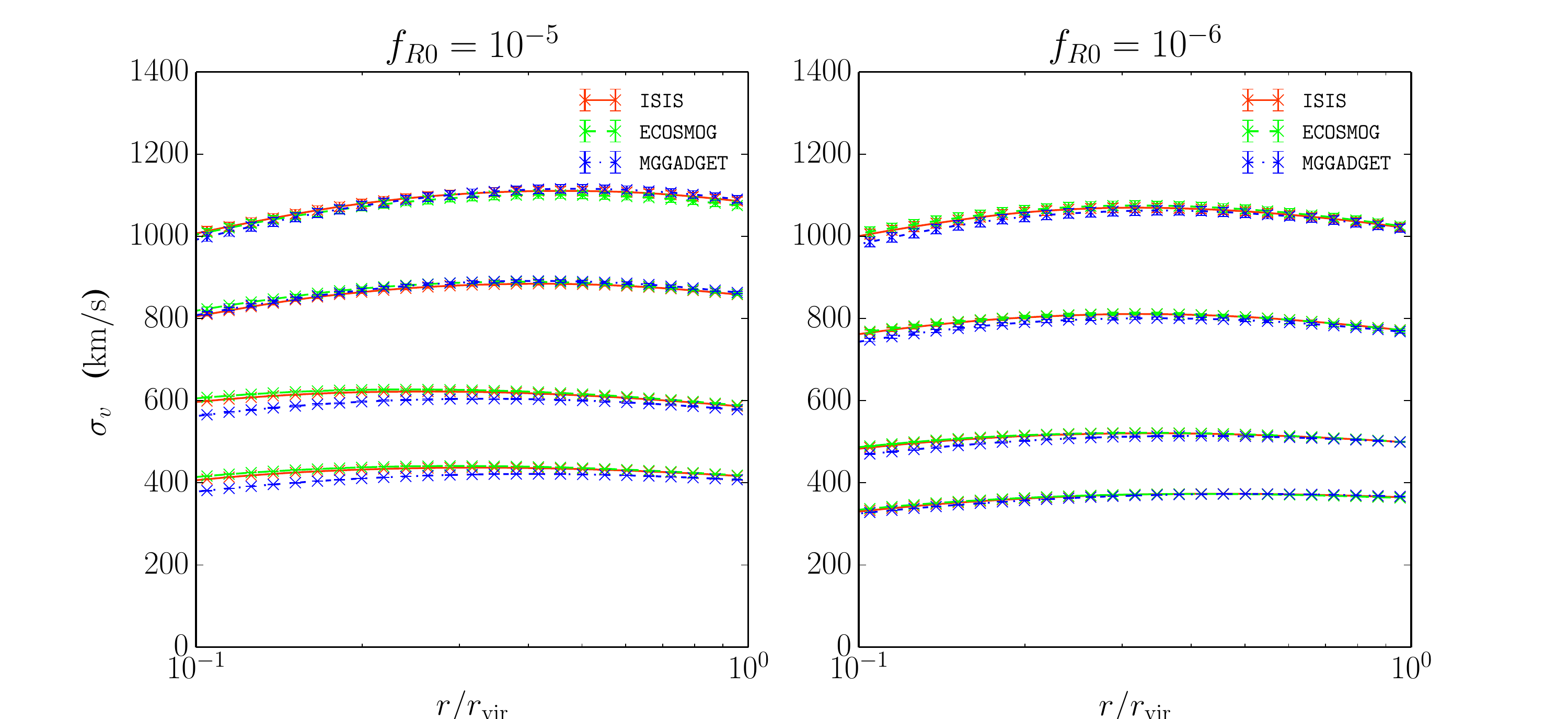}
\caption{Radial profiles of the density (upper) and velocity dispersion (bottom) for haloes in the F5 (left) and F6 (right) simulations, performed with the \ecosmog{} (CIC), \mggadget{} and \isis{} codes, as labelled. The result is shown by splitting the haloes into the same mass bins used in Figs.~\ref{fig:phi_fofr} and \ref{fig:force_fofr}. Note that in the lower panels the profile is only shown up to $r/r_{\rm vir} = 1$.}
\label{fig:velocity_fofr}
\end{figure*}

We plot the profiles of the scalar field in Fig.~\ref{fig:phi_fofr}, of the force modulus in Fig.~\ref{fig:force_fofr}, and of the density and velocity dispersion in Fig.~\ref{fig:velocity_fofr} for the haloes found in the F5 and F6 \ecosmog{}, \mggadget{} and \isis{} simulations. The \ecosmog{} results in these figures correspond to the runs performed with the CIC interpolation scheme (as in \isis{}). For \mggadget{}, the force profiles shown are those calculated by interpolating the gradient of the scalar field from the grid to the particle positions instead of the effective density method (recall the discussion about Fig.~\ref{fig:ratio_mggadget_comp} in \ref{sec:fr_algorithms}). The gradient method captures more accurately the suppression effects of the screening in the inner regions of the haloes.

Overall, the \isis{} and \ecosmog{} codes agree very well, with any deviations typically lying within the error bars in the profiles depicted in Figs.~\ref{fig:phi_fofr}, \ref{fig:force_fofr} and \ref{fig:velocity_fofr}. This is reassuring but not very surprising, considering that they are both based on \ramses{} and are run with the same settings. It is therefore more interesting to compare \isis{} and \ecosmog{} with \mggadget{}, for which some differences exist. For instance, although at large radii there is good agreement between the scalar field profiles obained by the three codes, in the inner regions of the haloes there is not. This is particularly noticeable in the F6 model at $r/r_{\rm vir} \lesssim 1$ (right panels of Fig.~\ref{fig:phi_fofr}), for which \mggadget{} overpredicts the values of $f_R$ compared to \ecosmog{} and \isis{}. In the F5 model the discrepancies soften considerably, especially for the low mass haloes. These differences in the $f_R$ profiles naturally translate into differences in the amplitude of the fifth force on small radial scales, as seen in the lower four panels of Fig.~\ref{fig:force_fofr}. However, the lower panels of Fig.~\ref{fig:force_fofr} show that the discrepancies between \mggadget{} and the two \ramses{}-based codes are only appreciable when the fifth force is a small fraction ($\lesssim 0.1$) of the total force (due to screening). This means that, although the codes may disagree on their exact predictions for the amplitude of the modifications to gravity, they only do so in regimes where the fifth force is not very important anyway. This is confirmed by the fact that the density and velocity dispersion profiles shown in Fig.~\ref{fig:velocity_fofr} agree very well (there are differences of order $5\%$, which we note are likely to come from differences in the base codes.)

At large radii ($r/r_{\rm vir} \gtrsim 1$), \mggadget{} overpredicts the amplitude of the Newtonian force compared to \ecosmog{} and \isis{}. Here, we stress that what we average in a given halo mass bin is the force modulus and not any directional component of the force (e.g. radial). This is why the force profiles do not keep decaying torwards large radii but instead level off due to the matter distribution that surrounds the haloes\footnote{If one would average the radial component of the force for a large number of haloes, then the contribution from the surrounding structure would cancel out.}. This can explain the mismatch in the Newtonian force at large radii because the matter distribution around haloes in {\sc gadget} and \ramses{} simulations is not exactly the same, and the base code algorithms that compute the Newtonian force are also different. What is important here is that, at these large radii, the three codes agree very well in their fifth force predictions, which are the corrections to normal gravity we are interested in testing in this paper.

\subsubsection{DGP}

\begin{figure*}
\includegraphics[width=1.75\columnwidth]{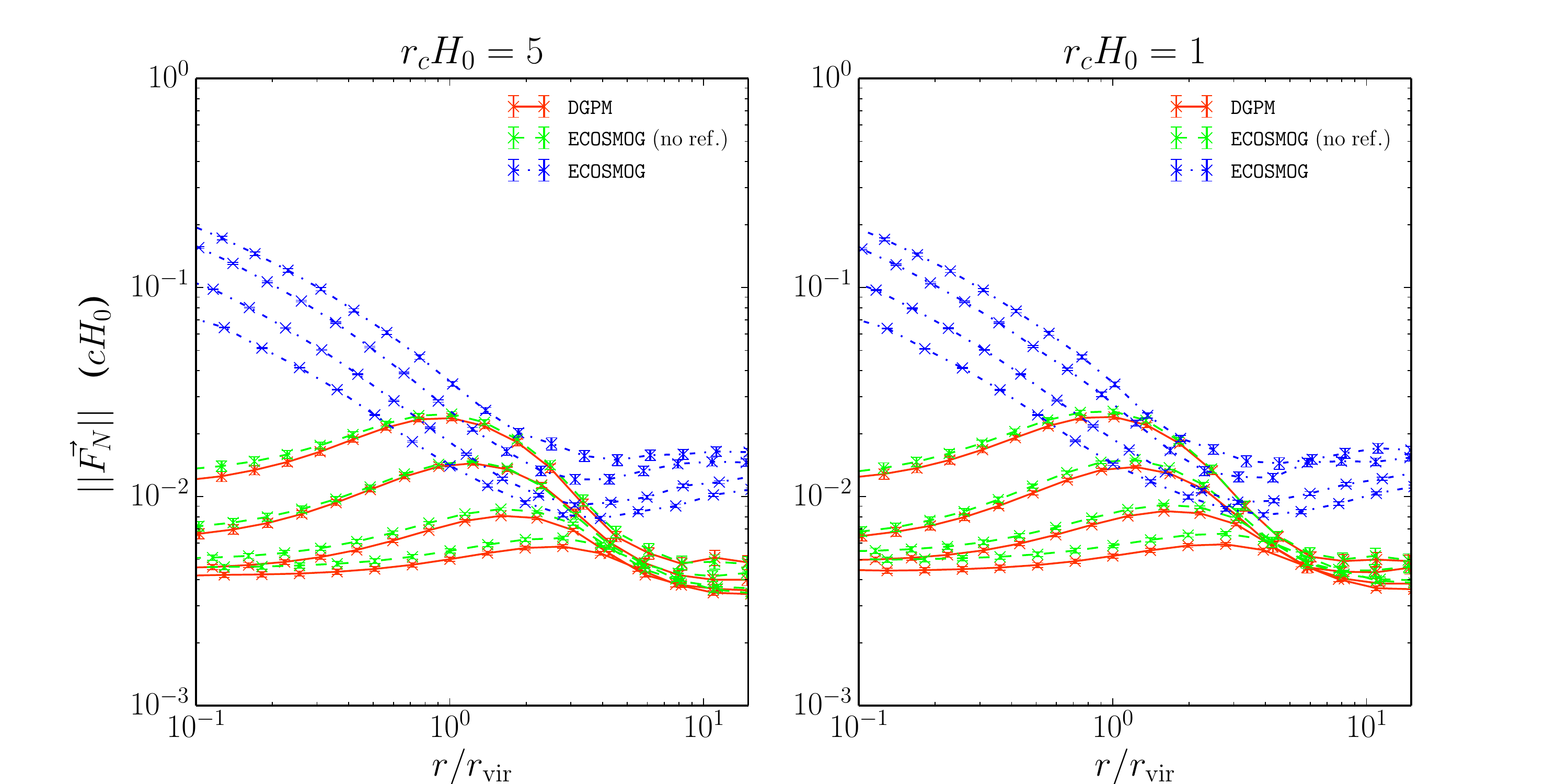}
\includegraphics[width=1.75\columnwidth]{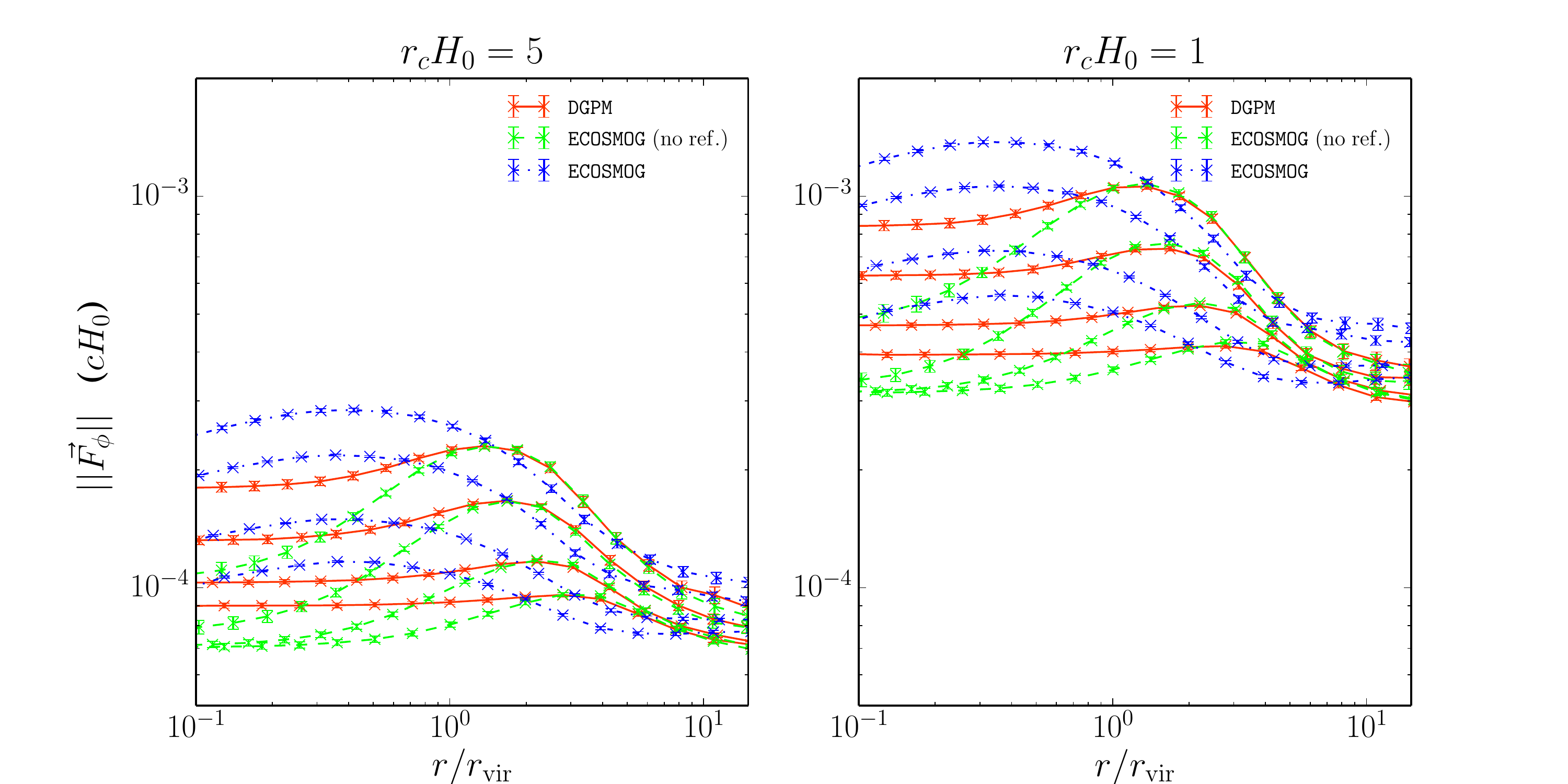}
\includegraphics[width=1.75\columnwidth]{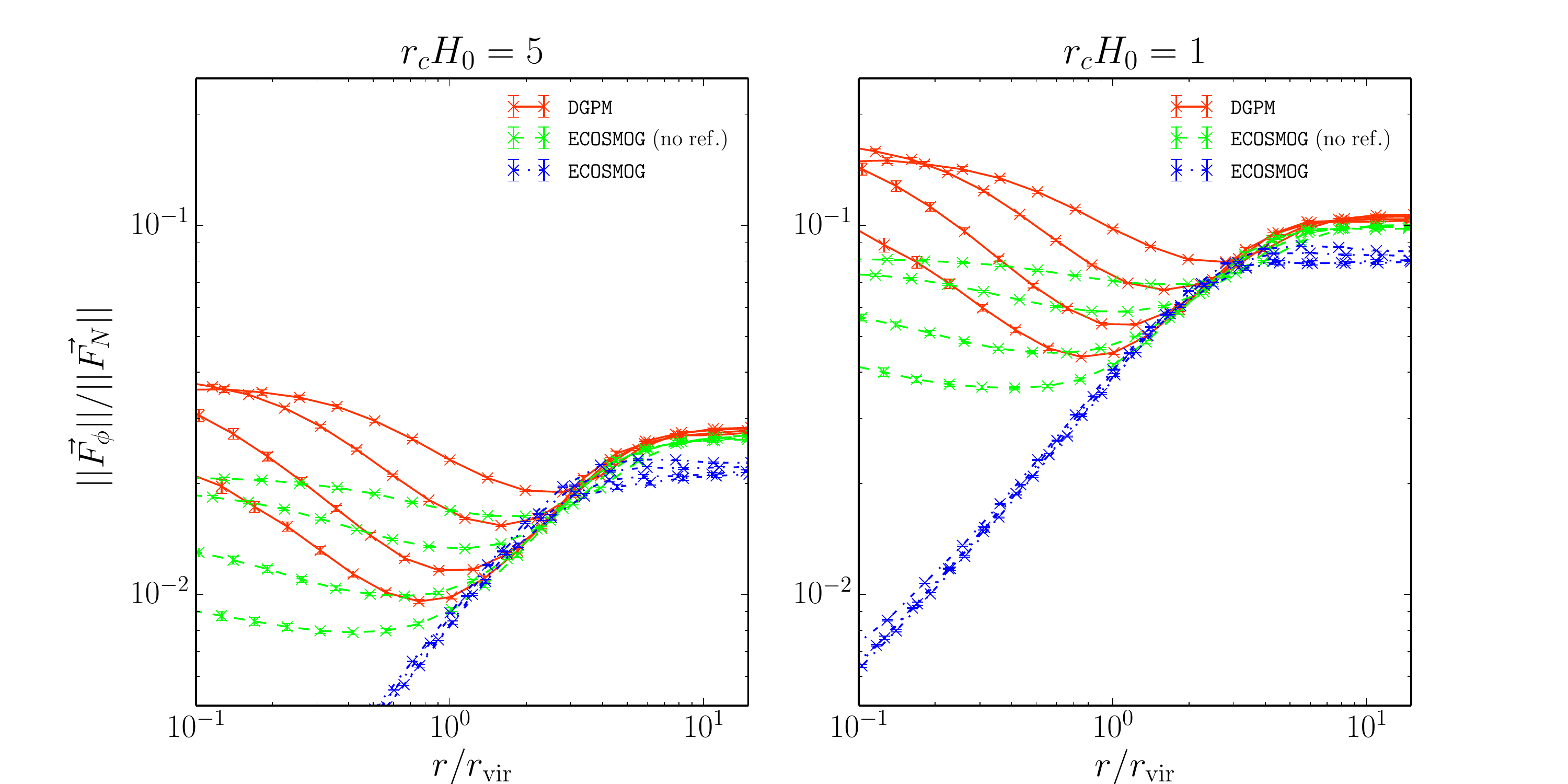}
\caption{Radial profiles of the Newtonian force (upper), fifth force (middle) and fifth to Newtonian force ratio (bottom) for haloes in the $r_cH_0=5$ (left) and $r_cH_0=1$ (right) simulations, performed with the \dgpm{} and \ecosmog{} codes. The \ecosmog{} results are shown for both the fixed and refined grid simulations, as labelled. The result is shown by splitting haloes into the same mass bins as used in Fig.~\ref{fig:force_fofr}. When determining the force profiles, what is averaged is the force modulii, and not its radial component. Note also that for all mass scales, the AMR nature of the grid plays a key role in the measured force profiles.}
\label{fig:force_dgp}
\end{figure*}

\begin{figure*}
\includegraphics[width=1.95\columnwidth]{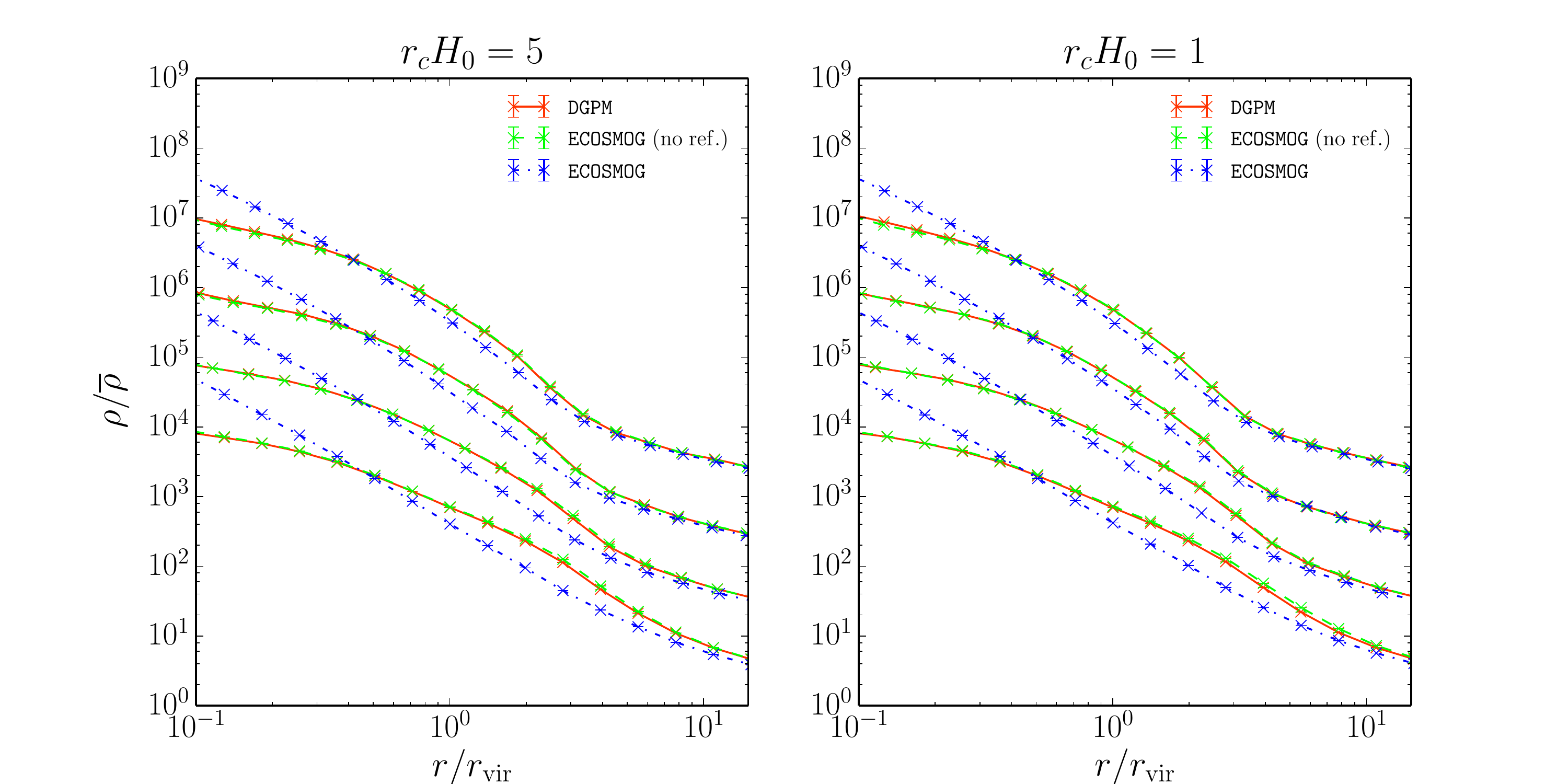}
\includegraphics[width=1.95\columnwidth]{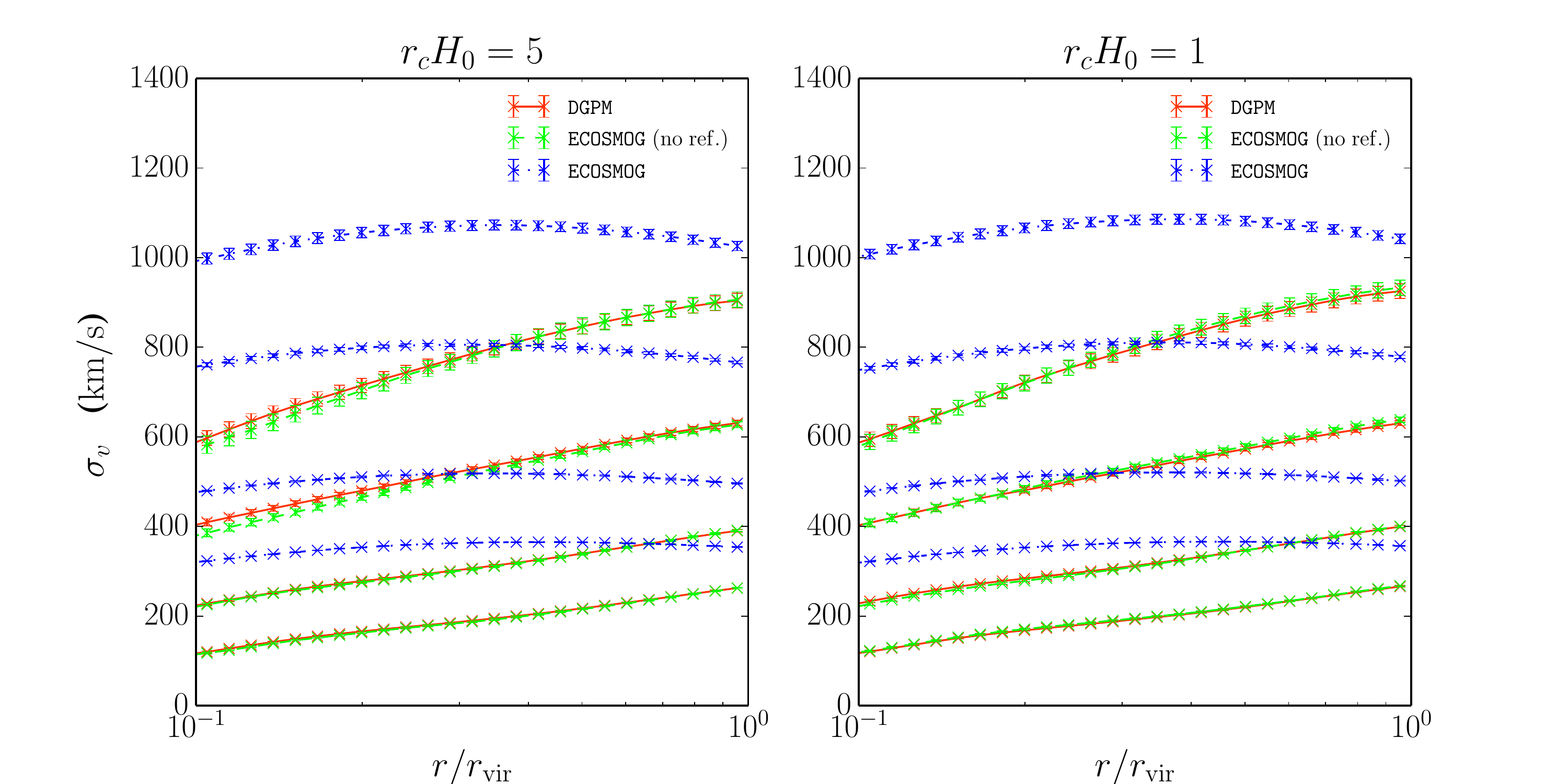}
\caption{Radial profiles of the density (upper) and velocity dispersion (bottom) for haloes in the $r_cH_0=5$ (left) and $r_cH_0=1$ (right) simulations, performed with the \dgpm{} and \ecosmog{} codes. The \ecosmog{} results are shown for both the fixed and refined grid simulations, as labelled. The result is shown by splitting the haloes into the same mass bins used in Fig.~\ref{fig:force_dgp}. For clarity, the result for the different mass bins has been displaced vertically in the density panels. In these, the differences between the two codes for fixed grid are kept below $\approx 5\%$ for all mass bins and scales shown. Note that in the lower panels the profile is only shown up to $r/r_{\rm vir} = 1$.}
\label{fig:velocity_dgp}
\end{figure*}

Figs.~\ref{fig:force_dgp} and \ref{fig:velocity_dgp} show the same as Figs.~\ref{fig:force_fofr} and \ref{fig:velocity_fofr}, but for the DGP simulations performed with the \dgpm{} and \ecosmog{} (with and without refinements) codes. The absolute value of the scalar field in the DGP model is irrelevant as the equations of the model contain only its derivatives. For this reason, we do not show the scalar field profiles and prefer to plot the gradient of the field (which, up to a factor of $1/2$, is the fifth force as seen in equation \ref{eq:psi}).

The Newtonian force profiles of the two codes are in very good agreement when \ecosmog{} is run without refinements (like \dgpm{}). There are marked differences in the solutions of the two codes for the size of the fifth force in the inner regions of the haloes, but since the amplitude of the fifth force is small there anyway, the difference does not translate into the density and velocity dispersion profiles (as seen in Fig.~\ref{fig:velocity_fofr}). However, when analysing these results, it is important to bear in mind that the grid size of the nonrefined simulations ($\Delta r \simeq 0.5 h^{-1} {\rm Mpc}$) is half of the typical halo size scales of $\approx 1 h^{-1} {\rm Mpc}$. As a result, one should not attempt to draw any physically meaningful conclusions from the results depicted by the red and green lines. We showed these lines simply to illustrate that the \dgpm{} and \ecosmog{} (nonrefined) codes agree in their total force profiles, even though their solutions may not be accurate.

When \ecosmog{} is run with refinements on the grid, the code is able to better resolve the matter distribution in and around the haloes. The top panels of Fig.~\ref{fig:velocity_dgp} show that the density profiles of the nonrefined simulations only agree with those of the refined simulation for $r/r_{\rm vir} \gtrsim 5-10$. This provides a measure for the radial scales below which one should not trust the nonrefined simulations. It is also interesting to note that the ratio of the fifth to Newtonian force does not depend on halo mass, as seen in the lower panels of Fig.~\ref{fig:force_dgp}. This illustrates that the efficiency of the Vainshtein mechanism in the DGP model is independent of the mass of the haloes \citep{2010PhRvD..81j3002S,2014JCAP...07..058F,2015arXiv150306673F}, which is different from what is seen in the lower panels of Fig.~\ref{fig:force_fofr} for the $f(R)$ models.


\subsubsection{Symmetron}

\begin{figure*}
\includegraphics[width=2.0\columnwidth]{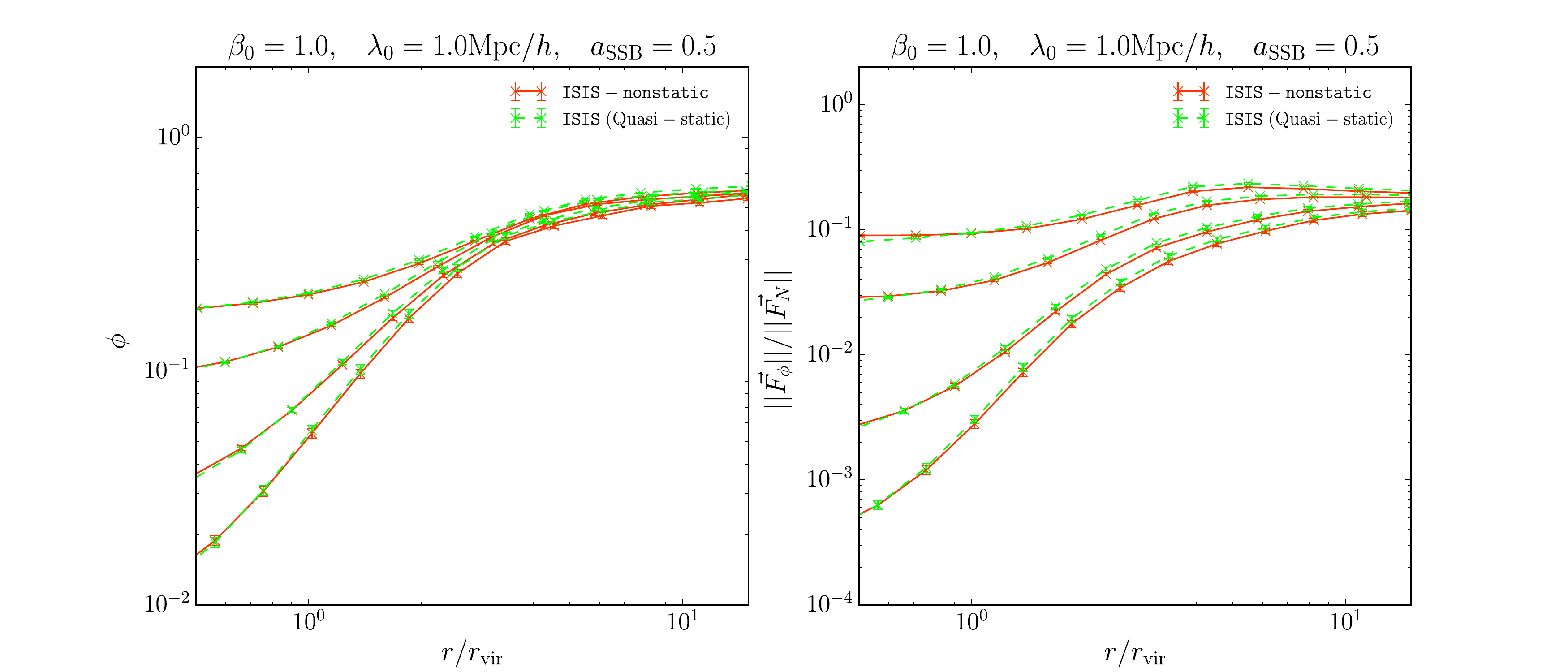}
\caption{Radial profiles of the scalar field (left) and of the fifth to Newtonian force ratio (right) for haloes in the Symmetron model simulations, performed with the \isis{} and \isisns{} codes. The result is shown by splitting the haloes into the same mass bins used in the figures of the $f(R)$ and DGP models. As in Fig.~\ref{fig:pofk_qsa}, the quasi-static limit remains an extremely good approximation.}
\label{fig:profile_Symmetron}
\end{figure*}

Fig.~\ref{fig:profile_Symmetron} shows the radial profiles of the scalar field and of the fifth to Newtonian force ratio obtained from the \isis{} and \isisns{} simulations of the Symmetron model. The scalar field in the \isisns{} simulations oscillates very rapidly with time. For this case, the profiles we show represent the mean value averaged over several oscillations close to $z = 0$. We note also that since these simulations were performed on a fixed grid, they suffer from the same resolution problems as the non-refined DGP runs. This prevents the simulations from fully capturing a number of effects such as the efficiency of the screening. Nevertheless, for the sake of determining the impact of the quasi-static limit these issues can be ignored. The result of Fig.~\ref{fig:profile_Symmetron} reinforces the conclusions drawn previously that relaxing the quasi-static limit has little impact on the simulations of the Symmetron model. 


\section{Summary and conclusions}
\label{sec:concl}

{\it N}-body simulations of modified gravity play a key role in cosmological tests of gravity using large-scale structure observations. In this work, we have performed an extensive comparison of the simulation results of $f(R)$, DGP and Symmetron gravity using five modified gravity {\it N}-body codes, which were the \dgpm{}, \ecosmog{}, \mggadget{}, \isis{} and \isisns{} codes. In the models we simulated, the gravitational law is modified due to the presence of a fifth force mediated by a scalar field. The algorithms in these codes differ from those for standard gravity by having extra modules that iteratively relax the equation of the scalar field on a grid/mesh to determine the fifth force. The \isisns{} code is a version of the \isis{} code that goes beyond the quasi-static approximation that is often employed in modified gravity studies. We used this code to test the validity of this assumption in the case of the Symmetron model.

The modified gravity routines included in the modified gravity codes are typically installed in existing and well tested {\it N}-body codes for GR (such as \ramses{} or {\sc p-gadget3}). Since these main codes can show some level of disagreement in their results for standard \lcdm{}, then these differences would naturally propagate into the modified gravity results. To overcome this, in this paper we have run also simulations of standard \lcdm{}, which were used as a reference to measure the effects of the modifications to gravity. Hence, our main concern in this paper was to compare the code predictions for the fractional difference with respect to \lcdm{} of a given quantity, and not so much its absolute values.

All our simulations of the different models and codes start from the same set of initial conditions, which allows a direct comparison between the results. We analysed the code results for the power spectrum of the matter density fluctuations and peculiar velocity divergence. We have also compared results for the abundances of dark matter haloes and for their density, velocity dispersion and force profiles.  

\emph{We have generally found agreement at the few-percent level in the properties of the matter density and velocity fields.  This means that modified gravity simulations satisfy the accuracy requirements of currently planned large-scale structure surveys,} provided that the absolute calibration of GR predictions is of comparable accuracy.

In what follows, we recap the main results of this comparison project in more detail.

\begin{itemize}[leftmargin=*]
\setlength\itemsep{1em}
\item {\bf Matter power spectrum} 

Given its immediate relation to galaxy redshift and lensing observables, the matter power spectrum $P(k)$ is one of the most important observables for tests of gravity. For \lcdm{}, we found that the modified gravity codes (run with the routines for modified gravity switched off) agree up to $1\%$ for $k\lesssim 1h {\rm Mpc}^{-1}$, but start differing on smaller scales ($\sim 5\%$ at $k \sim 5h {\rm Mpc}^{-1}$) due to different density and force assignments schemes and intrinsic algorithmic differences in the force calculation in AMR (for \ecosmog{} and \isis{}) and TreePM (for \mggadget{}) codes (cf.~Fig.~\ref{fig:pofk_LCDM_F5_F6}).

For the $f(R)$ simulations, which were performed with the \ecosmog, \mggadget{} and \isis{} codes, we find that any differences in $P(k)$ are driven almost exclusively by the differences in the base codes (\ramses{} and {\sc p-gadget3} respectively). In terms of the relative difference to \lcdm{}, all code results for $f(R)$ agree to better than $1\%$ for $k \lesssim 7h {\rm Mpc}^{-1}$ (cf.~Fig.~\ref{fig:dpofk_F5_F6}).  

While \ecosmog{} and \isis{} are based on the same GR code and use very similar algorithms, \mggadget{} is sufficiently different to make this a nontrivial consistency test.

The simulations of the DGP model were performed with the \dgpm{} and \ecosmog{} codes. The former code does not have an adaptive mesh, but \ecosmog{} can be run on both a fixed and refined mesh. For the fixed grid case, the two codes are in very good agreement in their predictions for the relative difference to \lcdm{} ($\lesssim 1\%$ for $k \lesssim 7h {\rm Mpc}^{-1}$) (cf.~Fig.~\ref{fig:pofk_DGP_dPP}). This is a nontrivial test since the DGP scalar field solvers in \dgpm{} and \ecosmog{} are substantially different. The fixed grid simulations start to differ from the refined grid simulations on small scales, but this is expected since the latter is able to resolve small scales much better.

For the test of the validity of the quasi-static limit in the Symmetron model, we have seen that the results from \isis{} and \isisns{} are nearly indistinguishable (cf.~Fig.~\ref{fig:pofk_qsa}). We conclude that the impact of adding the time derivative terms to the Symmetron field equation of motion is negligible ($<0.5\%$ on all scales and redshifts shown). We note, however, that in the current implementation of the time derivatives in the \isisns{} code, the particles do not feel the details of the rapid oscillations of the scalar field, due to the difference in the particle and scalar field time steps \citep[see][for more details]{2014PhRvD..89h4023L}. In the future, it would be of interest to clarify whether using the same time steps for particles and the scalar field has an impact on the results presented here.

\item {\bf Velocity power spectrum}

In terms of the velocity divergence power spectrum, $P_{\theta\theta}$, we have found that \ecosmog{}, \mggadget{} and \isis{} are again in very good agreement in their predictions for the fractional difference with respect to~\lcdm{} in $f(R)$ models. At $z = 0$, the three codes agree to better than $1\%$ for $k \lesssim 3h {\rm Mpc}^{-1}$, for both the F5 and F6 models (cf.~Fig.~\ref{fig:dptt_F5_F6}). For the DGP simulations, the \dgpm{} and \ecosmog{} codes are also in very good agreeent, with their predictions differing by $\lesssim 2\%$ for both models simulated and for all times and scales shown (cf.~Fig.~\ref{fig:dptt_DGP}). Interestingly, we have also seen that the $P_{\theta\theta}$ results of the DGP simulations on small scales are not critically affected by the use of a fixed or refined grid. For brevity, we did not show the impact of the quasi-static limit on $P_{\theta\theta}$ in the Symmetron model, but we have checked that it is negligible (just like for $P(k)$).

These results are encouraging as the precise modelling of cosmic velocity fields is a crucial ingredient in connecting the theoretically predicted clustering statistics in real space with the observed galaxy redshift space power spectrum.

\item {\bf Halo mass function}

The good agreement outlined  above for the matter and peculiar velocity divergence power spectrum holds also for the halo mass function. Given the relatively small simulation volume, the accuracy of the mass function comparison is limited by cosmic variance.  However, we are still able to conclude that the codes agree to a satisfactory level. In particular, for $f(R)$, the code predictions for $\Delta n/n_{\Lambda{\rm CDM}}$ agree up to $\approx 4\%$ for all mass scales probed by our simulation box (cf.~Fig.~\ref{fig:dnofm_F5_F6}). For the DGP simulations the agreement is at the $5-10\%$ level (cf.~Fig.~\ref{fig:nofm_DGP}). Like for $P(k)$, the inclusion of the time derivative terms in the Symmetron model equations leads to negligible differences only ($\lesssim 0.5\%$).

\item {\bf Halo profiles}

Understanding dark matter halo profiles in modified gravity is crucial to devise tests of gravity using galaxy clusters \citep{2012PhRvL.109e1301L, Terukina:2012ji, Lam:2013kma, Zu:2013joa, 2012PhRvD..85j2001L, Wilcox:2015kna, 2015arXiv150503468B, 2015arXiv150503692T}. For the $f(R)$ simulations, although there are significant differences in the fifth force profiles of the \ramses{}-based codes and \mggadget{} in the inner regions of the haloes (cf.~Fig.~\ref{fig:force_fofr}), these do not translate into differences in the resulting density and velocity divergence profiles because they only appear in regions where the force modification is highly suppressed by chameleon screening (cf.~Fig.~\ref{fig:velocity_fofr}).

Similarly, the fifth force halo profiles of the \dgpm{} and \ecosmog{} (fixed grid) DGP simulations also differ, but only in the inner regions of the haloes, where the fifth force is already weak (cf.~Fig.~\ref{fig:force_dgp}). We note, however, that although it is reassuring that the two codes agree in their halo density and velocity dispersion profiles (which is a cross check of the validity of the algorithms), the resolution of the fixed grid prevents us from trusting their physical results on halo size scales. Indeed, the DGP simulations run with \ecosmog{} (refined grid) show very different results, due to the gain in resolution (cf.~Fig.~\ref{fig:velocity_dgp}).

The halo profiles in the Symmetron simulations performed with the \isis{} and \isisns{} codes are nearly indistinguishable, which reinforces further the validity of the quasi-static limit (cf.~Fig.~\ref{fig:profile_Symmetron}).
\end{itemize}

Tests of gravity on large scales using galaxy power spectra, galaxy dynamics, cluster abundance, cluster profiles and so on are one of the main science drivers of many upcoming large surveys. Accurate simulations of nonlinear structure formation in these models are therefore crucial to ensure robust theoretical predictions that can be compared with the observational data. In this paper, we have seen that the {\it N}-body codes that have been developed so far are in very good agreement for different models. This dedicated comparison project constitutes an important validity check of the different codes, which brings us one step closer to performing larger and more expensive modified gravity simulations to be used to prepare for several future observational efforts.


\section*{Acknowledgements}
The authors are grateful to Volker Springel and Pedro G. Ferreira for helpful discussions.

HAW was supported by the BIPAC and the Oxford Martin School.  AB is supported by FCT-Portugal through grant SFRH/BD/75791/2011. SB is
supported by STFC through grant ST/K501979/1.  CLL and DFM are supported by the Research Council of Norway through grant 216756. GBZ is supported by the 1000 Young Talents program in China, and by the Strategic Priority Research Program of the Chinese Academy of Sciences, Grant No. XDB09000000. BF and KK are supported by the Science and Technology Facilities Council [grant number K00090X/1]. EP gratefully acknowledges support by the FP7 ERC Advanced Grant Emergence-320596. C.A. acknowledges support from the Deutsche
Forschungsgemeinschaft (DFG) through Transregio 33, ``The Dark Universe''. Part of this work was supported by the Science and Technology Facilities Council [grant number ST/F001166/1]. AB, SB, BL and CLL acknowledge support from the STFC consolidated grant ST/L00075X/1. WAH is grateful for support from Carlos S. Frenk's ERC Advanced Investigator grant COSMIWAY [grant number GA 267291].

This work used the DiRAC Data Centric system at Durham University, operated by the Institute for Computational Cosmology on behalf of the STFC DiRAC HPC Facility (www.dirac.ac.uk). This equipment was funded by BIS National E-infrastructure capital grant ST/K00042X/1, STFC capital grant ST/H008519/1, and STFC DiRAC Operations grant ST/K003267/1 and Durham University. This work used the DiRAC Complexity system, operated by the University of Leicester IT Services, which forms part of the STFC DiRAC HPC Facility (www.dirac.ac.uk ). This equipment is funded by BIS National E-Infrastructure capital grant ST/K000373/1 and  STFC DiRAC Operations grant ST/K0003259/1. DiRAC is part of the National E-Infrastructure. This work used the DiRAC Facility at the University of Oxford, jointly funded by STFC and the Large Facilities Capital Fund of BIS. The Symmetron simulations were performed on the NOTUR cluster HEXAGON, the computing facilities at the University of Bergen.



\clearpage
\bibliography{bibfile}

\begin{thebibliography}{}
\makeatletter
\relax
\def\mn@urlcharsother{\let\do\@makeother \do\$\do\&\do\#\do\^\do\_\do\%\do\~}
\def\mn@doi{\begingroup\mn@urlcharsother \@ifnextchar [ {\mn@doi@}
  {\mn@doi@[]}}
\def\mn@doi@[#1]#2{\def\@tempa{#1}\ifx\@tempa\@empty \href
  {http://dx.doi.org/#2} {doi:#2}\else \href {http://dx.doi.org/#2} {#1}\fi
  \endgroup}
\def\mn@eprint#1#2{\mn@eprint@#1:#2::\@nil}
\def\mn@eprint@arXiv#1{\href {http://arxiv.org/abs/#1} {{\tt arXiv:#1}}}
\def\mn@eprint@dblp#1{\href {http://dblp.uni-trier.de/rec/bibtex/#1.xml}
  {dblp:#1}}
\def\mn@eprint@#1:#2:#3:#4\@nil{\def\@tempa {#1}\def\@tempb {#2}\def\@tempc
  {#3}\ifx \@tempc \@empty \let \@tempc \@tempb \let \@tempb \@tempa \fi \ifx
  \@tempb \@empty \def\@tempb {arXiv}\fi \@ifundefined
  {mn@eprint@\@tempb}{\@tempb:\@tempc}{\expandafter \expandafter \csname
  mn@eprint@\@tempb\endcsname \expandafter{\@tempc}}}

\bibitem[\protect\citeauthoryear{{Adelberger}}{{Adelberger}}{2002}]{Adelberger2002cls..conf....9A}
{Adelberger} E.~G.,  2002, in {Kosteleck{\'y}} V.~A.,  ed., CPT and Lorentz
  Symmetry. pp 9--15 (\mn@eprint {} {hep-ex/0202008}),
  \mn@doi{10.1142/9789812778123_0002}

\bibitem[\protect\citeauthoryear{Amendola \& Tsujikawa}{Amendola \&
  Tsujikawa}{2010}]{Amendola2010}
Amendola L.,  Tsujikawa S.,  2010, {Dark energy: Theory and observations}.
Cambridge University Press

\bibitem[\protect\citeauthoryear{{Amendola} et~al.,}{{Amendola}
  et~al.}{2013}]{Amendola2013}
{Amendola} L.,  et~al., 2013, \mn@doi [Living Reviews in Relativity]
  {10.12942/lrr-2013-6}, \href
  {http://adsabs.harvard.edu/abs/2013LRR....16....6A} {16, 6}

\bibitem[\protect\citeauthoryear{{Arnold}, {Puchwein}  \& {Springel}}{{Arnold}
  et~al.}{2014}]{Arnold2014MNRAS.440..833A}
{Arnold} C.,  {Puchwein} E.,   {Springel} V.,  2014, \mn@doi [\mnras]
  {10.1093/mnras/stu332}, \href
  {http://adsabs.harvard.edu/abs/2014MNRAS.440..833A} {440, 833}

\bibitem[\protect\citeauthoryear{{Arnold}, {Puchwein}  \& {Springel}}{{Arnold}
  et~al.}{2015}]{2015MNRAS.448.2275A}
{Arnold} C.,  {Puchwein} E.,   {Springel} V.,  2015, \mn@doi [\mnras]
  {10.1093/mnras/stv146}, \href
  {http://adsabs.harvard.edu/abs/2015MNRAS.448.2275A} {448, 2275}

\bibitem[\protect\citeauthoryear{{Babichev}, {Deffayet}  \& {Ziour}}{{Babichev}
  et~al.}{2009}]{2009IJMPD..18.2147B}
{Babichev} E.,  {Deffayet} C.,   {Ziour} R.,  2009, \mn@doi [International
  Journal of Modern Physics D] {10.1142/S0218271809016107}, \href
  {http://adsabs.harvard.edu/abs/2009IJMPD..18.2147B} {18, 2147}

\bibitem[\protect\citeauthoryear{{Baldi}, {Villaescusa-Navarro}, {Viel},
  {Puchwein}, {Springel}  \& {Moscardini}}{{Baldi}
  et~al.}{2014}]{2014MNRAS.440...75B}
{Baldi} M.,  {Villaescusa-Navarro} F.,  {Viel} M.,  {Puchwein} E.,  {Springel}
  V.,   {Moscardini} L.,  2014, \mn@doi [\mnras] {10.1093/mnras/stu259}, \href
  {http://adsabs.harvard.edu/abs/2014MNRAS.440...75B} {440, 75}

\bibitem[\protect\citeauthoryear{{Barreira}, {Li}, {Hellwing}, {Baugh}  \&
  {Pascoli}}{{Barreira} et~al.}{2013a}]{2013JCAP...10..027B}
{Barreira} A.,  {Li} B.,  {Hellwing} W.~A.,  {Baugh} C.~M.,   {Pascoli} S.,
  2013a, \mn@doi [\jcap] {10.1088/1475-7516/2013/10/027}, \href
  {http://adsabs.harvard.edu/abs/2013JCAP...10..027B} {10, 27}

\bibitem[\protect\citeauthoryear{{Barreira}, {Li}, {Baugh}  \&
  {Pascoli}}{{Barreira} et~al.}{2013b}]{2013JCAP...11..056B}
{Barreira} A.,  {Li} B.,  {Baugh} C.~M.,   {Pascoli} S.,  2013b, \mn@doi
  [\jcap] {10.1088/1475-7516/2013/11/056}, \href
  {http://adsabs.harvard.edu/abs/2013JCAP...11..056B} {11, 56}

\bibitem[\protect\citeauthoryear{{Barreira}, {Li}, {Jennings}, {Merten},
  {King}, {Baugh}  \& {Pascoli}}{{Barreira} et~al.}{2015}]{2015arXiv150503468B}
{Barreira} A.,  {Li} B.,  {Jennings} E.,  {Merten} J.,  {King} L.,  {Baugh} C.,
    {Pascoli} S.,  2015, preprint, \href
  {http://adsabs.harvard.edu/abs/2015arXiv150503468B} {} (\mn@eprint {arXiv}
  {1505.03468})

\bibitem[\protect\citeauthoryear{{Bennett} et~al.,}{{Bennett}
  et~al.}{2013}]{2013ApJS..208...20B}
{Bennett} C.~L.,  et~al., 2013, \mn@doi [\apjs] {10.1088/0067-0049/208/2/20},
  \href {http://adsabs.harvard.edu/abs/2013ApJS..208...20B} {208, 20}

\bibitem[\protect\citeauthoryear{{Bernardeau} \& {van de
  Weygaert}}{{Bernardeau} \& {van de Weygaert}}{1996}]{Bernardeau1996}
{Bernardeau} F.,  {van de Weygaert} R.,  1996, \mnras, \href
  {http://adsabs.harvard.edu/abs/1996MNRAS.279..693B} {279, 693}

\bibitem[\protect\citeauthoryear{{Berti} et~al.,}{{Berti}
  et~al.}{2015}]{2015arXiv150107274B}
{Berti} E.,  et~al., 2015, preprint, \href
  {http://adsabs.harvard.edu/abs/2015arXiv150107274B} {} (\mn@eprint {arXiv}
  {1501.07274})

\bibitem[\protect\citeauthoryear{{Bertotti}, {Iess}  \& {Tortora}}{{Bertotti}
  et~al.}{2003}]{Berotti2003Natur.425..374B}
{Bertotti} B.,  {Iess} L.,   {Tortora} P.,  2003, \mn@doi [\nat]
  {10.1038/nature01997}, \href
  {http://adsabs.harvard.edu/abs/2003Natur.425..374B} {425, 374}

\bibitem[\protect\citeauthoryear{{Bose}, {Hellwing}  \& {Li}}{{Bose}
  et~al.}{2015}]{2015JCAP...02..034B}
{Bose} S.,  {Hellwing} W.~A.,   {Li} B.,  2015, \mn@doi [\jcap]
  {10.1088/1475-7516/2015/02/034}, \href
  {http://adsabs.harvard.edu/abs/2015JCAP...02..034B} {2, 34}

\bibitem[\protect\citeauthoryear{{Brandt}}{{Brandt}}{1977}]{Brandt77}
{Brandt} A.,  1977, Math. of Comp., 31, 333

\bibitem[\protect\citeauthoryear{{Brax} \& {Valageas}}{{Brax} \&
  {Valageas}}{2014}]{2014PhRvD..90b3507B}
{Brax} P.,  {Valageas} P.,  2014, \mn@doi [\prd] {10.1103/PhysRevD.90.023507},
  \href {http://adsabs.harvard.edu/abs/2014PhRvD..90b3507B} {90, 023507}

\bibitem[\protect\citeauthoryear{{Brax}, {van de Bruck}, {Davis}  \&
  {Shaw}}{{Brax} et~al.}{2008}]{2008PhRvD..78j4021B}
{Brax} P.,  {van de Bruck} C.,  {Davis} A.-C.,   {Shaw} D.~J.,  2008, \mn@doi
  [\prd] {10.1103/PhysRevD.78.104021}, \href
  {http://adsabs.harvard.edu/abs/2008PhRvD..78j4021B} {78, 104021}

\bibitem[\protect\citeauthoryear{{Brax}, {van de Bruck}, {Davis}  \&
  {Shaw}}{{Brax} et~al.}{2010}]{2010PhRvD..82f3519B}
{Brax} P.,  {van de Bruck} C.,  {Davis} A.-C.,   {Shaw} D.,  2010, \mn@doi
  [\prd] {10.1103/PhysRevD.82.063519}, \href
  {http://adsabs.harvard.edu/abs/2010PhRvD..82f3519B} {82, 063519}

\bibitem[\protect\citeauthoryear{{Brax}, {van de Bruck}, {Davis}, {Li}  \&
  {Shaw}}{{Brax} et~al.}{2011a}]{2011PhRvD..83j4026B}
{Brax} P.,  {van de Bruck} C.,  {Davis} A.-C.,  {Li} B.,   {Shaw} D.~J.,
  2011a, \mn@doi [\prd] {10.1103/PhysRevD.83.104026}, \href
  {http://adsabs.harvard.edu/abs/2011PhRvD..83j4026B} {83, 104026}

\bibitem[\protect\citeauthoryear{{Brax}, {van de Bruck}, {Davis}, {Li},
  {Schmauch}  \& {Shaw}}{{Brax} et~al.}{2011b}]{2011PhRvD..84l3524B}
{Brax} P.,  {van de Bruck} C.,  {Davis} A.-C.,  {Li} B.,  {Schmauch} B.,
  {Shaw} D.~J.,  2011b, \mn@doi [\prd] {10.1103/PhysRevD.84.123524}, \href
  {http://adsabs.harvard.edu/abs/2011PhRvD..84l3524B} {84, 123524}

\bibitem[\protect\citeauthoryear{{Brax}, {Davis}, {Li}, {Winther}  \&
  {Zhao}}{{Brax} et~al.}{2012a}]{2012JCAP...10..002B}
{Brax} P.,  {Davis} A.-C.,  {Li} B.,  {Winther} H.~A.,   {Zhao} G.-B.,  2012a,
  \mn@doi [\jcap] {10.1088/1475-7516/2012/10/002}, \href
  {http://adsabs.harvard.edu/abs/2012JCAP...10..002B} {10, 2}

\bibitem[\protect\citeauthoryear{{Brax}, {Davis}, {Li}  \& {Winther}}{{Brax}
  et~al.}{2012b}]{2012PhRvD..86d4015B}
{Brax} P.,  {Davis} A.-C.,  {Li} B.,   {Winther} H.~A.,  2012b, \mn@doi [\prd]
  {10.1103/PhysRevD.86.044015}, \href
  {http://adsabs.harvard.edu/abs/2012PhRvD..86d4015B} {86, 044015}

\bibitem[\protect\citeauthoryear{Brax, Davis, Li, Winther  \& Zhao}{Brax
  et~al.}{2013}]{Brax2013a}
Brax P.,  Davis A.,  Li B.,  Winther H.,   Zhao G.,  2013, preprint
  (arXiv:1303.0007v2), pp 1--16

\bibitem[\protect\citeauthoryear{{Brito}, {Terrana}, {Johnson}  \&
  {Cardoso}}{{Brito} et~al.}{2014}]{2014PhRvD..90l4035B}
{Brito} R.,  {Terrana} A.,  {Johnson} M.~C.,   {Cardoso} V.,  2014, \mn@doi
  [\prd] {10.1103/PhysRevD.90.124035}, \href
  {http://adsabs.harvard.edu/abs/2014PhRvD..90l4035B} {90, 124035}

\bibitem[\protect\citeauthoryear{{Burrage} \& {Khoury}}{{Burrage} \&
  {Khoury}}{2014}]{Burrage2014arXiv1403.6120B}
{Burrage} C.,  {Khoury} J.,  2014, preprint (arXiv:1403.6120), \href
  {http://adsabs.harvard.edu/abs/2014arXiv1403.6120B} {}

\bibitem[\protect\citeauthoryear{{Carroll}, {Duvvuri}, {Trodden}  \&
  {Turner}}{{Carroll} et~al.}{2004}]{Carroll/etal}
{Carroll} S.~M.,  {Duvvuri} V.,  {Trodden} M.,   {Turner} M.~S.,  2004, \mn@doi
  [\prd] {10.1103/PhysRevD.70.043528}, \href
  {http://adsabs.harvard.edu/abs/2004PhRvD..70d3528C} {70, 043528}

\bibitem[\protect\citeauthoryear{{Cautun} \& {van de Weygaert}}{{Cautun} \&
  {van de Weygaert}}{2011}]{2011arXiv1105.0370C}
{Cautun} M.~C.,  {van de Weygaert} R.,  2011, preprint (arXiv:1105.0370), \href
  {http://adsabs.harvard.edu/abs/2011arXiv1105.0370C} {}

\bibitem[\protect\citeauthoryear{{Chan} \& {Scoccimarro}}{{Chan} \&
  {Scoccimarro}}{2009}]{2009PhRvD..80j4005C}
{Chan} K.~C.,  {Scoccimarro} R.,  2009, \mn@doi [\prd]
  {10.1103/PhysRevD.80.104005}, \href
  {http://adsabs.harvard.edu/abs/2009PhRvD..80j4005C} {80, 104005}

\bibitem[\protect\citeauthoryear{{Chuang} et~al.,}{{Chuang}
  et~al.}{2014}]{2014arXiv1412.7729C}
{Chuang} C.-H.,  et~al., 2014, preprint, \href
  {http://adsabs.harvard.edu/abs/2014arXiv1412.7729C} {} (\mn@eprint {arXiv}
  {1412.7729})

\bibitem[\protect\citeauthoryear{Clifton, Ferreira, Padilla  \&
  Skordis}{Clifton et~al.}{2012}]{Clifton2012}
Clifton T.,  Ferreira P.~G.,  Padilla A.,   Skordis C.,  2012, \mn@doi
  [PhysRep] {10.1016/j.physrep.2012.01.001}, 513, 1

\bibitem[\protect\citeauthoryear{{Colberg} et~al.,}{{Colberg}
  et~al.}{2008}]{2008MNRAS.387..933C}
{Colberg} J.~M.,  et~al., 2008, \mn@doi [\mnras]
  {10.1111/j.1365-2966.2008.13307.x}, \href
  {http://adsabs.harvard.edu/abs/2008MNRAS.387..933C} {387, 933}

\bibitem[\protect\citeauthoryear{{Colombi}, {Jaffe}, {Novikov}  \&
  {Pichon}}{{Colombi} et~al.}{2009}]{2009MNRAS.393..511C}
{Colombi} S.,  {Jaffe} A.,  {Novikov} D.,   {Pichon} C.,  2009, \mn@doi
  [\mnras] {10.1111/j.1365-2966.2008.14176.x}, \href
  {http://adsabs.harvard.edu/abs/2009MNRAS.393..511C} {393, 511}

\bibitem[\protect\citeauthoryear{{Crocce}, {Pueblas}  \&
  {Scoccimarro}}{{Crocce} et~al.}{2006}]{2006MNRAS.373..369C}
{Crocce} M.,  {Pueblas} S.,   {Scoccimarro} R.,  2006, \mn@doi [\mnras]
  {10.1111/j.1365-2966.2006.11040.x}, \href
  {http://adsabs.harvard.edu/abs/2006MNRAS.373..369C} {373, 369}

\bibitem[\protect\citeauthoryear{{Davis}, {Li}, {Mota}  \& {Winther}}{{Davis}
  et~al.}{2012}]{2012ApJ...748...61D}
{Davis} A.-C.,  {Li} B.,  {Mota} D.~F.,   {Winther} H.~A.,  2012, \mn@doi
  [\apj] {10.1088/0004-637X/748/1/61}, \href
  {http://adsabs.harvard.edu/abs/2012ApJ...748...61D} {748, 61}

\bibitem[\protect\citeauthoryear{{Dvali}, {Gabadadze}  \& {Porrati}}{{Dvali}
  et~al.}{2000}]{2000PhLB..485..208D}
{Dvali} G.,  {Gabadadze} G.,   {Porrati} M.,  2000, \mn@doi [Physics Letters B]
  {10.1016/S0370-2693(00)00669-9}, \href
  {http://adsabs.harvard.edu/abs/2000PhLB..485..208D} {485, 208}

\bibitem[\protect\citeauthoryear{{Eisenstein} et~al.,}{{Eisenstein}
  et~al.}{2005}]{2005ApJ...633..560E}
{Eisenstein} D.~J.,  et~al., 2005, \mn@doi [\apj] {10.1086/466512}, \href
  {http://adsabs.harvard.edu/abs/2005ApJ...633..560E} {633, 560}

\bibitem[\protect\citeauthoryear{{Elahi} et~al.,}{{Elahi}
  et~al.}{2013}]{2013MNRAS.433.1537E}
{Elahi} P.~J.,  et~al., 2013, \mn@doi [\mnras] {10.1093/mnras/stt825}, \href
  {http://adsabs.harvard.edu/abs/2013MNRAS.433.1537E} {433, 1537}

\bibitem[\protect\citeauthoryear{{Falck}, {Koyama}, {Zhao}  \& {Li}}{{Falck}
  et~al.}{2014}]{2014JCAP...07..058F}
{Falck} B.,  {Koyama} K.,  {Zhao} G.-b.,   {Li} B.,  2014, \mn@doi [\jcap]
  {10.1088/1475-7516/2014/07/058}, \href
  {http://adsabs.harvard.edu/abs/2014JCAP...07..058F} {7, 58}

\bibitem[\protect\citeauthoryear{{Falck}, {Koyama}  \& {Zhao}}{{Falck}
  et~al.}{2015}]{2015arXiv150306673F}
{Falck} B.,  {Koyama} K.,   {Zhao} G.-b.,  2015, preprint, \href
  {http://adsabs.harvard.edu/abs/2015arXiv150306673F} {} (\mn@eprint {arXiv}
  {1503.06673})

\bibitem[\protect\citeauthoryear{{Fang}, {Wang}, {Hu}, {Haiman}, {Hui}  \&
  {May}}{{Fang} et~al.}{2008}]{2008PhRvD..78j3509F}
{Fang} W.,  {Wang} S.,  {Hu} W.,  {Haiman} Z.,  {Hui} L.,   {May} M.,  2008,
  \mn@doi [\prd] {10.1103/PhysRevD.78.103509}, \href
  {http://adsabs.harvard.edu/abs/2008PhRvD..78j3509F} {78, 103509}

\bibitem[\protect\citeauthoryear{{Gill}, {Knebe}  \& {Gibson}}{{Gill}
  et~al.}{2004}]{MHF}
{Gill} S.~P.~D.,  {Knebe} A.,   {Gibson} B.~K.,  2004, \mn@doi [\mnras]
  {10.1111/j.1365-2966.2004.07786.x}, \href
  {http://adsabs.harvard.edu/abs/2004MNRAS.351..399G} {351, 399}

\bibitem[\protect\citeauthoryear{Gronke, Llinares  \& Mota}{Gronke
  et~al.}{2014}]{Gronke:2013mea}
Gronke M.~B.,  Llinares C.,   Mota D.~F.,  2014, \mn@doi [Astron.Astrophys.]
  {10.1051/0004-6361/201322403}, 562, A9

\bibitem[\protect\citeauthoryear{{Gronke}, {Mota}  \& {Winther}}{{Gronke}
  et~al.}{2015a}]{2015arXiv150507129G}
{Gronke} M.,  {Mota} D.~F.,   {Winther} H.~A.,  2015a, preprint, \href
  {http://adsabs.harvard.edu/abs/2015arXiv150507129G} {} (\mn@eprint {arXiv}
  {1505.07129})

\bibitem[\protect\citeauthoryear{Gronke, Llinares, Mota  \& Winther}{Gronke
  et~al.}{2015b}]{Gronke:2014gaa}
Gronke M.,  Llinares C.,  Mota D.~F.,   Winther H.~A.,  2015b, \mn@doi
  [Mon.Not.Roy.Astron.Soc.] {10.1093/mnras/stv496}, 449, 2837

\bibitem[\protect\citeauthoryear{{Hagala}, {Llinares}  \& {Mota}}{{Hagala}
  et~al.}{2015}]{2015arXiv150407142H}
{Hagala} R.,  {Llinares} C.,   {Mota} D.~F.,  2015, preprint, \href
  {http://adsabs.harvard.edu/abs/2015arXiv150407142H} {} (\mn@eprint {arXiv}
  {1504.07142})

\bibitem[\protect\citeauthoryear{{Hammami} \& {Mota}}{{Hammami} \&
  {Mota}}{2015}]{Hammami:2015ela}
{Hammami} A.,  {Mota} D.~F.,  2015, preprint, \href
  {http://adsabs.harvard.edu/abs/2015arXiv150506803H} {} (\mn@eprint {arXiv}
  {1505.06803})

\bibitem[\protect\citeauthoryear{{Hammami}, {Llinares}, {Mota}  \&
  {Winther}}{{Hammami} et~al.}{2015}]{2015MNRAS.449.3635H}
{Hammami} A.,  {Llinares} C.,  {Mota} D.~F.,   {Winther} H.~A.,  2015, \mn@doi
  [\mnras] {10.1093/mnras/stv529}, \href
  {http://adsabs.harvard.edu/abs/2015MNRAS.449.3635H} {449, 3635}

\bibitem[\protect\citeauthoryear{{Hellwing} \& {Juszkiewicz}}{{Hellwing} \&
  {Juszkiewicz}}{2009}]{2009PhRvD..80h3522H}
{Hellwing} W.~A.,  {Juszkiewicz} R.,  2009, \mn@doi [\prd]
  {10.1103/PhysRevD.80.083522}, \href
  {http://adsabs.harvard.edu/abs/2009PhRvD..80h3522H} {80, 083522}

\bibitem[\protect\citeauthoryear{{Hellwing}, {Knollmann}  \&
  {Knebe}}{{Hellwing} et~al.}{2010}]{2010MNRAS.408L.104H}
{Hellwing} W.~A.,  {Knollmann} S.~R.,   {Knebe} A.,  2010, \mn@doi [\mnras]
  {10.1111/j.1745-3933.2010.00940.x}, \href
  {http://adsabs.harvard.edu/abs/2010MNRAS.408L.104H} {408, L104}

\bibitem[\protect\citeauthoryear{{Hellwing}, {Cautun}, {Knebe}, {Juszkiewicz}
  \& {Knollmann}}{{Hellwing} et~al.}{2013a}]{Hellwing2013}
{Hellwing} W.~A.,  {Cautun} M.,  {Knebe} A.,  {Juszkiewicz} R.,   {Knollmann}
  S.,  2013a, \mn@doi [\jcap] {10.1088/1475-7516/2013/10/012}, \href
  {http://adsabs.harvard.edu/abs/2013JCAP...10..012H} {10, 12}

\bibitem[\protect\citeauthoryear{{Hellwing}, {Li}, {Frenk}  \&
  {Cole}}{{Hellwing} et~al.}{2013b}]{Hellwing2013_clustering}
{Hellwing} W.~A.,  {Li} B.,  {Frenk} C.~S.,   {Cole} S.,  2013b, \mn@doi
  [\mnras] {10.1093/mnras/stt1430}, \href
  {http://adsabs.harvard.edu/abs/2013MNRAS.435.2806H} {435, 2806}

\bibitem[\protect\citeauthoryear{{Hellwing}, {Barreira}, {Frenk}, {Li}  \&
  {Cole}}{{Hellwing} et~al.}{2014}]{2014PhRvL.112v1102H}
{Hellwing} W.~A.,  {Barreira} A.,  {Frenk} C.~S.,  {Li} B.,   {Cole} S.,  2014,
  \mn@doi [\prl] {10.1103/PhysRevLett.112.221102}, \href
  {http://adsabs.harvard.edu/abs/2014PhRvL.112v1102H} {112, 221102}

\bibitem[\protect\citeauthoryear{{Hinterbichler} \& {Khoury}}{{Hinterbichler}
  \& {Khoury}}{2010}]{2010PhRvL.104w1301H}
{Hinterbichler} K.,  {Khoury} J.,  2010, \mn@doi [Physical Review Letters]
  {10.1103/PhysRevLett.104.231301}, \href
  {http://adsabs.harvard.edu/abs/2010PhRvL.104w1301H} {104, 231301}

\bibitem[\protect\citeauthoryear{{Hinterbichler}, {Khoury}, {Levy}  \&
  {Matas}}{{Hinterbichler} et~al.}{2011}]{2011PhRvD..84j3521H}
{Hinterbichler} K.,  {Khoury} J.,  {Levy} A.,   {Matas} A.,  2011, \mn@doi
  [\prd] {10.1103/PhysRevD.84.103521}, \href
  {http://adsabs.harvard.edu/abs/2011PhRvD..84j3521H} {84, 103521}

\bibitem[\protect\citeauthoryear{{Hoffmann} et~al.,}{{Hoffmann}
  et~al.}{2014}]{2014MNRAS.442.1197H}
{Hoffmann} K.,  et~al., 2014, \mn@doi [\mnras] {10.1093/mnras/stu933}, \href
  {http://adsabs.harvard.edu/abs/2014MNRAS.442.1197H} {442, 1197}

\bibitem[\protect\citeauthoryear{{Hu} \& {Sawicki}}{{Hu} \&
  {Sawicki}}{2007}]{2007PhRvD..76f4004H}
{Hu} W.,  {Sawicki} I.,  2007, \mn@doi [\prd] {10.1103/PhysRevD.76.064004},
  \href {http://adsabs.harvard.edu/abs/2007PhRvD..76f4004H} {76, 064004}

\bibitem[\protect\citeauthoryear{{Jain} \& {Khoury}}{{Jain} \&
  {Khoury}}{2010}]{Jain/Khoury}
{Jain} B.,  {Khoury} J.,  2010, \mn@doi [Annals of Physics]
  {10.1016/j.aop.2010.04.002}, \href
  {http://adsabs.harvard.edu/abs/2010AnPhy.325.1479J} {325, 1479}

\bibitem[\protect\citeauthoryear{{Jennings}, {Baugh}, {Li}, {Zhao}  \&
  {Koyama}}{{Jennings} et~al.}{2012}]{2012MNRAS.425.2128J}
{Jennings} E.,  {Baugh} C.~M.,  {Li} B.,  {Zhao} G.-B.,   {Koyama} K.,  2012,
  \mn@doi [\mnras] {10.1111/j.1365-2966.2012.21567.x}, \href
  {http://adsabs.harvard.edu/abs/2012MNRAS.425.2128J} {425, 2128}

\bibitem[\protect\citeauthoryear{{Joyce}, {Jain}, {Khoury}  \&
  {Trodden}}{{Joyce} et~al.}{2015}]{2015PhR...568....1J}
{Joyce} A.,  {Jain} B.,  {Khoury} J.,   {Trodden} M.,  2015, \mn@doi [\physrep]
  {10.1016/j.physrep.2014.12.002}, \href
  {http://adsabs.harvard.edu/abs/2015PhR...568....1J} {568, 1}

\bibitem[\protect\citeauthoryear{Khoury \& Weltman}{Khoury \&
  Weltman}{2004b}]{Khoury}
Khoury J.,  Weltman A.,  2004b, \prd

\bibitem[\protect\citeauthoryear{Khoury \& Weltman}{Khoury \&
  Weltman}{2004a}]{Khourya}
Khoury J.,  Weltman A.,  2004a, \prl

\bibitem[\protect\citeauthoryear{{Khoury} \& {Wyman}}{{Khoury} \&
  {Wyman}}{2009}]{2009PhRvD..80f4023K}
{Khoury} J.,  {Wyman} M.,  2009, \mn@doi [\prd] {10.1103/PhysRevD.80.064023},
  \href {http://adsabs.harvard.edu/abs/2009PhRvD..80f4023K} {80, 064023}

\bibitem[\protect\citeauthoryear{{Kitching} \& {Taylor}}{{Kitching} \&
  {Taylor}}{2011}]{2011MNRAS.416.1717K}
{Kitching} T.~D.,  {Taylor} A.~N.,  2011, \mn@doi [\mnras]
  {10.1111/j.1365-2966.2011.18772.x}, \href
  {http://adsabs.harvard.edu/abs/2011MNRAS.416.1717K} {416, 1717}

\bibitem[\protect\citeauthoryear{{Knebe}, {Green}  \& {Binney}}{{Knebe}
  et~al.}{2001}]{2001MNRAS.325..845K}
{Knebe} A.,  {Green} A.,   {Binney} J.,  2001, \mn@doi [\mnras]
  {10.1046/j.1365-8711.2001.04532.x}, \href
  {http://adsabs.harvard.edu/abs/2001MNRAS.325..845K} {325, 845}

\bibitem[\protect\citeauthoryear{{Knebe} et~al.,}{{Knebe}
  et~al.}{2011}]{2011MNRAS.415.2293K}
{Knebe} A.,  et~al., 2011, \mn@doi [\mnras] {10.1111/j.1365-2966.2011.18858.x},
  \href {http://adsabs.harvard.edu/abs/2011MNRAS.415.2293K} {415, 2293}

\bibitem[\protect\citeauthoryear{{Knebe} et~al.,}{{Knebe}
  et~al.}{2013a}]{2013MNRAS.428.2039K}
{Knebe} A.,  et~al., 2013a, \mn@doi [\mnras] {10.1093/mnras/sts173}, \href
  {http://adsabs.harvard.edu/abs/2013MNRAS.428.2039K} {428, 2039}

\bibitem[\protect\citeauthoryear{{Knebe} et~al.,}{{Knebe}
  et~al.}{2013b}]{2013MNRAS.435.1618K}
{Knebe} A.,  et~al., 2013b, \mn@doi [\mnras] {10.1093/mnras/stt1403}, \href
  {http://adsabs.harvard.edu/abs/2013MNRAS.435.1618K} {435, 1618}

\bibitem[\protect\citeauthoryear{{Knollmann} \& {Knebe}}{{Knollmann} \&
  {Knebe}}{2009}]{Amiga}
{Knollmann} S.~R.,  {Knebe} A.,  2009, \mn@doi [\apjs]
  {10.1088/0067-0049/182/2/608}, \href
  {http://adsabs.harvard.edu/abs/2009ApJS..182..608K} {182, 608}

\bibitem[\protect\citeauthoryear{Koivisto, Mota  \& Zumalacarregui}{Koivisto
  et~al.}{2012}]{Koivisto:2012za}
Koivisto T.~S.,  Mota D.~F.,   Zumalacarregui M.,  2012, \mn@doi
  [Phys.Rev.Lett.] {10.1103/PhysRevLett.109.241102}, 109, 241102

\bibitem[\protect\citeauthoryear{{Koyama}}{{Koyama}}{2007}]{2007CQGra..24R.231K}
{Koyama} K.,  2007, \mn@doi [Classical and Quantum Gravity]
  {10.1088/0264-9381/24/24/R01}, \href
  {http://adsabs.harvard.edu/abs/2007CQGra..24R.231K} {24, 231}

\bibitem[\protect\citeauthoryear{{Koyama}}{{Koyama}}{2015}]{2015arXiv150404623K}
{Koyama} K.,  2015, preprint, \href
  {http://adsabs.harvard.edu/abs/2015arXiv150404623K} {} (\mn@eprint {arXiv}
  {1504.04623})

\bibitem[\protect\citeauthoryear{{Koyama} \& {Silva}}{{Koyama} \&
  {Silva}}{2007}]{2007PhRvD..75h4040K}
{Koyama} K.,  {Silva} F.~P.,  2007, \mn@doi [\prd]
  {10.1103/PhysRevD.75.084040}, \href
  {http://adsabs.harvard.edu/abs/2007PhRvD..75h4040K} {75, 084040}

\bibitem[\protect\citeauthoryear{{Lam}, {Nishimichi}, {Schmidt}  \&
  {Takada}}{{Lam} et~al.}{2012}]{2012PhRvL.109e1301L}
{Lam} T.~Y.,  {Nishimichi} T.,  {Schmidt} F.,   {Takada} M.,  2012, \mn@doi
  [\prl] {10.1103/PhysRevLett.109.051301}, \href
  {http://adsabs.harvard.edu/abs/2012PhRvL.109e1301L} {109, 051301}

\bibitem[\protect\citeauthoryear{Lam, Schmidt, Nishimichi  \& Takada}{Lam
  et~al.}{2013}]{Lam:2013kma}
Lam T.~Y.,  Schmidt F.,  Nishimichi T.,   Takada M.,  2013, \mn@doi [Phys.Rev.]
  {10.1103/PhysRevD.88.023012}, D88, 023012

\bibitem[\protect\citeauthoryear{{Laureijs} et~al.,}{{Laureijs}
  et~al.}{2011}]{2011arXiv1110.3193L}
{Laureijs} R.,  et~al., 2011, preprint, \href
  {http://adsabs.harvard.edu/abs/2011arXiv1110.3193L} {} (\mn@eprint {arXiv}
  {1110.3193})

\bibitem[\protect\citeauthoryear{{Li} \& {Barrow}}{{Li} \&
  {Barrow}}{2011}]{2011PhRvD..83b4007L}
{Li} B.,  {Barrow} J.~D.,  2011, \mn@doi [\prd] {10.1103/PhysRevD.83.024007},
  \href {http://adsabs.harvard.edu/abs/2011PhRvD..83b4007L} {83, 024007}

\bibitem[\protect\citeauthoryear{{Li} \& {Zhao}}{{Li} \&
  {Zhao}}{2009}]{2009PhRvD..80d4027L}
{Li} B.,  {Zhao} H.,  2009, \mn@doi [\prd] {10.1103/PhysRevD.80.044027}, \href
  {http://adsabs.harvard.edu/abs/2009PhRvD..80d4027L} {80, 044027}

\bibitem[\protect\citeauthoryear{{Li} \& {Zhao}}{{Li} \&
  {Zhao}}{2010}]{2010PhRvD..81j4047L}
{Li} B.,  {Zhao} H.,  2010, \mn@doi [\prd] {10.1103/PhysRevD.81.104047}, \href
  {http://adsabs.harvard.edu/abs/2010PhRvD..81j4047L} {81, 104047}

\bibitem[\protect\citeauthoryear{Li, Mota  \& Barrow}{Li
  et~al.}{2011}]{Li:2010zw}
Li B.,  Mota D.~F.,   Barrow J.~D.,  2011, \mn@doi [Astrophys.J.]
  {10.1088/0004-637X/728/2/109}, 728, 109

\bibitem[\protect\citeauthoryear{{Li}, {Zhao}, {Teyssier}  \& {Koyama}}{{Li}
  et~al.}{2012}]{2012JCAP...01..051L}
{Li} B.,  {Zhao} G.-B.,  {Teyssier} R.,   {Koyama} K.,  2012, \mn@doi [\jcap]
  {10.1088/1475-7516/2012/01/051}, \href
  {http://adsabs.harvard.edu/abs/2012JCAP...01..051L} {1, 51}

\bibitem[\protect\citeauthoryear{{Li}, {Zhao}  \& {Koyama}}{{Li}
  et~al.}{2013a}]{2013JCAP...05..023L}
{Li} B.,  {Zhao} G.-B.,   {Koyama} K.,  2013a, \mn@doi [\jcap]
  {10.1088/1475-7516/2013/05/023}, \href
  {http://adsabs.harvard.edu/abs/2013JCAP...05..023L} {5, 23}

\bibitem[\protect\citeauthoryear{{Li}, {Barreira}, {Baugh}, {Hellwing},
  {Koyama}, {Pascoli}  \& {Zhao}}{{Li} et~al.}{2013b}]{Li2013JCAP...11..012L}
{Li} B.,  {Barreira} A.,  {Baugh} C.~M.,  {Hellwing} W.~A.,  {Koyama} K.,
  {Pascoli} S.,   {Zhao} G.-B.,  2013b, \mn@doi [\jcap]
  {10.1088/1475-7516/2013/11/012}, \href
  {http://adsabs.harvard.edu/abs/2013JCAP...11..012L} {11, 12}

\bibitem[\protect\citeauthoryear{{Li}, {Hellwing}, {Koyama}, {Zhao}, {Jennings}
   \& {Baugh}}{{Li} et~al.}{2013c}]{2013MNRAS.428..743L}
{Li} B.,  {Hellwing} W.~A.,  {Koyama} K.,  {Zhao} G.-B.,  {Jennings} E.,
  {Baugh} C.~M.,  2013c, \mn@doi [\mnras] {10.1093/mnras/sts072}, \href
  {http://adsabs.harvard.edu/abs/2013MNRAS.428..743L} {428, 743}

\bibitem[\protect\citeauthoryear{{Llinares} \& {Mota}}{{Llinares} \&
  {Mota}}{2013}]{2013PhRvL.110p1101L}
{Llinares} C.,  {Mota} D.~F.,  2013, \mn@doi [\prl]
  {10.1103/PhysRevLett.110.161101}, \href
  {http://adsabs.harvard.edu/abs/2013PhRvL.110p1101L} {110, 161101}

\bibitem[\protect\citeauthoryear{{Llinares} \& {Mota}}{{Llinares} \&
  {Mota}}{2014}]{2014PhRvD..89h4023L}
{Llinares} C.,  {Mota} D.~F.,  2014, \mn@doi [\prd]
  {10.1103/PhysRevD.89.084023}, \href
  {http://adsabs.harvard.edu/abs/2014PhRvD..89h4023L} {89, 084023}

\bibitem[\protect\citeauthoryear{{Llinares} \& {Mota}}{{Llinares} \&
  {Mota}}{2015}]{llinares_disformal}
{Llinares} C.,  {Mota} D.~F.,  2015, \mn@doi [(In preparation)] {0}, \href
  {http://adsabs.harvard.edu/abs/2014PhRvD..90l4041L} {p.~0}

\bibitem[\protect\citeauthoryear{{Llinares} \& {Pogosian}}{{Llinares} \&
  {Pogosian}}{2014}]{2014PhRvD..90l4041L}
{Llinares} C.,  {Pogosian} L.,  2014, \mn@doi [\prd]
  {10.1103/PhysRevD.90.124041}, \href
  {http://adsabs.harvard.edu/abs/2014PhRvD..90l4041L} {90, 124041}

\bibitem[\protect\citeauthoryear{{Llinares}, {Mota}  \& {Winther}}{{Llinares}
  et~al.}{2014}]{2014AA...562A..78L}
{Llinares} C.,  {Mota} D.~F.,   {Winther} H.~A.,  2014, \mn@doi [\aap]
  {10.1051/0004-6361/201322412}, \href
  {http://adsabs.harvard.edu/abs/2014A%26A...562A..78L} {562, A78}

\bibitem[\protect\citeauthoryear{{Lombriser}, {Schmidt}, {Baldauf},
  {Mandelbaum}, {Seljak}  \& {Smith}}{{Lombriser}
  et~al.}{2012}]{2012PhRvD..85j2001L}
{Lombriser} L.,  {Schmidt} F.,  {Baldauf} T.,  {Mandelbaum} R.,  {Seljak} U.,
  {Smith} R.~E.,  2012, \mn@doi [PRD] {10.1103/PhysRevD.85.102001}, \href
  {http://adsabs.harvard.edu/abs/2012PhRvD..85j2001L} {85, 102001}

\bibitem[\protect\citeauthoryear{{Lombriser}, {Koyama}  \& {Li}}{{Lombriser}
  et~al.}{2014}]{2014JCAP...03..021L}
{Lombriser} L.,  {Koyama} K.,   {Li} B.,  2014, \mn@doi [\jcap]
  {10.1088/1475-7516/2014/03/021}, \href
  {http://adsabs.harvard.edu/abs/2014JCAP...03..021L} {3, 21}

\bibitem[\protect\citeauthoryear{Lue \& Starkman}{Lue \&
  Starkman}{2004}]{LueStarkman}
Lue A.,  Starkman G.~D.,  2004, \mn@doi [Phys. Rev. D]
  {10.1103/PhysRevD.70.101501}, 70, 101501

\bibitem[\protect\citeauthoryear{{Luty}, {Porrati}  \& {Rattazzi}}{{Luty}
  et~al.}{2003}]{2003JHEP...09..029L}
{Luty} M.~A.,  {Porrati} M.,   {Rattazzi} R.,  2003, \mn@doi [Journal of High
  Energy Physics] {10.1088/1126-6708/2003/09/029}, \href
  {http://adsabs.harvard.edu/abs/2003JHEP...09..029L} {9, 29}

\bibitem[\protect\citeauthoryear{{Mota} \& {Shaw}}{{Mota} \&
  {Shaw}}{2006}]{Mota2006}
{Mota} D.~F.,  {Shaw} D.~J.,  2006, \mn@doi [\prl]
  {10.1103/PhysRevLett.97.151102}, \href
  {http://adsabs.harvard.edu/abs/2006PhRvL..97o1102M} {97, 151102}

\bibitem[\protect\citeauthoryear{{Nicolis} \& {Rattazzi}}{{Nicolis} \&
  {Rattazzi}}{2004}]{2004JHEP...06..059N}
{Nicolis} A.,  {Rattazzi} R.,  2004, \mn@doi [Journal of High Energy Physics]
  {10.1088/1126-6708/2004/06/059}, \href
  {http://adsabs.harvard.edu/abs/2004JHEP...06..059N} {6, 59}

\bibitem[\protect\citeauthoryear{{Nicolis}, {Rattazzi}  \&
  {Trincherini}}{{Nicolis} et~al.}{2009}]{Nicolis2009PhRvD..79f4036N}
{Nicolis} A.,  {Rattazzi} R.,   {Trincherini} E.,  2009, \mn@doi [\prd]
  {10.1103/PhysRevD.79.064036}, \href
  {http://adsabs.harvard.edu/abs/2009PhRvD..79f4036N} {79, 064036}

\bibitem[\protect\citeauthoryear{{Old} et~al.,}{{Old}
  et~al.}{2014}]{2014MNRAS.441.1513O}
{Old} L.,  et~al., 2014, \mn@doi [\mnras] {10.1093/mnras/stu545}, \href
  {http://adsabs.harvard.edu/abs/2014MNRAS.441.1513O} {441, 1513}

\bibitem[\protect\citeauthoryear{{Olive} \& {Pospelov}}{{Olive} \&
  {Pospelov}}{2008}]{2008PhRvD..77d3524O}
{Olive} K.~A.,  {Pospelov} M.,  2008, \mn@doi [\prd]
  {10.1103/PhysRevD.77.043524}, \href
  {http://adsabs.harvard.edu/abs/2008PhRvD..77d3524O} {77, 043524}

\bibitem[\protect\citeauthoryear{{Onions} et~al.,}{{Onions}
  et~al.}{2012}]{2012MNRAS.423.1200O}
{Onions} J.,  et~al., 2012, \mn@doi [\mnras]
  {10.1111/j.1365-2966.2012.20947.x}, \href
  {http://adsabs.harvard.edu/abs/2012MNRAS.423.1200O} {423, 1200}

\bibitem[\protect\citeauthoryear{{Onions} et~al.,}{{Onions}
  et~al.}{2013}]{2013MNRAS.429.2739O}
{Onions} J.,  et~al., 2013, \mn@doi [\mnras] {10.1093/mnras/sts549}, \href
  {http://adsabs.harvard.edu/abs/2013MNRAS.429.2739O} {429, 2739}

\bibitem[\protect\citeauthoryear{{Oyaizu}}{{Oyaizu}}{2008}]{2008PhRvD..78l3523O}
{Oyaizu} H.,  2008, \mn@doi [\prd] {10.1103/PhysRevD.78.123523}, \href
  {http://adsabs.harvard.edu/abs/2008PhRvD..78l3523O} {78, 123523}

\bibitem[\protect\citeauthoryear{{Pearson}}{{Pearson}}{2014}]{2014PhRvD..90l5011P}
{Pearson} J.~A.,  2014, \mn@doi [\prd] {10.1103/PhysRevD.90.125011}, \href
  {http://adsabs.harvard.edu/abs/2014PhRvD..90l5011P} {90, 125011}

\bibitem[\protect\citeauthoryear{{Pietroni}}{{Pietroni}}{2005}]{2005PhRvD..72d3535P}
{Pietroni} M.,  2005, \mn@doi [\prd] {10.1103/PhysRevD.72.043535}, \href
  {http://adsabs.harvard.edu/abs/2005PhRvD..72d3535P} {72, 043535}

\bibitem[\protect\citeauthoryear{{Planck Collaboration} et~al.,}{{Planck
  Collaboration} et~al.}{2015}]{2015arXiv150201589P}
{Planck Collaboration} et~al., 2015, preprint, \href
  {http://adsabs.harvard.edu/abs/2015arXiv150201589P} {} (\mn@eprint {arXiv}
  {1502.01589})

\bibitem[\protect\citeauthoryear{{Puchwein}, {Baldi}  \& {Springel}}{{Puchwein}
  et~al.}{2013}]{2013MNRAS.436..348P}
{Puchwein} E.,  {Baldi} M.,   {Springel} V.,  2013, \mn@doi [\mnras]
  {10.1093/mnras/stt1575}, \href
  {http://adsabs.harvard.edu/abs/2013MNRAS.436..348P} {436, 348}

\bibitem[\protect\citeauthoryear{{Pujol} et~al.,}{{Pujol}
  et~al.}{2014}]{2014MNRAS.438.3205P}
{Pujol} A.,  et~al., 2014, \mn@doi [\mnras] {10.1093/mnras/stt2446}, \href
  {http://adsabs.harvard.edu/abs/2014MNRAS.438.3205P} {438, 3205}

\bibitem[\protect\citeauthoryear{{Riess} et~al.,}{{Riess}
  et~al.}{1998}]{1998AJ....116.1009R}
{Riess} A.~G.,  et~al., 1998, \mn@doi [AJ] {10.1086/300499}, \href
  {http://adsabs.harvard.edu/abs/1998AJ....116.1009R} {116, 1009}

\bibitem[\protect\citeauthoryear{{Sahni} \& {Shtanov}}{{Sahni} \&
  {Shtanov}}{2003}]{SahniShtanov}
{Sahni} V.,  {Shtanov} Y.,  2003, \mn@doi [Journal of Cosmology and
  Astro-Particle Physics] {10.1088/1475-7516/2003/11/014}, \href
  {http://adsabs.harvard.edu/abs/2003JCAP...11..014S} {11, 14}

\bibitem[\protect\citeauthoryear{{Scannapieco} et~al.,}{{Scannapieco}
  et~al.}{2012}]{2012MNRAS.423.1726S}
{Scannapieco} C.,  et~al., 2012, \mn@doi [\mnras]
  {10.1111/j.1365-2966.2012.20993.x}, \href
  {http://adsabs.harvard.edu/abs/2012MNRAS.423.1726S} {423, 1726}

\bibitem[\protect\citeauthoryear{{Schaap} \& {van de Weygaert}}{{Schaap} \&
  {van de Weygaert}}{2000}]{Schaap_vdWeygaert2000}
{Schaap} W.~E.,  {van de Weygaert} R.,  2000, \aap, \href
  {http://adsabs.harvard.edu/abs/2000A%26A...363L..29S} {363, L29}

\bibitem[\protect\citeauthoryear{{Schmidt}}{{Schmidt}}{2009a}]{2009PhRvD..80d3001S}
{Schmidt} F.,  2009a, \mn@doi [\prd] {10.1103/PhysRevD.80.043001}, \href
  {http://adsabs.harvard.edu/abs/2009PhRvD..80d3001S} {80, 043001}

\bibitem[\protect\citeauthoryear{{Schmidt}}{{Schmidt}}{2009b}]{2009PhRvD..80l3003S}
{Schmidt} F.,  2009b, \mn@doi [\prd] {10.1103/PhysRevD.80.123003}, \href
  {http://adsabs.harvard.edu/abs/2009PhRvD..80l3003S} {80, 123003}

\bibitem[\protect\citeauthoryear{{Schmidt}}{{Schmidt}}{2010}]{2010PhRvD..81j3002S}
{Schmidt} F.,  2010, \mn@doi [\prd] {10.1103/PhysRevD.81.103002}, \href
  {http://adsabs.harvard.edu/abs/2010PhRvD..81j3002S} {81, 103002}

\bibitem[\protect\citeauthoryear{{Schmidt}, {Lima}, {Oyaizu}  \&
  {Hu}}{{Schmidt} et~al.}{2009}]{2009PhRvD..79h3518S}
{Schmidt} F.,  {Lima} M.,  {Oyaizu} H.,   {Hu} W.,  2009, \mn@doi [\prd]
  {10.1103/PhysRevD.79.083518}, \href
  {http://adsabs.harvard.edu/abs/2009PhRvD..79h3518S} {79, 083518}

\bibitem[\protect\citeauthoryear{{Schmidt}, {Hu}  \& {Lima}}{{Schmidt}
  et~al.}{2010}]{2010PhRvD..81f3005S}
{Schmidt} F.,  {Hu} W.,   {Lima} M.,  2010, \mn@doi [\prd]
  {10.1103/PhysRevD.81.063005}, \href
  {http://adsabs.harvard.edu/abs/2010PhRvD..81f3005S} {81, 063005}

\bibitem[\protect\citeauthoryear{{Schneider} et~al.,}{{Schneider}
  et~al.}{2015}]{2015arXiv150305920S}
{Schneider} A.,  et~al., 2015, preprint, \href
  {http://adsabs.harvard.edu/abs/2015arXiv150305920S} {} (\mn@eprint {arXiv}
  {1503.05920})

\bibitem[\protect\citeauthoryear{{Shi}, {Li}, {Han}, {Gao}  \&
  {Hellwing}}{{Shi} et~al.}{2015}]{2015arXiv150301109S}
{Shi} D.,  {Li} B.,  {Han} J.,  {Gao} L.,   {Hellwing} W.~A.,  2015, preprint,
  \href {http://adsabs.harvard.edu/abs/2015arXiv150301109S} {} (\mn@eprint
  {arXiv} {1503.01109})

\bibitem[\protect\citeauthoryear{{Springel}}{{Springel}}{2005}]{2005MNRAS.364.1105S}
{Springel} V.,  2005, \mn@doi [\mnras] {10.1111/j.1365-2966.2005.09655.x},
  \href {http://adsabs.harvard.edu/abs/2005MNRAS.364.1105S} {364, 1105}

\bibitem[\protect\citeauthoryear{{Springel} \& {Hernquist}}{{Springel} \&
  {Hernquist}}{2002}]{2002MNRAS.333..649S}
{Springel} V.,  {Hernquist} L.,  2002, \mn@doi [\mnras]
  {10.1046/j.1365-8711.2002.05445.x}, \href
  {http://adsabs.harvard.edu/abs/2002MNRAS.333..649S} {333, 649}

\bibitem[\protect\citeauthoryear{{Srisawat} et~al.,}{{Srisawat}
  et~al.}{2013}]{2013MNRAS.436..150S}
{Srisawat} C.,  et~al., 2013, \mn@doi [\mnras] {10.1093/mnras/stt1545}, \href
  {http://adsabs.harvard.edu/abs/2013MNRAS.436..150S} {436, 150}

\bibitem[\protect\citeauthoryear{Terukina \& Yamamoto}{Terukina \&
  Yamamoto}{2012}]{Terukina:2012ji}
Terukina A.,  Yamamoto K.,  2012, \mn@doi [Phys.Rev.]
  {10.1103/PhysRevD.86.103503}, D86, 103503

\bibitem[\protect\citeauthoryear{{Terukina}, {Yamamoto}, {Okabe}, {Matsushita}
  \& {Sasaki}}{{Terukina} et~al.}{2015}]{2015arXiv150503692T}
{Terukina} A.,  {Yamamoto} K.,  {Okabe} N.,  {Matsushita} K.,   {Sasaki} T.,
  2015, preprint, \href {http://adsabs.harvard.edu/abs/2015arXiv150503692T} {}
  (\mn@eprint {arXiv} {1505.03692})

\bibitem[\protect\citeauthoryear{{Teyssier}}{{Teyssier}}{2002}]{2002A&A...385..337T}
{Teyssier} R.,  2002, \mn@doi [\aap] {10.1051/0004-6361:20011817}, \href
  {http://adsabs.harvard.edu/abs/2002A%26A...385..337T} {385, 337}

\bibitem[\protect\citeauthoryear{{Trottenberg}, {Oosterlee}  \&
  {Scholler}}{{Trottenberg} et~al.}{2000}]{Trottenberg}
{Trottenberg} U.,  {Oosterlee} C.,   {Scholler} A.,  2000, {Multigrid}.
Academic Press

\bibitem[\protect\citeauthoryear{{Vainshtein}}{{Vainshtein}}{1972}]{Vainshtein1972PhLB...39..393V}
{Vainshtein} A.~I.,  1972, \mn@doi [Physics Letters B]
  {10.1016/0370-2693(72)90147-5}, \href
  {http://adsabs.harvard.edu/abs/1972PhLB...39..393V} {39, 393}

\bibitem[\protect\citeauthoryear{{Viel}, {Haehnelt}  \& {Springel}}{{Viel}
  et~al.}{2010}]{2010JCAP...06..015V}
{Viel} M.,  {Haehnelt} M.~G.,   {Springel} V.,  2010, \mn@doi [JCAP]
  {10.1088/1475-7516/2010/06/015}, \href
  {http://adsabs.harvard.edu/abs/2010JCAP...06..015V} {6, 15}

\bibitem[\protect\citeauthoryear{{Wesseling}}{{Wesseling}}{1992}]{Wesseling92}
{Wesseling} P.,  1992, {An Introduction to Multigrid Methods}.
John Wiley and Sons Inc, |c1992, 2nd ed.

\bibitem[\protect\citeauthoryear{Wilcox, Bacon, Nichol, Rooney, Terukina
  et~al.}{Wilcox et~al.}{2015}]{Wilcox:2015kna}
Wilcox H.,  Bacon D.,  Nichol R.~C.,  Rooney P.~J.,  Terukina A.,   et~al.,
  2015, arXiv:1504.03937

\bibitem[\protect\citeauthoryear{{Will}}{{Will}}{2014}]{2014LRR....17....4W}
{Will} C.~M.,  2014, \mn@doi [Living Reviews in Relativity]
  {10.12942/lrr-2014-4}, \href
  {http://adsabs.harvard.edu/abs/2014LRR....17....4W} {17, 4}

\bibitem[\protect\citeauthoryear{{Williams}, {Turyshev}  \& {Boggs}}{{Williams}
  et~al.}{2004}]{Williams2004PhRvL..93z1101W}
{Williams} J.~G.,  {Turyshev} S.~G.,   {Boggs} D.~H.,  2004, \mn@doi [\prl]
  {10.1103/PhysRevLett.93.261101}, \href
  {http://adsabs.harvard.edu/abs/2004PhRvL..93z1101W} {93, 261101}

\bibitem[\protect\citeauthoryear{{Winther} \& {Ferreira}}{{Winther} \&
  {Ferreira}}{2014}]{2014arXiv1403.6492W}
{Winther} H.~A.,  {Ferreira} P.~G.,  2014, preprint, \href
  {http://adsabs.harvard.edu/abs/2014arXiv1403.6492W} {} (\mn@eprint {arXiv}
  {1403.6492})

\bibitem[\protect\citeauthoryear{{Winther} \& {Ferreira}}{{Winther} \&
  {Ferreira}}{2015}]{2015arXiv150503539W}
{Winther} H.~A.,  {Ferreira} P.~G.,  2015, preprint, \href
  {http://adsabs.harvard.edu/abs/2015arXiv150503539W} {} (\mn@eprint {arXiv}
  {1505.03539})

\bibitem[\protect\citeauthoryear{{Winther}, {Mota}  \& {Li}}{{Winther}
  et~al.}{2012}]{2012ApJ...756..166W}
{Winther} H.~A.,  {Mota} D.~F.,   {Li} B.,  2012, \mn@doi [\apj]
  {10.1088/0004-637X/756/2/166}, \href
  {http://adsabs.harvard.edu/abs/2012ApJ...756..166W} {756, 166}

\bibitem[\protect\citeauthoryear{{Zhao}, {Li}  \& {Koyama}}{{Zhao}
  et~al.}{2011}]{2011PhRvD..83d4007Z}
{Zhao} G.-B.,  {Li} B.,   {Koyama} K.,  2011, \mn@doi [\prd]
  {10.1103/PhysRevD.83.044007}, \href
  {http://adsabs.harvard.edu/abs/2011PhRvD..83d4007Z} {83, 044007}

\bibitem[\protect\citeauthoryear{Zu, Weinberg, Jennings, Li  \& Wyman}{Zu
  et~al.}{2014}]{Zu:2013joa}
Zu Y.,  Weinberg D.,  Jennings E.,  Li B.,   Wyman M.,  2014, \mn@doi
  [Mon.Not.Roy.Astron.Soc.] {10.1093/mnras/stu1739}, 445, 1885

\makeatother
\end{thebibliography}
\bsp

\label{lastpage}
\end{document}